\documentclass[prb,showpacs,twocolumn,floats,10pt,aps,citeautoscript,longbibliography,superscriptaddress,nofootinbib]{revtex4-2}

\usepackage[dvipsnames]{xcolor}
\usepackage{tabularx}
\newcolumntype{Y}{>{\centering\arraybackslash}X}

\newcommand{\Eq}[1]{Eq.~\eqref{#1}}
\newcommand{\Eqs}[1]{Eqs.~\eqref{#1}}
\newcommand{\Equ}[1]{Equation~\eqref{#1}}
\newcommand{\Equs}[1]{Equations~\eqref{#1}}
\newcommand{\Sec}[1]{Sec.~\ref{#1}}

\newcommand{\Fig}[1]{Fig.~\ref{#1}}
\newcommand{\Figs}[1]{Figs.~\ref{#1}}

\newcommand{\mr}[1]{ \mathrm{#1} }
\newcommand{\mc}[1]{ \mathcal{#1} }

\newcommand{\nint}{\!\int\!} % narrow integral without boundaries
\newcommand{\nbint}[2]{\!\int_{#1}^{#2}\!\!} % narrow integral with boundaries
\newcommand{\nsint}[1]{\!\int_{#1}\!\!} % narrow integral with subscript
\newcommand{\pdag}{{\vphantom{dagger}}}  % phantom dagger, 
\usepackage{oubraces}
\usepackage{bm,bbm,mathbbol}
\usepackage{physics}
\usepackage{mathtools}
\usepackage{amsmath, amsthm, amssymb, amsfonts}
\usepackage{bbold}
\usepackage{esvect}
\usepackage{slashed}

\usepackage{epsfig}
\usepackage{array}
\usepackage{multirow}
\usepackage{graphicx}
\usepackage[normalem]{ulem}

\usepackage{float}
\usepackage[section]{placeins}
\usepackage{afterpage}

\allowdisplaybreaks  % allow displayed equations to be pagebroken, set by jvd

\usepackage{lipsum}

\usepackage{xifthen}
\usepackage[export]{adjustbox}
\usepackage{cancel}
\usepackage[colorlinks=true,citecolor=blue,linkcolor=blue,urlcolor=blue]{hyperref}

\usepackage{orcidlink}

\makeatletter
\newcommand\footnoteref[1]{\protected@xdef\@thefnmark{\ref{#1}}\@footnotemark}
\makeatother

\newcommand{\mb}[1]{{\bm{#1}}}

\newcommand{\nnnl}{\nonumber \\}

\newcommand{\veps}[0]{\varepsilon}
\newcommand{\ul}[1]{ \uline{#1} }

\newcommand{\one}[0]{\mathbbm{1}}
\newcommand{\opd}{d^\pdag}
\newcommand{\opdd}{d^\dagger}
\newcommand{\opq}{q^\pdag}
\newcommand{\opqd}{q^\dagger}
\newcommand{\opc}[0]{c^\pdag}
\newcommand{\opcd}[0]{c^\dagger}
\newcommand{\opp}[0]{\psi^\pdag}
\newcommand{\oppd}[0]{\psi^\dagger}
\newcommand{\opH}[0]{H}

\newcommand{\mcC}[0]{{\mathcal{C}}}
\newcommand{\mcG}[0]{{\mathcal{G}}}
\newcommand{\mcF}[0]{{\mathcal{F}}}
\newcommand{\mcK}[0]{{\mathcal{K}}}

\newcommand{\mcO}[0]{{\mathcal{O}}}
\newcommand{\mcT}[0]{{\mathcal{T}}}

\newcommand{\lp}[0]{{$\ell$p}}
\newcommand{\lpmath}[0]{{\ell\mathrm{p}}}
\newcommand{\rmv}[1]{{\slashed{#1}}}
\newcommand{\lpartial}[0]{{\overleftarrow{\partial}\!}}
\newcommand{\lpartialp}[0]{{\overleftarrow{\partial'}\!\!}}
\newcommand{\rmvsum}[2]{{\sum_{#1}}^{\rmv{#2}}}

\newcommand{\ccdots}{{{\cdot}{\cdot}{\cdot}}} % more compact cdots
\newcommand{\mmmm}{{{-}{-}{-}{-}}} % uniform spacing
\newcommand{\stkout}[1]{\ifmmode\text{\sout{\ensuremath{#1}}}\else\sout{#1}\fi}
\newcommand{\Bigexpval}[1]{{\Bigl\langle {#1} \Bigr\rangle}}
\newcommand{\smallexpval}[1]{{\langle {#1} \rangle}}

\usepackage{scalefnt}  %to scale font sizes arbirarily
\newcommand{\medfrac}[2]{{\scalefont{0.8} \mbox{${\displaystyle \frac{#1}{#2}}$}}}

\begin{document}

\title{Symmetric improved estimators for multipoint vertex functions}

\author{Jae-Mo~Lihm\,\orcidlink{0000-0003-0900-0405}}
\affiliation{Department of Physics and Astronomy, Seoul National University, Seoul 08826, Korea}
\affiliation{Center for Correlated Electron Systems, Institute for Basic Science, Seoul 08826, Korea}
\affiliation{Center for Theoretical Physics, Seoul National University, Seoul 08826, Korea}
\author{Johannes~Halbinger\,\orcidlink{0000-0002-6286-2736}}
\affiliation{Arnold Sommerfeld Center for Theoretical Physics, 
Center for NanoScience,\looseness=-1\,  and 
Munich Center for \\ Quantum Science and Technology,
Ludwig-Maximilians-Universit\"at M\"unchen, 80333 Munich, Germany}
\author{Jeongmin~Shim\,\orcidlink{0000-0002-0589-0001}}
\affiliation{Arnold Sommerfeld Center for Theoretical Physics, 
Center for NanoScience,\looseness=-1\,  and 
Munich Center for \\ Quantum Science and Technology,
Ludwig-Maximilians-Universit\"at M\"unchen, 80333 Munich, Germany}
\author{Jan~von~Delft\,\orcidlink{0000-0002-8655-0999}}
\affiliation{Arnold Sommerfeld Center for Theoretical Physics, 
Center for NanoScience,\looseness=-1\,  and 
Munich Center for \\ Quantum Science and Technology,
Ludwig-Maximilians-Universit\"at M\"unchen, 80333 Munich, Germany}
\author{Fabian~B.~Kugler\,\orcidlink{0000-0002-3108-6607}}
\thanks{These authors contributed equally.}
\affiliation{Center for Computational Quantum Physics, Flatiron Institute, 162 5th Avenue, New York, NY 10010, USA}
\author{Seung-Sup~B.~Lee\,\orcidlink{0000-0003-0715-5964}}
\thanks{These authors contributed equally.}
\affiliation{Department of Physics and Astronomy, Seoul National University, Seoul 08826, Korea}
\affiliation{Center for Theoretical Physics, Seoul National University, Seoul 08826, Korea}

\date{\today}

\begin{abstract}
Multipoint vertex functions, and the four-point vertex in particular, are crucial ingredients in many-body theory.
Recent years have seen significant algorithmic progress toward numerically computing their dependence on multiple frequency arguments.
However, such computations remain challenging and are prone to suffer from numerical artifacts, especially in the real-frequency domain.
Here, we derive estimators for multipoint vertices that are numerically more robust than those previously available.
We show that the two central steps for extracting vertices from correlators, namely the subtraction of disconnected contributions and the amputation of external legs, can be achieved accurately through repeated application of equations of motion, in a manner that is symmetric with respect to all frequency arguments and involves only fully renormalized objects.
The symmetric estimators express the core part of the vertex and all asymptotic contributions through separate expressions that can be computed independently, without subtracting the large-frequency limits of various terms with different asymptotic behaviors.
Our strategy is general and applies equally to the Matsubara formalism, the real-frequency zero-temperature formalism, and the Keldysh formalism.
We demonstrate the advantages of the symmetric improved estimators by computing the Keldysh four-point vertex of the single-impurity Anderson model using the numerical renormalization group.
\end{abstract}

\maketitle

% =======================================================================================

\section{Introduction}
Two-particle correlators and vertices play a crucial role in many-body physics. They encode the effective interaction between two particles, altered from their bare value due to the many-body environment.
Understanding and calculating two-particle correlators is essential for studying collective modes, instabilities, and response properties.
The two-particle or four-point (4p) vertex is also a key ingredient for extending methods based on quantum impurity models,
like dynamical mean-field theory (DMFT)~\cite{1996GeorgesRMP,2006KotliarRMP}, to treat nonlocal correlations~\cite{2018RohringerRMP}.
Owing to the dependence of the 4p vertex on multiple frequency, spin, and orbital degrees of freedom, analytic treatments are limited to only the simplest models~\cite{2016Ribic, 2018Thunstrom}.
Thus, there has been a longstanding interest
in developing efficient and accurate computational methods for evaluating these quantities~\cite{2012Rohringer}.
Indeed, recent years have brought significant algorithmic progress toward numerically computing the dependence of multipoint functions on their multiple frequency arguments, 
using, e.g., quantum Monte Carlo (QMC)~\cite{2011Kunes,2012HafermannEstimator,2013Schafer,2019KaufmannEstimator}
or the numerical renormalization group (NRG) for solving quantum impurity models~\cite{2008BullaRMP,2021KuglerPRX,2021LeePRX}.

The present paper addresses the following question: given a reliable numerical method for computing multipoint correlators in the frequency domain, such as QMC or NRG,
how can it best be harnessed to extract the corresponding vertex?
This extraction involves subtracting disconnected parts and amputating external legs.
Naive implementations of such subtractions and amputations are prone to numerical artifacts.
To minimize their effects, various improved estimators have been proposed.
Such estimators are expressions for the quantity of interest (e.g.\ a self-energy or vertex) that are formally equivalent to the original definition but more robust against numerical artifacts~\cite{1992Jarrell,1983SweenyEstimator,1988WolffEstimator,1995BietenholzEstimator,1998BullaEstimator,1998AmmonEstimator,1999AssarafEstimator,2000MishchenkoEstimator}.
Such artifacts are, e.g., statistical noise in QMC or discretization effects in NRG. 

A fruitful approach for deriving improved estimators is to utilize equations of motion (EOMs).
In 1998, Bulla, Hewson, and Pruschke~\cite{1998BullaEstimator} used EOMs to derive an improved estimator for the self-energy of quantum impurity models.
This estimator is constructed from the usual 2p propagator and an auxiliary 2p correlator involving a certain composite operator generated by the EOM.
It is an asymmetric improved estimator (aIE), since it was derived via an EOM acting on only one of the two time arguments of the propagator.
The resulting aIE for the self-energy has been widely used for NRG computations ever since.

The EOM strategy of Ref.~\cite{1998BullaEstimator} was generalized to the case of 4p vertices by Hafermann, Patton, and Werner~\cite{2012HafermannEstimator}.
They derived an aIE for the 4p vertex that contains an additional, 4p auxiliary correlator, again involving a composite operator.
The terms are then combined in such a manner that the disconnected parts cancel and one external leg is amputated.
Their aIE is asymmetric in the frequency arguments, since it was derived via EOMs acting on only one of the four time arguments.
For some applications, this is a serious limitation.
An example is the 4p vertex in the real-frequency Keldysh formalism.
The Keldysh vertex of the Anderson impurity model (AIM) was recently computed by three of the present authors using NRG~\cite{2021KuglerPRX,2021LeePRX}.  
There, the aIE of Ref.~\cite{2012HafermannEstimator} was used, but it was pointed out that this yields improvements for only 4 of the 16 components of the Keldysh vertices.
To improve all 16, a \textit{symmetric} improved estimator (sIE) is needed. 

A sIE was derived through repeated use of EOMs by Kaufmann, Gunacker, Kowalski, Sangiovanni, and Held~\cite{2019KaufmannEstimator}, and found to be significantly less prone to numerical artifacts than the aIE of Ref.~\cite{2012HafermannEstimator}.
Yet, their sIE involves not only various full (interacting) multipoint correlators, but also the bare (noninteracting) 2p propagator.
It was noted before \cite{1998BullaEstimator,2021LeePRX,2022KuglerEstimator} that this is not ideal for methods where bare and full correlators stem from different numerical settings and would compromise the accuracy of some intended cancellations.
NRG is an example of such a method: there, bare and full correlators are typically computed without or with energy discretization, respectively.
Consequently, the sIE of Ref.~\cite{2019KaufmannEstimator} was not used in the NRG computations of Refs.~\cite{2021KuglerPRX,2021LeePRX}.
The occurrence of bare propagators is also unfavorable in scenarios where they qualitatively differ very strongly from the full ones, as, e.g., in a Mott insulating state.

For the self-energy, one of us recently derived an improved estimator that is (i) symmetric in all operators and (ii) involves only fully renormalized correlators.
The combination of both properties sets Kugler's sIE~\cite{2022KuglerEstimator} apart from previous results ((i) from Ref.~\cite{1998BullaEstimator} and (ii) from Ref.~\cite{2019KaufmannEstimator}).
The aforementioned literature on improved estimators is summarized in Tab.~\ref{tab:literature}.

\begin{table}
\centering
\begin{tabular}{c|cc}
\hline
& 2p self-energy $\Sigma$ & 4p vertex $\Gamma$ \\
\hline
asymmetric, \textbf{full} & \citet{1998BullaEstimator}  &  $\quad$\citet{2012HafermannEstimator}
\\
\textbf{symmetric}, bare & 
\multicolumn{2}{c}{\citet{2019KaufmannEstimator}}
\\
\textbf{symmetric}, \textbf{full} & \citet{2022KuglerEstimator} & \textbf{this work}
\\
\hline
\end{tabular}
\caption{
Summary of different improved estimators for self-energies and vertex functions derived using EOMs.
}
\label{tab:literature}
\end{table}

In the present paper, we generalize Kugler's approach to derive sIEs for multipoint vertices.
For these, properties (i) and (ii) are particularly useful, since the division by full propagators is required for the amputation of external legs. 
We also show how the asymptotic and core contributions to the vertex can be isolated and computed separately via estimators of their own, all expressed through combinations of auxiliary correlators involving composite operators.
Asymptotic contributions depend on only a subset of all frequency arguments and remain finite if the complementary frequencies are sent to infinity; the core contribution, by contrast, depends on all frequency arguments but decays in all directions.
Separate estimators for asymptotic and core contributions are numerically advantageous, as they directly yield the desired quantities, without the need for subtracting the large-frequency limits of various terms with different asymptotic behaviors.

\begin{figure}[t]
    \centering
    \includegraphics[width=1.0\columnwidth]{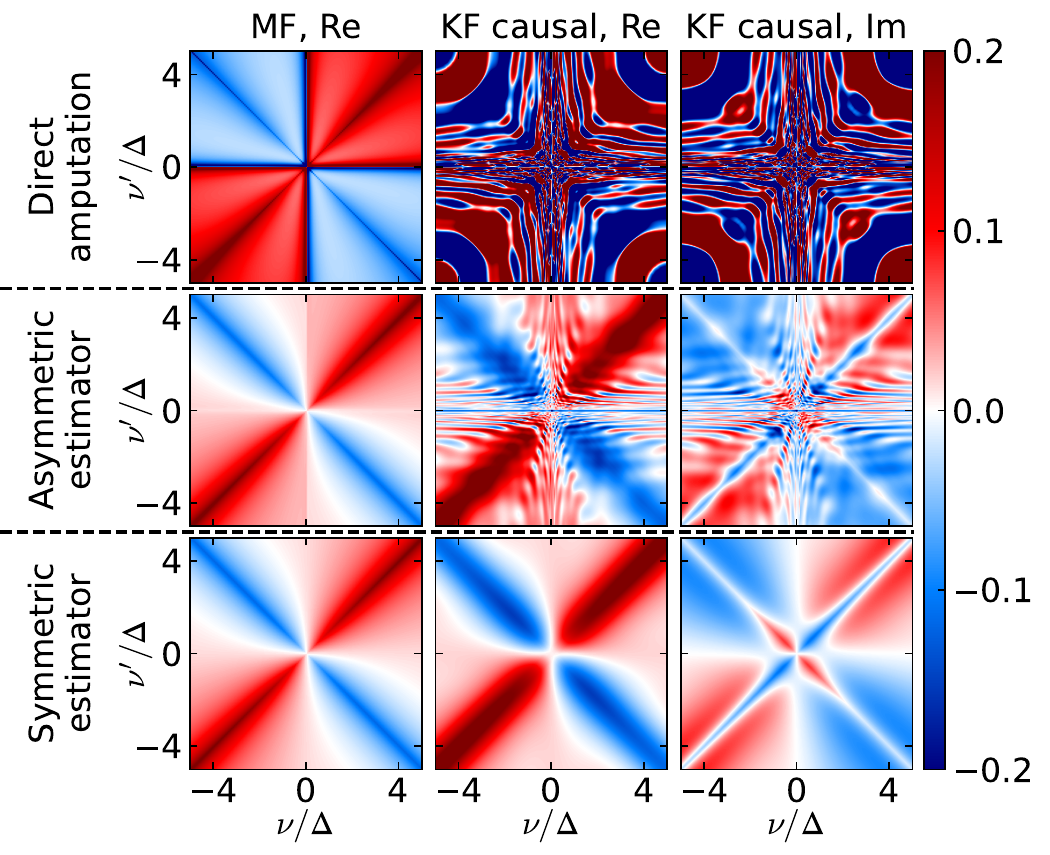}
    \caption{%
    MF and KF 4p vertices $U^{-1}[\Gamma_{\uparrow \downarrow} - \Gamma_{\mathrm{bare;} \uparrow \downarrow}](\nu, \nu', \omega=0)$ of the AIM at weak interaction.
    Here, $\nu$, $\nu'$ are fermionic frequencies, $\omega$ a bosonic transfer frequency, $U$ is the interaction strength and $\Delta$ the hybridization strength, chosen the same as in \Fig{fig:weak_vertex_overview}. [See \Eq{eq:ph_representation} for a concrete definition of $\Gamma$.]
    The three rows show the results obtained by direct amputation of the correlator, by using an asymmetric improved estimator (aIE)~\cite{2012HafermannEstimator,2021KuglerPRX,2021LeePRX}, and by using the symmetric improved estimator (sIE) derived in this work.
    The first, second, and third columns show the MF vertex (which is purely real), the real part of the causal ($\mb{c}=\mmmm$) component of the KF vertex, and its imaginary part, respectively.
    }
    \label{fig:weak_direct_vs_asym_vs_sym}
\end{figure}

Our derivation utilizes the framework and notation for multipoint correlators developed in Ref.~\cite{2021KuglerPRX}. 
There, as here, the overall strategy of the derivation applies equally (modulo some technical differences) in all three of the commonly used many-body frameworks: the imaginary-frequency Matsubara formalism (MF), the real-frequency zero-temperature formalism (ZF), and the real-frequency Keldysh formalism (KF).
We illustrate the utility of our sIE for vertices by NRG computations of the 4p vertex of the AIM.
We find dramatic improvements relative to results obtained using direct amputation or an aIE.
Typical examples of such improvements are shown in \Fig{fig:weak_direct_vs_asym_vs_sym}, serving as a preview for results presented later on.
 
The rest of the paper is organized as follows.
In Sec.~\ref{sec:2p_review}, we set the scene by concisely recapitulating the derivation of the symmetric self-energy estimator~\cite{2022KuglerEstimator} in the MF. 
In Sec.~\ref{sec:eom}, we formulate EOMs for multipoint correlators in the MF, ZF, and KF.
In Sec.~\ref{sec:estimator}, we use these EOMs to derive sIEs for the self-energy, the 3p and the 4p vertices, and discuss why the proposed estimators are expected to be more robust against numerical artifacts.
We present numerical results for the Keldysh 4p vertex of the AIM in Sec.~\ref{sec:results} and conclude with an outlook in Sec.~\ref{sec:conclusion}.

% ======================================================
\section{Symmetric self-energy estimator in the Matsubara formalism} \label{sec:2p_review}

To provide context, this section reviews the derivation of the symmetric self-energy estimator presented in Ref.~\cite{2022KuglerEstimator}.
For concreteness, we do so in the MF, for a very simple quantum impurity model, 
the AIM.
We first define 2p correlators (\Sec{sec:Definition2pCorrelators}) and derive general EOMs for them (\Sec{sec:GeneralEOM2p}).
We then specialize to the AIM, define various auxiliary correlators (\Sec{sec:AuxiliaryCorrelatorsAIM}), and finally derive a symmetric self-energy estimator (\Sec{sec:self-energy-estimator}).

\subsection{Definition of 2p correlators} \label{sec:Definition2pCorrelators}
We begin by introducing some notation.
We write
\begin{equation}
    [\mcO^1\!, \mcO^2]_{\zeta} = \mcO^1 \mcO^2  - \zeta \mcO^2 \mcO^1 ,
\end{equation}
with $\zeta=1$ for commutators or $\zeta = -1$ for anticommutators of two operators.
Given a Hamiltonian $H$, thermal expectation values at temperature $1/\beta$ are defined as 
\begin{align}
\label{eq:ThermalAverage}
\expval{\mcO^m} & = \frac{\Tr(\mcO^m e^{-\beta \opH})}{\Tr e^{-\beta \opH}} 
\end{align}
and Heisenberg time evolution in imaginary time as
\begin{align} \label{eq:Heisenberg_def}
    \mcO^m(m) & = e^{i \opH z_m} \mcO^m  e^{-i \opH z_m}.
\end{align}
Here, $(m) \!=\! (z_m) \!=\! (-i \tau_m)$, with $\tau_m$ real,
is a short\-hand for the MF imaginary time argument (this notation ensures consistency with ZF and KF formulas later, where
$z_m = t_m$).
A MF 2p correlator of operators $\mcO^1$ and $\mcO^2$ at
times $(1,2) =  (-i\tau_1, -i \tau_2) = -i \mb{\tau}$ is defined as
\begin{align}
\label{eq:2p_correlator_def}
\mcG[\mcO^1\!,\mcO^2](1,2)
= -i \expval{\mcT \bigl[\mcO^1\!, \mcO^2\bigr] (1,2)} 
.
\end{align}
Here, $\mcT$ denotes $\tau$ ordering,
\begin{align}
\mcT \bigl[\mcO^1\!, \mcO^2\bigr] (1,2) 
& = \theta(\tau_1 \! - \! \tau_2) \expval{\mcO^1(1)\mcO^2(2)} 
\nonumber \\
& \quad + \zeta^2_1 \theta(\tau_2 \! - \! \tau_1) \expval{\mcO^2(2)\mcO^1(1)} ,  
\label{eq:2pTimeOrdering}
\end{align}
and $\zeta_1^2=\zeta_2^1$ is the sign arising when permuting $\mcO^1$ past $\mcO^2$: $\zeta_1^2=-1$ if both are fermionic, $\zeta_1^2 = +1$ otherwise.

The corresponding transformation to the Matsubara frequency domain is
\begin{subequations}
\begin{align}
    \mcG[\mcO^1\!,\mcO^2](i\mb{\omega})
    & = -i \nbint{0}{\beta} d^2\tau \, e^{i\mb{\omega} \cdot \mb{\tau}} \mcG[\mcO^1\!,\mcO^2](-i\mb{\tau})
    \\ & =
    \beta \delta_{\omega_{12},0} \, G[\mcO^1\!,\mcO^2](i\mb{\omega}),
\end{align}
\end{subequations}
with $\omega_{12} \!=\! \omega_1 \!+\! \omega_2$. Here, time-translational invariance was exploited to factor out a Kronecker delta expressing energy conservation, $\omega_{12} = 0$. 
We take this constraint to be implicitly understood for the frequency arguments of $G[\mcO^1,\mcO^2](i\mb{\omega})$ and thus omit the second one, writing
\begin{equation} \label{eq:green_2p_with_1_frequency}
     G[\mcO^1\!,\mcO^2](i\omega, -i\omega) = G[\mcO^1\!,\mcO^2](i\omega) .
\end{equation}
For brevity, we will often omit the operator arguments  $[\mcO^1\!,\mcO^2]$ when they can be inferred from the context.

By analogy, an equilibrium expectation value may be viewed as
a (constant) 1p function, $\mcG[\mcO^1](1)  = \expval{\mcO^1}$.
Its Matsubara transform, 
\begin{align}
    \mcG[\mcO^1](i\omega) & = \nbint{0}{\beta} d \tau \,
    e^{i \omega \tau} \mcG[\mcO^1] (-i \tau) = \beta \delta_{\omega,0} \, G[\mcO^1] , 
\end{align}
is nonzero only for zero frequency, with $G[\mcO^1] = \expval{\mcO^1}$ being independent of frequency.

\subsection{General EOMs for 2p correlators}
\label{sec:GeneralEOM2p}
Next, we recall the derivation of EOMs for 2p correlators. 
We write derivatives as $\partial_m =  \partial/\partial z_m = 
\partial/\partial (-i \tau_m)$, and $\delta(1,2)= 
\delta(z_1 - z_2) = i\delta(\tau_1-\tau_2)$ for delta functions,
such that $ \partial_1 \theta(\tau_1-\tau_2) = - \partial_2 \theta(\tau_1-\tau_2) = \delta(1,2)$.

The derivatives of a time-ordered product read
\begin{subequations}
\begin{align} 
    \label{eq:EOM-O1-O2a}
    \partial_{1} \mcT \bigl[ \mcO^1 \! , \mcO^2 \bigr] (1,2)
    &= \mcT \bigl[  \partial_1 \mcO^1 \!, \mcO^2 \bigr] (1,2) \nnnl
    & \quad + \delta(1, 2) \, \comm{\mcO^1}{\mcO^2}_{\zeta_1^2}(2),
    \\
    \label{eq:EOM-O1-O2b}
    \partial_{2} \mcT \bigl[ \mcO^1 \! , \mcO^2 \bigr] (1,2)
    &= \mcT \bigl[  \mcO^1 \!,  \partial_2 \mcO^2 \bigr] (1,2)  \nnnl
    & \quad +  \delta(2,1) \, {\zeta_1^2} \, \comm{\mcO^2}{\mcO^1}_{\zeta_1^2}(1) .
\end{align}
\end{subequations}
Here, $\delta(1, 2)$ arises from differentiating the time ordering step functions. 
For the last term of \Eq{eq:EOM-O1-O2b}, we used 
$\mcT \bigl[\mcO^1(1) \mcO^2(2) \bigr] = \zeta^2_1 \mcT \bigl[\mcO^2(2) \mcO^1(1) \bigr]$ to 
move $\mcO^2(2)$ to the left of $\mcO^1(1)$ within $\mcT[ \mb{\cdot} ]$ before differentiating the step functions. As a result, the (anti)commutator obtained from $\partial_2$  is ``flipped'' relative to that from $\partial_1$ and multiplied by an extra $\zeta^2_1$.
A similar feature will occur later on in our discussion of multipoint EOMs.

Using the EOM for Heisenberg operators ($m=1,2$),
\begin{equation} \label{eq:eom_heisenberg}
    i \partial_m \mcO^m(m) = \comm{\mcO^m(m)}{H} ,
\end{equation}
we obtain the following EOMs for 2p correlators:
\begin{subequations} \label{eq:2p_eom_general}
\begin{align}
    i\partial_1 \mcG[\mcO^1\!, \mcO^2](1, 2)
    &= \mcG\bigl[\comm{\mcO^1\!}{H},\mcO^{2} \bigr](1, 2) \\
    \nonumber
    & \quad + \delta(1, 2) \, \mcG\bigl[ \comm{\mcO^1}{\mcO^2}_{\zeta_1^2} \bigr](2),
    \\
    i\partial_2 \mcG[\mcO^1\!, \mcO^2](1, 2)
    &= \mcG\bigl[\mcO^{1}\!, \comm{\mcO^2\!}{H} \bigr](1, 2) \\
    & \quad + \delta(2,1) \, \zeta_1^2 
    \mcG\bigl[ \comm{\mcO^2}{\mcO^1}_{\zeta_1^2} \bigr](1).
    \hspace{0.5cm} \nonumber
\end{align}
\end{subequations}
Their Matsubara Fourier transforms read
\begin{subequations}
\begin{align}
    \label{eq:2p_eom_1} 
    i\omega G[\mcO^1\!, \mcO^2](i\omega)
    &= G\bigl[\comm{\mcO^1\!}{H},\mcO^{2} \bigr] (i\omega) \\
\nonumber 
    & \quad + \big\langle \comm{\mcO^1\!}{\mcO^2}_{\zeta_1^2} \big\rangle, 
     \\ 
    \label{eq:2p_eom_2}
    -i\omega G[\mcO^1\!, \mcO^2](i\omega)
    &= G\bigl[\mcO^{1}\!,\comm{\mcO^2\!}{H} \bigr](i\omega) \\
    \nonumber
    & \quad + \zeta_1^2 \big\langle \comm{\mcO^2\!}{\mcO^1}_{\zeta_1^2} \big\rangle.
    \end{align} 
\label{eq:2p_eom}
\end{subequations}
These two general EOMs, obtained by differentiating $\mcG(1,2)$ using $\partial_1$ or $\partial_2$, will be used repeatedly below.

\subsection{Full, bare, and auxiliary correlators of the AIM}
\label{sec:AuxiliaryCorrelatorsAIM}

For concreteness, we frame the following discussion within the context of an SU(2)-symmetric AIM.
Its Hamiltonian has the form $\opH = \opH^0 + \opH_{\rm int}$, with
\begin{align} \label{eq:2p_AIM}
    \opH^0 & = \sum_\sigma \varepsilon_d \opdd_\sigma \opd_\sigma
    + \sum_{\sigma b} \varepsilon_{b} \opcd_{b\sigma} \opc_{b\sigma}
    + \sum_{\sigma b}  (V_b \opdd_\sigma \opc_{b\sigma} + {\rm H.c.}), 
    \nonumber 
    \\
    \opH_{\rm int} & = U n_{\uparrow} n_{\downarrow}, 
    \quad n_\sigma = \opdd_\sigma \opd_\sigma \, ,
\end{align}
with $\sigma \in \{ \uparrow, \downarrow \}$.
$H$ describes a two-flavor impurity with impurity operators $\opd_{\sigma}, \opdd_{\sigma}$ with an energy $\varepsilon_d$
experiencing a local, flavor-off-diagonal interaction $U$, and hybridizing with a two-flavor bath with bath operators $\opc_{b\sigma}, \opcd_{b\sigma}$ ($b = 1, \dots, N_\mathrm{bath})$ and energies $\varepsilon_b$.
We focus on the conventional AIM where these single-particle operators are all fermionic ($\zeta = -1$) and the flavor corresponds to the spin, but we keep track of the sign factor $\zeta$ for generality.

We denote the 2p correlator of $\opd$ and $\opdd$ by
\begin{equation}
    g(1,2) =  \mcG[\opd, \opdd](1,2) , \quad 
    g(i\omega) = G[\opd, \opdd](i\omega)
\end{equation}
and call it the ``propagator'', in distinction to other 2p correlators encountered below.
It is flavor-diagonal, hence we omit flavor indices.
The full propagator $g(i\omega)$, its bare ($U=0$) version $g^0(i \omega)$, and the self-energy $\Sigma(i\omega)$ satisfy the Dyson equation
\begin{equation} \label{eq:2p_selfen_dyson}
    g - g^0 = g^0 \Sigma g = g \Sigma g^0.
\end{equation}
The bare propagator can be obtained by setting up the EOMs for $g^0$ and $G[\opd,\opcd_b]$ and eliminating the latter (``integrating out the bath''). 
As shown below, one finds
\begin{equation} \label{eq:2p_green_hyb}
    g^0(i\omega) = \frac{1}{i\omega - \varepsilon_d - \Delta(i\omega)},\quad
    \Delta(i\omega) = \sum_b \frac{\abs{V_b}^2}{i\omega - \varepsilon_b} .
\end{equation}
The hybridization function $\Delta(i\omega)$ fully characterizes the impurity-bath coupling.

Next, we consider the EOMs for the full $g$.
When setting them up using \Eq{eq:2p_eom}, the equal-time commutators $[\opd_\sigma,H_{\rm int}]$ and $[\opdd_\sigma,H_{\rm int}]$ yield ``composite operators'' which we denote as follows, for short:
\begin{subequations}
\begin{align}
    \label{eq:q_def}
    \opq_\sigma & = \big[ \opd_\sigma, \opH_{\rm int} \big],
    \quad \opqd_\sigma = \big[ \opH_{\rm int}, \opdd_\sigma \big], \\
     \label{eq:q2_def}
     \opq_{\sigma\sigma'}
        &= \big[\opq_\sigma, \opdd_{\sigma'}\big]_{\zeta}
        = \big[\big[ \opd_\sigma, \opH_{\rm int} \big], \opdd_{\sigma'}\big]_{\zeta} \\
        &= \big[\opd_\sigma, \opqd_{\sigma'}\big]_{\zeta}
        = \big[\opd_\sigma, \big[ \opH_{\rm int}, \opdd_{\sigma'} \big] \big]_{\zeta} .
        \label{eq:q2_defc}
\end{align}
\end{subequations}
Here, \Eq{eq:q2_defc} equals \eqref{eq:q2_def} due to 
the identity
\begin{equation}
\nonumber
    \big[\big[ \mcO^1\!, \mcO^2\big], \mcO^3\big]_{\zeta}
    = \big[\mcO^1\!, \big[ \mcO^2\!, \mcO^3\big] \big]_{\zeta} + 
    \big[\big[ \mcO^1\!, \mcO^3\big]_\zeta, \mcO^2 \big].
\end{equation}
The composite operator $\opq_{\sigma\sigma'}$ carries the composite index $\sigma\sigma'$ but has just a single time argument.
For the AIM [\Eq{eq:2p_AIM}], the composite operators take the form
\begin{subequations}
\begin{align}
    \label{eq:q_AIM}
    q^{(\dagger)}_\sigma & = U d^{(\dagger)}_{\sigma} n_{-\sigma}, \\
    \label{eq:q2_AIM}
    \opq_{\sigma \sigma'} & = U \left( \delta_{\sigma,\sigma'} n_{-\sigma} + \zeta \delta_{\sigma,-\sigma'} \opdd_{-\sigma} \opd_\sigma \right).
\end{align}
\end{subequations}
When discussing multipoint correlators later on, we will encounter further composite operators, defined via multiple equal-time commutators and labeled by longer composite indices.
All correlators involving at least one composite operator will be called ``auxiliary correlators''.

\begin{figure}[t]
    \centering
    \includegraphics[width=1.0\columnwidth]{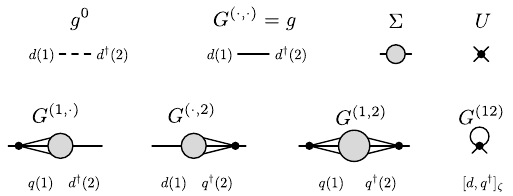}
    \caption{%
    Diagrammatic representations of the bare and full propagators
    $g^0$ and $g$, the self-energy $\Sigma$, the bare vertex $U$ for a quartic interaction, and the auxiliary correlators of \Eqs{eq:2p_clamped} (and \Eqs{eq:est_clamped_2p} in Sec.~\ref{sec:AuxilaryCorrelators}).
    A grey circle with $\ell$ lines attached represents an \lp\ correlator of $\ell$ elementary $\opd$ and $\opdd$ operators.
    Long legs represent $g$'s that can be amputated, and short legs indicate completed amputations.
    An \lp\ vertex is obtained from a grey circle with $\ell$ long legs by subtracting all disconnected parts and amputating all $\ell$ legs (see Fig.~\ref{fig:def_boxes}).}
    \label{fig:def_clamp_2p}
\end{figure}

Below, we need the propagator $g$, three auxiliary 2p correlators and one auxiliary 1p correlator, defined as follows and depicted diagrammatically in Fig.~\ref{fig:def_clamp_2p}:
\begin{subequations} \label{eq:2p_clamped}
\begin{align}
    G^{(\cdot,\cdot)}(i\omega) &= G[\opd_\sigma, \opdd_\sigma](i\omega) = g(i\omega),
    \\
    G^{(1,\cdot)}(i\omega) &= G[\opq_\sigma, \opdd_\sigma](i\omega), \label{eq:2p_clamped_2p_1}
    \\
    G^{(\cdot,2)}(i\omega) &= G[\opd_\sigma, \opqd_\sigma](i\omega),
    \\
    G^{(1,2)}(i\omega) &= G[\opq_\sigma, \opqd_\sigma](i\omega), 
    \\
    G^{(12)}
    & = G[\opq_{\sigma\sigma}]
    =\expval{\opq_{\sigma\sigma}} = U \expval{n_{-\sigma}} = \Sigma^{\rm H} \, . 
    \label{eq:2p_hartree}
\end{align}
\end{subequations}
The shorthand notation on the left distinguishes them 
using superscripts, with a single argument, 12, for the 1p correlator and two comma-separated arguments for 2p correlators. 
For the latter, `$\cdot$' serves as placeholder for $\opd$ or $\opdd$, while $1$ or $2$ signal their replacement by $\opq$ or $\opqd$, respectively.
The auxiliary correlators $G^{(1,\cdot)}$, $G^{(\cdot,2)}$, and $G^{(1,2)}$ are called $F^{\rm L}$, $F^{\rm R}$, and $I$ in Ref.~\cite{2022KuglerEstimator}.
For correlators diagonal in $\sigma$, as here, $G^{(1,\cdot)} \!=\! G^{(\cdot,2)}$ \cite{2022KuglerEstimator};
they are denoted $F$ in Refs.~\cite{1998BullaEstimator,2012HafermannEstimator}.
They are 2p correlators of single-particle and composite operators, but involve four single-particle operators since $\opq \!\sim \! \opd \opdd \opd$.
The 1p auxi\-liary correlator $G^{(12)}$ 
equals the Hartree self-energy $\Sigma^{\rm H}$; its diagrammatic representation in Fig.~\ref{fig:def_clamp_2p} reflects this fact.

\subsection{Self-energy estimators for the AIM}
\label{sec:self-energy-estimator}

Finally, we are ready to derive estimators for $\Sigma$, exploiting various EOMs.
These are obtained using the general EOMs, \Eq{eq:2p_eom},
and the commutators
\begin{equation}
    \comm{\opd_\sigma}{\opH} = \varepsilon_d d_\sigma + V_b \opc_{b\sigma} + \opq_\sigma, \quad
    \bigl[\opc_{b \sigma}, \opH\bigr] = \varepsilon_b \opc_{b\sigma} + V_b^* \opd_{\sigma}.
\end{equation}
We henceforth omit flavor subscripts $\sigma$ and frequency arguments $(i\omega)$.
Setting up the first general EOM, \Eq{eq:2p_eom_1}, for $G^{(\cdot,\cdot)}$ and $G[\opc_b, \opdd]$, we find
\begin{subequations} \label{eq:2p_eom_dc}
\begin{align}
    \label{eq:2p_eom_d}
    (i\omega - \varepsilon_d) G^{(\cdot,\cdot)}
    &= {\textstyle \sum_b} V_b G[\opc_b, \opdd] + G^{(1,\cdot)} + 1, \\
    \label{eq:2p_eom_c}
    (i\omega - \varepsilon_b) G[\opc_b, \opdd]
    &= V_b^* G^{(\cdot,\cdot)}.
\end{align}
\end{subequations}
By using \Eq{eq:2p_eom_c} to eliminate $G[\opc_b, \opdd]$ from \Eq{eq:2p_eom_d}, we obtain an EOM involving only impurity correlators,
\begin{subequations}
\begin{align} 
    \bigl(i\omega - \varepsilon_d - \Delta(i\omega)\bigr) G^{(\cdot,\cdot)}
   &  = G^{(1,\cdot)} + 1 , 
    \\
    G^{(\cdot,\cdot)}  \bigl(i\omega - \varepsilon_d - \Delta(i\omega)\bigr)
    & = G^{(\cdot, 2)} + 1.
\end{align}
\label{eq:2p_eom_G_long}
\end{subequations}
The second equation can be found analogously to the first, starting 
from the second general EOM, \eqref{eq:2p_eom_2}.
The bare $g^0 =  G^{(\cdot,\cdot)}|_{U=0}$ follows by setting 
$G^{(1,\cdot)}=G^{(\cdot,2)}=0$ in \Eq{eq:2p_eom_G_long}, yielding \Eq{eq:2p_green_hyb}.
Hence, the factors multiplying $ G^{(\cdot,\cdot)}$ on the left of \Eq{eq:2p_eom_G_long} equal $(g^0(i\omega))^{-1}$, implying
\begin{equation} \label{eq:2p_eom_G}
    g = G^{(\cdot,\cdot)} = g^0 \, ( G^{(1,\cdot)} + 1 ) 
    =  ( G^{(\cdot,2)} + 1 ) g^0 .
\end{equation}
One may also derive \Eq{eq:2p_eom_G} by writing \Eq{eq:2p_eom_dc} in the matrix form
\begin{flalign} \label{eq:2p_eom_matrix}
\!    \begin{pmatrix}
        i\omega \!-\! \varepsilon_d \!& -V_1 & -V_2 & \cdots \\
        -V_1^*  & \!i\omega \!-\! \varepsilon_1\!& 0 & \cdots \\
        -V_2^* & 0 & \!i\omega \!-\! \varepsilon_2 & \cdots \\
        \vdots & \vdots & \vdots & \ddots \\
    \end{pmatrix} \!\!
    \begin{pmatrix}
        G^{(\cdot,\cdot)} \\ \!G[\opc_1,\opdd]\! \\ \!G[\opc_2, \opdd] \!\\ \vdots
    \end{pmatrix}
    = \begin{pmatrix}
        G^{(1,\cdot)} \!+\! 1 \\ 0 \\ 0 \\ \vdots
    \end{pmatrix} \!\!. & 
\end{flalign}
The matrix on the left is the inverse bare propagator $i\omega - H^0$.
By inverting it using the block-matrix identity
\begin{equation}
    \begin{pmatrix} A & B \\ C & D \end{pmatrix}^{-1}
    = \begin{pmatrix} (A - BD^{-1}C)^{-1} & \cdots \\ \cdots & \cdots \end{pmatrix}
\end{equation}
and solving for the first element $G^{(\cdot, \cdot)}$, we find
\begin{align} \label{eq:2p_eom_matrix_result}
    G^{(\cdot,\cdot)}
    &= \big( i\omega - \varepsilon_d - \sum_b V_b \frac{1}{i\omega - \varepsilon_b} V_b^* \big)^{-1} (G^{(1,\cdot)} + 1) \nnnl
    &= \big( i\omega - \varepsilon_d - \Delta(i\omega) \big)^{-1} (G^{(1,\cdot)} + 1),
\end{align}
which equals \Eq{eq:2p_eom_G}.
Equating \Eq{eq:2p_eom_G} to the Dyson equation \eqref{eq:2p_selfen_dyson} and solving for $\Sigma$, we find the aIE for the self-energy first proposed in  Ref.~\cite{1998BullaEstimator}:
\begin{equation}
    \label{eq:2p_estimator_asym}
    \Sigma
    = G^{(1,\cdot)} g^{-1}
    = g^{-1} G^{(\cdot,2)}.
\end{equation}
This result corresponds to the Schwinger--Dyson equation for the self-energy.

\begin{figure}[t]
\centering
\includegraphics[width=1\columnwidth]{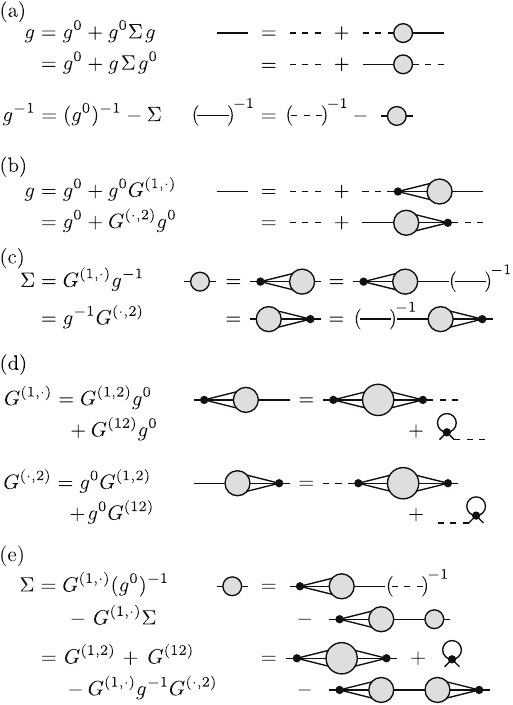}
\caption{%
Diagrammatic derivation of asymmetric and symmetric self-energy estimators.
(a) Inversion of the Dyson equation yields $g^{-1} = (g^0)^{-1} \! - \! \Sigma$.
(b) EOMs for $g$; by comparing their forms to the equations for $g$ shown
in (a), one identifies (c) aIEs for $\Sigma$, involving $g$-amputated auxiliary correlators, $G^{(1, \cdot)} g^{-1}$ or $g^{-1} G^{(\cdot,2)}$.
(d) $G^{(1, \cdot)}$ and  $G^{(\cdot,2)}$ satisfy EOMs themselves, involving the auxiliary correlators $G^{(1,2)}$ and $G^{(12)}$.
(e) Performing the $g$-amputation for $G^{(1,\cdot)} g^{-1}$ in (c) by using (a) for $g^{-1}$, one obtains terms containing $G^{(1,\cdot)} (g^0)^{-1}$; these are known from the EOMs from (d).
The last line gives the desired sIE for $\Sigma$.
Its first and third terms both contain one-particle-reducible contributions.
In their difference, however, such contributions cancel, ensuring that $\Sigma$ is one-particle irreducible (see Fig.~\ref{fig:pert_exp} for an example at order $\mcO(U^2)$). }
\label{fig:derivation_symm_Sigma_estimator}
\end{figure}

Next, we follow Ref.~\cite{2022KuglerEstimator} to obtain a sIE for $\Sigma$.
We use the second general EOM, \Eq{eq:2p_eom_2}, to obtain two more EOMs involving $G^{(1,\cdot)}$:
\begin{subequations}
\begin{align}
    & G^{(1,\cdot)} \; (i\omega - \varepsilon_d)
    = {\textstyle \sum_b} G[\opq, \opcd_b] \, V_b^* + G^{(1,2)} + G^{(12)}, \\
    & G[\opq, \opcd_b] \;  (i\omega - \varepsilon_b)
     = G^{(1,\cdot)}V_b.
\end{align}
\end{subequations}
The $G^{(12)}$ term, which is independent of $\omega$, comes from the last term of \Eq{eq:2p_eom_2}, involving an equal-time (anti)commutator that yields an expectation value, 
$\langle [d^\dagger_\sigma, q_\sigma]_\zeta\rangle = \langle q_{\sigma \sigma} \rangle$.
\begin{subequations}
\label{eq:2p_eom_G2}
We eliminate $G[\opq, \opcd_b]$ to obtain
\begin{equation} \label{eq:2p_eom_G2a}
    G^{(1,\cdot)} = ( G^{(1,2)} + G^{(12)} ) g^0 .
\end{equation}
In a similar manner, we obtain
\begin{equation} \label{eq:2p_eom_G2b}
    G^{(\cdot,2)} = g^0 ( G^{(1,2)} + G^{(12)} )  .
\end{equation}
\end{subequations}
Substituting $g^{-1} = (g^0)^{-1} - \Sigma$ in \Eq{eq:2p_estimator_asym}, we find
\begin{align} \label{eq:2p_estimator_sym}
    \Sigma
    &= G^{(1,\cdot)} [(g^0)^{-1} - \Sigma]  \nnnl
    &= G^{(1,2)} + G^{(12)} - G^{(1,\cdot)} \Sigma \nnnl
    &= G^{(1,2)} + \Sigma^{\rm H} - G^{(1,\cdot)} g^{-1} G^{(\cdot,2)}.
\end{align}
For the second equality, we used 
\Eq{eq:2p_eom_G2a} to 
eliminate $G^{(1,\cdot)}(g^0)^{-1}$.
For the third, we replaced $\Sigma$ by its aIE \eqref{eq:2p_estimator_asym} and used $G^{(12)} = \Sigma^{\rm H}$ [\Eq{eq:2p_hartree}].
\Equ{eq:2p_estimator_sym} is the sIE of Ref.~\cite{2022KuglerEstimator}.
It has the desirable properties of being (i) symmetric w.r.t.\ both frequency arguments and (ii) expressed purely through full correlators.

\begin{figure}[t]
    \centering
    \includegraphics[width=1.0\columnwidth]{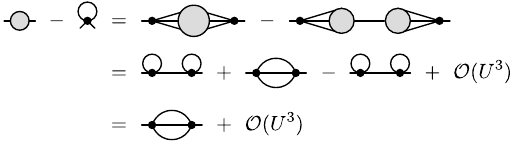}
    \caption{Expansion of the symmetric self-energy estimator to second order in the interaction $U$. The one-particle-reducible contributions to the diagrams on the right of the first line cancel, guaranteeing the one-particle irreducibility of $\Sigma$.}
    \label{fig:pert_exp}
\end{figure}

Figure~\ref{fig:derivation_symm_Sigma_estimator} diagrammatically summarizes the derivation of the sIE \eqref{eq:2p_estimator_sym} for the self-energy.
It rests on two key insights. 
First, the external $g$ legs of any $G^{(\ccdots)}$ 
(e.g.\ $G^{(1,\cdot)}$ or $G^{(\cdot,2)}$ in  \Fig{fig:derivation_symm_Sigma_estimator}) can be amputated 
through multiplication by  $g^{-1} = (g^0)^{-1} - \Sigma$; this
yields terms of the form $G^{(\ccdots)} (g^0)^{-1}$ or $(g^0)^{-1} G^{(\ccdots)}$. 
Second, each such product corresponds to the left side of an EOM; the right side of that EOM contains other auxiliary correlators and frequency-indepedent constants arising from (anti)commutators, thus $(g^0)^{-1}$ can be eliminated altogether. Moreover, this can be done in a fashion symmetric w.r.t.\ frequency arguments by combining EOMs derived using either $\partial_1$ or $\partial_2$.
The resulting symmetric self-energy estimator [\Fig{fig:derivation_symm_Sigma_estimator}(e)] is one-particle irreducible, as illustrated at order $\mcO(U^2)$ in \Fig{fig:pert_exp}.

In the following sections, our goal is to use a similar strategy to derive similar sIEs for general \lp\ functions, for arbitrary flavor indices and in any of the MF, ZF, and KF.
For the above discussion in the MF, we worked directly in the frequency domain.
In the KF, one cannot use the same strategy as the bare propagator is matrix-valued and not simply given by $i\omega - \varepsilon_d - \Delta$ as in \Eq{eq:2p_eom_G_long}.
A more fundamental reason is that, in the KF, the EOM does not fully determine the correlator.
For example, the $\mb{k}=(1,1)$ and $(2,2)$ components of the bare KF propagator obey the same EOM [see \Eq{eq:eom_kf_2p_example_k}], while the former is zero and the latter is not. In the KF, the boundary condition of the correlators [see \Eq{eq:KMS} in App.~\ref{sec:boundary}] needs to be explicitly used, whereas in the MF it was implicitly imposed through the structure of Matsubara frequencies.
We will henceforth work in the time domain, where this boundary condition is formulated.
To that end, we need a general, compact notation of multipoint correlators, which we introduce next.

% =======================================================================================
\section{EOMs for multipoint correlators} \label{sec:eom}

In this section, we generalize the discussion of the previous section from 2p to \lp\ correlators.
We define them in \Sec{sec:eom_correlator} and derive their general EOMs in \Sec{sec:GeneralEOMellp}.
We then write the Hamiltonian as the sum of noninteracting and interacting parts and derive EOMs for full (interacting) \lp\ correlators involving bare (noninteracting) propagators and auxiliary \lp\ correlators in \Sec{sec:FullBareAuxiliaryGeneral}.
We discuss how the noninteracting bath degrees of freedom can be integrated out if needed in \Sec{sec:eom_impurity}.
In \Sec{sec:eom_freq}, we adopt specific formalisms (MF, ZF, KF) and derive EOMs in the frequency domain.
In \Sec{sec:eom_connected}, we show that the obtained EOMs also hold for the connected part of the correlator.
Finally, in \Sec{sec:eom_interacting}, we derive an EOM that involves full propagators instead of the bare ones.

% ======================================================
\subsection{Definition of \lp\ functions}
\label{sec:eom_correlator}

\begin{figure}[t]
    \centering
    \includegraphics[width=1.0\columnwidth]{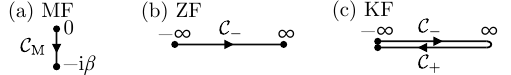}
    \caption{Time contour for each many-body formalism considered.
    }
    \label{fig:contours_simple}
\end{figure}

\begin{figure}[t]
\centering
\includegraphics[width=1.0\columnwidth]{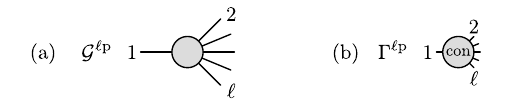}
\caption{%
Diagrammatic representations of
(a) the \lp\ correlator $\mc{G}^\lpmath$ and 
(b) the \lp\ vertex $\Gamma^\lpmath$, identified as the connected part of an \lp\ correlator after amputation of external legs.}
\label{fig:def_boxes}
\end{figure}

In the MF, ZF, and KF, \lp\ correlators of the list of operators $[\mb{\mc{O}}] = [\mc{O}{^1}, \ccdots,\mc{O}^\ell]$  are defined as 
\begin{subequations} \label{eq:green_def}
\begin{align}
    \label{eq:green_mf_expval}
    \mcG_{\rm M}[\mb{\mcO}](-i\mb{\tau})
    &= (-i)^{\ell - 1} \Bigexpval{ \mcT [\mb{\mcO}] (-i\mb{\tau}) } ,
    \\
    \label{eq:green_zf_expval}
    \mcG_{\rm Z}[\mb{\mcO}](\mb{t})
    &= (-i)^{\ell - 1} \Bigexpval{ \mcT [\mb{\mcO}] (\mb{t}) },
    \\
    \label{eq:green_kf_adia_expval}
    \mcG_{\rm K}^{\mb{c}}[\mb{\mcO}](\mb{t})
    &= (-i)^{\ell - 1} \Bigexpval{ \mcT [\mb{\mcO}] (\mb{t}^\mb{c}) } ,
\end{align}
\end{subequations}
respectively. Here, $(-i\mb{\tau}) = (-i\tau_1, \ccdots, -i\tau_\ell)$, and analogously for $\mb{c}$, $\mb{t}$ and $\mb{t}^\mb{c}$;
$\expval{A}$ denotes thermal averaging w.r.t.\ the full Hamiltonian $\opH$ according to \Eq{eq:ThermalAverage}; $\mcO(z)= e^{i \opH z} \mcO e^{-i \opH z}$ as in \Eq{eq:Heisenberg_def},  
and $\mcT$ denotes time ordering along one of the contours
\begin{subequations} \label{eq:contour_contours_simple}
\begin{alignat}{2}
&{\rm MF:}\ -i \tau_m \in\ && \mcC_{\rm M}= [0, -i\beta]
, \\
&{\rm ZF:}\qquad t_m \in\ &&  \mcC_- 
, \\
&{\rm KF:}\quad\ \ t_m^{c_m} \in\ && \mcC_- \oplus \, \mcC_+
,
\end{alignat}
\end{subequations}
shown in Fig.~\ref{fig:contours_simple}.
Here, $\mcC_- = [-\infty, \infty]$ and $\mcC_+ = [\infty, -\infty]$
are the forward and backward branches of the Keldysh contour.
In the KF, each contour variable is specified by a real-valued time argument and a contour index, i.e., $z = t^{c}$
with $c=-$ ($c=+$) for the forward (backward) branch, and
contour variables follow time ordering (anti-time ordering) on the forward (backward) branch.
In this work, we focus only on systems in the ground state (ZF) or thermal equilibrium (MF, KF), and do not consider the nonequilibrium KF.
We generically denote \lp\ correlators by $\mcG^\lpmath$, but suppress the superscript if the number of arguments is clear from the context. For a diagrammatic depiction, see Fig.~\ref{fig:def_boxes}(a).

To analyze the boundary conditions of the correlators, one must attach a vertical branch to the ZF and KF contours.
The only step where the boundary condition affects the results is when deducing the EOM in integral form from its differential form via integration by parts [\Eq{eq:eom_integral_long}].
This derivation is carried out in App.~\ref{sec:boundary}.

As in the 2p case, Fourier-transformed \lp\ correlators are defined by factoring out 
a delta function arising from time-translational invariance. For the MF, ZF, and KF, we have
\begin{flalign} 
    \mcG_{\rm M}(i\mb{\omega})
    & = i (-i)^{\ell} \!
     \nbint{0}{\beta} \! d^\ell \tau \, e^{i\mb{\omega}\cdot\mb{\tau}} \mcG_{\rm M}(-i\mb{\tau})
    = \beta \delta_{\omega_{1\ccdots\ell}, 0} G_{\rm M}(i\mb{\omega}),  &
    \nonumber \\
    \mcG_{\rm Z}(\mb{\omega})
    & = \nint d^\ell t \, e^{i\mb{\omega}\cdot\mb{t}} \mcG_{\rm Z}(\mb{t}) 
    = 2\pi \delta(\omega_{1\ccdots\ell}) G_{\rm Z}(\mb{\omega}), & 
    \label{eq:Fourier} \\
    \mcG_{\rm K}^\mb{c}(\mb{\omega})
    & = \nint d^\ell t \, e^{i\mb{\omega}\cdot\mb{t}}
    \mcG^{\mb{c}}_{\rm K}(\mb{t})
    = 2\pi \delta(\omega_{1\ccdots\ell}) G_{\rm K}^\mb{c}(\mb{\omega}), &
    \nonumber
\end{flalign}
respectively, where $\omega_{1\ccdots\ell} = \omega_1 + \ccdots + \omega_\ell$.
We omitted the operator arguments $[\mb{\mcO}]$ for brevity.
Combining the MF prefactors in \Eqs{eq:green_def} and \eqref{eq:Fourier} yields $i (-i)^\ell (-i)^{\ell-1}=(-1)^{\ell-1}$, matching the choice in Ref.~\cite{2021KuglerPRX}.
The analytic properties of correlators $G$ in the frequency domain, 
such as the position of the poles in the complex plane, are determined by the causal structure of the corresponding correlators $\mcG$ in the time domain (see, e.g., Ref.~\cite{2021KuglerPRX}).

\subsection{General EOMs for \lp\ correlators}
\label{sec:GeneralEOMellp}

In this subsection, we derive general EOMs for \lp\ correlators in the time domain, employing a unified notation equally applicable to the MF, ZF, and KF.

We begin by introducing the contour variable $z = -i\tau$ for MF, $z = t$ for ZF, and $z = t^c$ for KF. Corresponding 
definitions for $\partial_z$, $\int dz$, $\delta(z,z')$ and 
$\theta(z,z')$ are given in Table~\ref{t:zVariables}. 
We further write $(m) = (z_m)$, $\partial_m = \partial_{z_m}$, and $\int_m = 
\int \! dz_m$, for short. 
In all three formalisms, we have
\begin{flalign}
    \int_m \! f(m)\delta(m,n)  & = f(n),  &
    \begin{rcases}
    \partial_{m} \! \\
    \partial_{n} \!
    \end{rcases} \theta(m,n)
    & = \pm \delta(m,n) . \hspace{-5mm}  & 
    \label{eq:DeltaThetamn} 
\end{flalign}

\vspace{-2mm}
\noindent 
In this unified notation, \Eqs{eq:green_def} are expressed as~\cite{StefanucciBook}
\begin{equation} \label{eq:green_contour}
    \mcG[\mb{\mcO}](\mb{z})
    = (-i)^{\ell - 1} \bigl \langle
    \mcT [\mb{\mcO}](\mb{z})
    \bigr \rangle .
\end{equation}
Here, $\mcT[\mb{\mcO}](\mb{z})= \mcT \prod_{m=1}^\ell \mcO^m(m)$ denotes contour ordering: it reorders the product such that ``larger'' times sit to the left (if $\theta(m,n) = 1$ (or 0), it puts $\mcO^m(m)$ to the left (or right) of $\mcO^n(n)$), and yields an overall sign of $+1$ ($-1$) if this reordering involves an even (odd) number of exchanges of fermionic operators.
To track such signs, we write $\zeta_m^n = \zeta_n^m$ or $\zeta_{m}^{i_1 \ccdots i_n} = \zeta_m^{i_1} \ccdots \, \zeta_m^{i_n}$ for the sign arising when moving $\mcO^m$ past $\mcO^n$ or $\mcO^{i_1} \ccdots \, \mcO^{i_n}$, respectively.

\begin{table}
\begin{tabular}{l@{\hspace{2mm}}c@{\hspace{2mm}}c@{\hspace{1mm}}c@{\hspace{3mm}}c@{\hspace{3mm}}c}
 &
$z$ &
$\partial_z$ & 
${\displaystyle \int \! dz}$ & 
 $\delta(z,z')$ & 
 $\theta(z,z')$ 
\\[3mm]
\hline
MF & 
$-i \tau$ & 
$\frac{\partial}{\partial(-i \tau)}$ & 
${\displaystyle -i \int_0^\beta \! d \tau}
\vrule width 0pt height 16pt$ & 
 $i \delta(\tau\!-\!\tau')$ & 
 $ \theta(\tau\!-\!\tau')$ 
\\[3.5mm]
ZF & 
$t$ & 
$\frac{\partial}{\partial t}$ & 
${\displaystyle \int_{-\infty}^\infty \! \! d t}$ & 
 $\delta(t\!-\!t')$ & 
 $\theta(t\!-\!t')$ 
\\[3.5mm]
KF 
&
$t^c$ 
&
$\frac{\partial}{\partial t}$ 
&
${\displaystyle \sum_{c} \!\! \int_{-\infty}^\infty \!\! d t \, Z^{cc}}$ 
&
$Z^{cc'}\!\delta(t\!-\!t')$
&
$\theta_{cc'}(t\!-\!t')$  
 \end{tabular} \vspace{-3mm}
\caption{%
    Definition of contour variables, differentiation, integration, delta functions and step functions, in the Matsubara, zero-temperature and Keldysh
    formalisms.
    In the KF, the matrix $Z$ has elements $Z^{cc'} = \delta_{cc'} (-1)^{\delta_{c,+}} = \begin{psmallmatrix} 1 & \phantom{-}0 \\ 0 & -1 \end{psmallmatrix}$, and the contour-ordering step function
    $\theta_{cc'}(t\!-\!t')$ by definition equals $\theta (t - t')$, 0, 1, or 
    $\theta (t'-t)$ for $cc' = --,-+,+- $ or $++$, respectively.
    These definitions readily lead to \Eq{eq:DeltaThetamn}.
}
\label{t:zVariables}
\end{table}%

Below, we set up EOMs for $i \partial_m \mcG^\lpmath$, generalizing
the procedure of \Sec{sec:GeneralEOM2p}. Just as there,
the EOMs for $i\partial_m \mcG^\lpmath$ contain further correlators of two types: auxiliary \lp\ correlators that  differ from $\mcG^\lpmath$ through the replacement of $\mcO^{m}$ by $i\partial_m \mcO^m = [\mc{O}^m,H]$,
and $(\ell\!-\!1)p$ correlators that differ from $\mcG^\lpmath$ through the removal of $\mcO^m$ from the argument list and the replacement of another operator, say $\mcO^{n(\neq m)}$, by the (anti)commutator $[\mcO^m,\mcO^n]_{\zeta^n_m}$.
To describe such objects, we introduce some shorthands:
we define two lists derived from $(\cdot)= (\mb{z}) = (z_1, \ccdots, z_\ell)$, 
\begin{align} \label{eq:list_slashed}
    (\rmv{m}) &= (\mb{z}^\rmv{m}) = (\ccdots, \cancel{z_m}, \ccdots) = (\ccdots, z_{m-1}, z_{m+1}, \ccdots), \nnnl
   (m',\rmv{m}) & = (\mb{z}^{\prime m})
   = (z_m', \mb{z}^\rmv{m}) = (\ccdots, z_{m-1}, z_m', z_{m+1}, \ccdots ),
\end{align}
obtained from the list $(\mb{z})$ by removing slot $m$ entirely, or by replacing
$z_m$ by $z'_m$ in slot $m$.
For example, if $\ell = 4$, then $(\rmv{2}) = (z_1,z_3,z_4)$ and
$(3',\rmv{3}) = (z_1,z_2,z'_3,z_4)$.
(Note that $m'$ is a shorthand for $z'_m$, not $z_{m'}$.)
Likewise, we define two lists derived from $[ \mb{\mcO} ] =  [ \mcO^1, \ccdots, \mcO^\ell ]$,
\begin{flalign}
     \label{eq:NewOperatorLists}
     [ \mcO^{\prime m},\mb{\mcO}^{\rmv{m}} ]  & = 
     [ \ccdots, \mcO^{m-1}, \mcO^{\prime m}, \mcO^{m+1}, \ccdots ], &
     \\
     [\mb{\mcO}^{\rmv{m}n} ]  & = 
     [\ccdots, \cancel{\mcO^m}, \ccdots, 
     \mcO^{n-1}, [\mcO^m, \mcO^n]_{\zeta^n_m}, \mcO^{n+1}, \ccdots ] .  & \nonumber
\end{flalign}
The superscript on $\mb{\mcO}^{\rmv{m}n}$ indicates that $\mc{O}^m$ has been removed 
and slot $n$ ``altered'' by replacing $\mc{O}^n$ by the
(anti)commutator $[\mcO^m, \mcO^n]_{\zeta^n_m}$.
A related list, derived from $\bm{\omega} = (\omega_1, \ccdots, \omega_\ell)$, is defined as
\begin{align}
    \label{eq:eom_omega_mn_def}
    (\mb{\omega}^{\rmv{m}n})
        & = (\ccdots, \cancel{\omega_m}, \ccdots, \omega_{n-1}, \omega_{mn}, \omega_{n+1}, \ccdots) ,
\end{align}
with $\omega_m$ removed and slot $n$ altered by replacing $\omega_n$ by $\omega_{mn} = \omega_m\!+\omega_n$.
$(\mb{\omega}^{\rmv{m}n})$ appears in Fourier transforms involving $\delta(m, n)$, e.g., $\int_m e^{i(\omega_m z_m + \omega_n z_n)} \delta(m,n) = e^{i\omega_{mn} z_n}$.

A trivial identity is $\mcT[\mb{\mcO}](\mb{z}) = \mcT[O^m,\mb{\mcO}^{\rmv{m}}](m, \rmv{m})$, 
useful when evaluating the action of $\partial_m$, where $O^m$ plays a special role. 
In the following, we omit arguments such as $(\mb{z})$ and $[\mb{\mcO}]$ when they can be inferred from context.

The action of $\partial_m$ on a contour-ordered product
$\mcT [\mb{\mcO}](\mb{z}) $ of $\ell$ operators can be compactly expressed as 
\begin{align} 
    \partial_{m} \mcT [\mb{\mcO}]
    & = \mcT \bigl[\partial_m \mcO^m ,\mb{\mcO}^{\rmv{m}}\bigr]
    + \rmvsum{n}{m} \delta(m, n) \mcT \bigl[ \mb{\mcO}^{\rmv{m}n}\bigr] (\rmv{m}) , 
    \nnnl
    \label{eq:eom_operators}
    \rmvsum{n}{m} \mb{\cdot}
    & = \sum_{n=1}^{m-1} \zeta_m^{n\ccdots m-1} \mb{\cdot}
    + \!\! \sum_{n=m+1}^{m+1} \!\mb{\cdot}
    + \!\! \sum_{n=m+2}^{\ell} \!\zeta_m^{m+1\ccdots n-1} \mb{\cdot}  \; .
\end{align}
Here, the signs in  $\sum_n^{\rmv{m}} $ arise from permuting $\mcO^m$ leftward if $n\!<\!m$ (first term) or rightward if $n\!>\!m\!+\!1$ (third term)
to sit next to $\mcO^n$ as either $\theta(m,n)\mcO^m \mcO^n$ or $\theta (n,m)\zeta^m_n \mcO^n \mcO^m$,
as required by contour time ordering. The action of $\partial_m$ on these step functions then yields $\delta(m,n)$ times $[\mcO^m, \mcO^n]_{\zeta^n_m}$ in the altered slot $n$, as encoded in $\mcT[\mb{\mcO}^{\rmv{m}n}]$.

Finally, general EOMs for the \lp\ correlator of 
\Eq{eq:green_contour} follow using \Eqs{eq:eom_operators} 
and the operator EOM \eqref{eq:eom_heisenberg}: 
\begin{align} \label{eq:eom_general}
    &i\partial_m \mcG[\mb{\mcO}]
    = \mcG\bigl[\comm{\mcO^m\!}{H},\mb{\mcO}^{\rmv{m}}\bigr]
    +\rmvsum{n}{m} \delta(m, n)
    \mcG [\mb{\mcO}^{\rmv{m}n}] (\rmv{m}). 
\end{align}
These generalize the 2p equations \eqref{eq:2p_eom_general} to arbitrarily $\ell$, expressing $i \partial_m G^{\lpmath}$
through $G^\lpmath$ and $G^{(\ell-1)\mathrm{p}}$ functions.

\subsection{Single-particle differentiated EOMs}
\label{sec:FullBareAuxiliaryGeneral}

We henceforth focus on ``single-particle differentiated'' EOMs, i.e., EOMs in which the operator $O^m$ being time-differentiated in $i \partial_m O^m$ is a single-particle operator.
More general EOMs, involving time derivatives of composite operators, are not needed in this work.

We consider a Hamiltonian of the form
\begin{equation} \label{eq:Hamiltonian_general}
    \opH
    = \opH^0 + \opH_{\rm int}
    = H^0_{aa'} \oppd_a \opp_{a'} + \opH_{\rm int} ,
\end{equation}
with summation over repeated indices implied. Here, $\opp_a$, $\oppd_a$ ($a = 1, \ccdots, N_{\rm tot}$) are single-particle operators,
such as the $\opd_\sigma$, $\opdd_\sigma$ and $\opc_{\sigma b}$, $\opcd_{\sigma b}$ 
of \Sec{sec:AuxiliaryCorrelatorsAIM}. We will call $a$ an ``orbital'' index,
though it may include spin.
In this section, we do not assume a specific form of interaction.
Later in \Sec{sec:estimator}, we apply the EOM to a quartic interaction.
The single-particle operators may be either bosonic ($\zeta=1$) or fermionic ($\zeta=-1$), but they should all have the same type, so that
\begin{align}
    \label{eq:singleparticleanticommutators}
    [\opp_a,\oppd_{a'}]_\zeta = \one_{aa'}, 
    \qquad 
    [\oppd_a,\opp_{a'}]_\zeta = -\zeta \one_{aa'}.
\end{align}
However, the correlators considered below can be of mixed type, i.e., involve both single-particle operators and composite ones, such as $n_a = \oppd_a \opp_a$.

For the bare propagator, defined for $H_\mr{int}=0$, we write
\begin{equation}
\label{eq:define-bare-propagator-general}
    g^{0}_{aa'}(m, m') = g^{0}[\opp_a, \oppd_{a'}](z_m, z_{m}').
\end{equation}
Using the general EOM \eqref{eq:eom_general} and the equal-time relations
\begin{equation}
    [\opp_a, \opH^0] = H^0_{aa'} \opp_{a'}, \quad
    [\opp_a, \oppd_{a'}]_\zeta = \one_{aa'},
\end{equation}
one finds  two bare-propagator EOMs,
\begin{subequations} \label{eq:eom_nonint_2p}
\begin{align}
    \label{eq:eom_nonint_2p_d}
    ( i\partial_m \one - \opH^0 )_{a\bar{a}} g^{0}_{\bar{a}a'}(m, m')
    &= \delta(m, m') \one_{aa'}, \\
    \label{eq:eom_nonint_2p_ddag}
    g^{0}_{a\bar{a}}(m, m') ( i \lpartialp_{m} \one + \opH^0 )_{\bar{a}a'}
    &= -\delta(m, m') \one_{aa'}.
\end{align}
\end{subequations}
$\lpartialp_m$ denotes a derivative w.r.t.\  $z_m'$, acting to the left.
According to \Eqs{eq:eom_nonint_2p}, $g^0$ serves as inverse for 
the ``bare time evolution'' expressions $(i\partial_m \one - \opH^0 )$
and $( i \lpartialp_{m} \one + \opH^0 )$.
Below, this will be exploited to remove such expressions from EOMs for general correlators.

\begin{figure*}[t]
    \centering
    \includegraphics[width=0.96\textwidth]{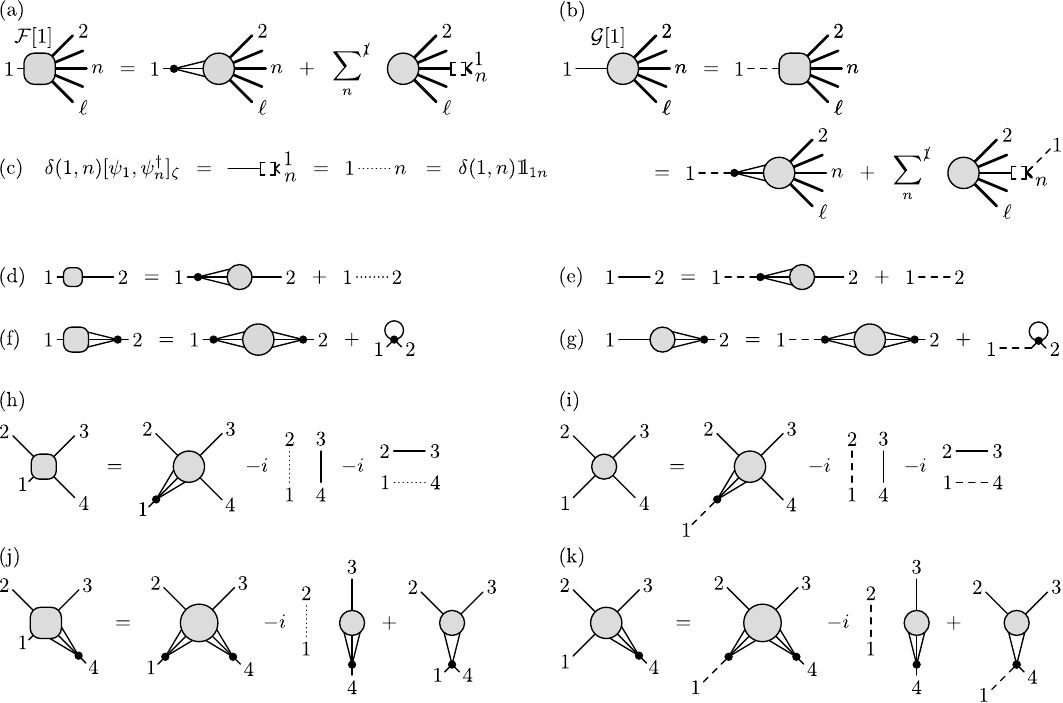}
    \caption{Diagrammatic depiction of $\mc{F}[1] = \mc{F}^1[\opp_1,\mb{\mcO}^\rmv{1}]$ [\Eq{eq:eom_F_def}, rounded squares] and $\mc{G}[1] = \mc{G}[\opp_1,\mb{\mcO}^\rmv{1}]$ [\Eq{eq:eom_integral}, circles]. 
    Operators from the argument list $\mb{\mcO}^{\rmv{1}}$ are represented by long lines, drawn thin for single-particle operators or thick for operators of unspecified type (single-particle or composite). 
    (a,b) General case.    
    In the $\sum_{n}^\rmv{1}$ sums, leg $n$ is decorated
    by tiny ${\scalefont{0.3} [ \; ]}$ brackets and two endpoints labeled $1$ and $n$, representing the replacement of $\mcO^n$ by $\delta(1,n) [\opp_1,\mcO^n]_{\zeta_1^n}$ [cf.\ \Eq{eq:eom_F_def}].
    (c) For $\mcO^n=\oppd_n$, the latter reduces to the identity contraction $\delta(1,n) \one_{1n}$, depicted as a dotted line.
    (d-k) Examples illustrating (a,b) for various choices of $[\opp_1, \mb{\mcO}^{\rmv{1}}]$,
    with (d-g) $\ell=2$ and (h-k) $\ell=4$:   
    (d,e) $[\opp_1,\oppd_2]$, (f,g) $[\opp_1,\opqd_2]$, (h,i) $[\opp_1,\oppd_2,\opp_3,\oppd_4]$, and (j,k) $[\opp_1,\oppd_2,\opp_3,\opqd_4]$.
    The 2p diagrams (d-g) illustrate how cases encountered in Sec.~\ref{sec:self-energy-estimator} on the self-energy (cf.\ Fig.~\ref{fig:derivation_symm_Sigma_estimator}) arise in the present formulation.
    (h) represents the first equation of \Eqs{eq:eom_F_example_4p}. The last diagram of (j) depicts the 3p correlator $\mc{G}\bigl[\oppd_2,\opp_3,[\opp_1,\opqd_4]_\zeta \bigr]$.    }
    \label{fig:diag_eom_bare}
\end{figure*}

Now, consider \lp\ correlators involving at least one 
single-particle operator, say $\mcO^m = \psi_a^{(\dagger)}$. 
Corresponding single-particle differentiated EOMs follow via \Eq{eq:eom_general}:
\begin{subequations} \label{eq:eom_int}
\begin{align}
    \label{eq:eom_int_d}
    ( i\partial_m \one - \opH^0 )_{aa'} \mcG[\opp_{a'},\mb{\mcO}^\rmv{m}]
    &= \mc{F}^m[\opp_a,\mb{\mcO}^\rmv{m}], \\
    \label{eq:eom_int_ddag}
    \mcG[\oppd_{a'},\mb{\mcO}^\rmv{m}] ( i \!\lpartial\!_{m} \one + \opH^0 )_{a'a}
    &= \mc{F}^m[\oppd_a,\mb{\mcO}^\rmv{m}].
\end{align}
\end{subequations}
For the correlators on the right, containing all contributions not involving $g^0$, we used the shorthand
\begin{align} \label{eq:eom_F_def}
    \mc{F}^m[\mb{\mcO}]
    = \mcG\bigl[\comm{\mcO^m\!}{H_{\rm int}},\mb{\mcO}^\rmv{m}\bigr] 
    + \! \rmvsum{n}{m} \delta(m, n)
   \mcG [ \mb{\mcO}^{\rmv{m}n}](\rmv{m}),
\end{align}
where the superscript  on $\mc{F}^m$ singles out $m$ for special treatment.
The first term on the right involves single-index composite operators, to be denoted
(cf.\ \Eq{eq:q_def})
\begin{align}
    \label{eq:q_def_general}
    \opq_a & = \big[ \opp_a, \opH_{\rm int} \big],
    \quad \opqd_a = - \big[ \oppd_a , \opH_{\rm int} \big]. 
\end{align}
Figure~\ref{fig:diag_eom_bare}(a) gives a diagrammatic representation of $\mc{F}^m$.

Let us exemplify \Eq{eq:eom_F_def} for the case that 
\textit{all} operators in 
$[\mb{\mc{O}}]$ are single-particle operators, $\mc{O}^n = \psi_a^{(\dagger)}$. Then, the (anti)commutator  in the altered slot $n$ of $[\mb{\mcO}^{\rmv{m}n}]$ is nonzero only for the cases 
listed in \Eqs{eq:singleparticleanticommutators}.
For $\mb{\mcO} = [\opp_1, \oppd_2]$, e.g., we obtain
\begin{equation}
\begin{aligned}
    \mc{F}^1 [\opp_1, \oppd_2] &= \phantom{-} \mcG[\opq_1, \oppd_2] + \delta(1, 2) \one_{12}, \\
    \mc{F}^2 [\opp_1, \oppd_2] &= - \mcG[\opp_1, \opqd_2] - \delta(2, 1) \one_{21} .
\end{aligned}
\label{eq:eom_F_example_2p}
\end{equation}
The second terms on the right were simplified using $\mc{G}\bigl[[\opp_1,\oppd_2]_\zeta \bigr] = \mc{G}[\one_{12}] = \one_{12}$.
We will call the resulting combinations $\delta(1,2)\one_{12}$ ``identity contractions'' and diagrammatically depict them using dotted lines (see Fig.~\ref{fig:diag_eom_bare}).
Similarly, for $\mb{\mcO} = [\opp_1, \oppd_2, \opp_3, \oppd_4]$, we have
\begin{alignat}{2} \label{eq:eom_F_example_4p}
    &\mc{F}^1[\mb{\mcO}] = \phantom{-}\mcG[\opq_1, \oppd_2, \opp_3, \oppd_4]
      && -i \delta(1, 2)\one_{12} \mcG[\opp_3, \oppd_4](3, 4) \nnnl
    & && -i \delta(1, 4)\one_{14} \mcG[\oppd_2, \opp_3](2, 3),\nnnl
    &\mc{F}^2[\mb{\mcO}] = -\mcG[\opp_1, \opqd_2, \opp_3, \oppd_4]
      && +i \delta(2, 1) \one_{21} \mcG[\opp_3, \oppd_4](3, 4) \nnnl
    & && +i \zeta \delta(2, 3) \one_{23} \mcG[\opp_1, \oppd_4](1, 4),\nnnl
    &\mc{F}^3[\mb{\mcO}] = \phantom{-} \mcG[\opp_1, \oppd_2, \opq_3, \oppd_4]
      && -i \zeta \delta(3, 2) \one_{32} \mcG[\opp_1, \oppd_4](1, 4) \nnnl
    & && -i \delta(3, 4) \one_{34} \mcG[\opp_1, \oppd_2](1, 2),\nnnl
    &\mc{F}^4[\mb{\mcO}] = - \mcG[\opp_1, \oppd_2, \opp_3, \opqd_4]
      && +i \delta(4, 1) \one_{41} \mcG[\oppd_2, \opp_3](2, 3) \nnnl
    & && +i \delta(4, 3) \one_{43} \mcG[\opp_1, \oppd_2](1, 2).
\end{alignat}
Again, identity contractions arise on the right from, e.g.,
$\mcG\bigl[[\opp_1,\oppd_2]_\zeta, \opp_3, \oppd_4\bigr](2, 3, 4) = (-i) \one_{12} \, \mcG[\opp_3, \oppd_4](3,4)$. The first lines of 
\Eqs{eq:eom_F_example_2p} and \eqref{eq:eom_F_example_4p} are illustrated in 
Figs.~\ref{fig:diag_eom_bare}(d) and \ref{fig:diag_eom_bare}(h), respectively.

We now eliminate the bare time evolution expressions from the EOMs (\ref{eq:eom_int}) for $\mcG[\opp_a,\mb{\mcO}^\rmv{m}]$. We begin with
\begin{equation} \label{eq:eom_integral_derivation_start}
    \mcG[\opp_a,\mb{\mcO}^\rmv{m}]
    = \nsint{m'} \delta(m,m') \one_{aa'} \mcG[\opp_{a'},\mb{\mcO}^\rmv{m}](m', \rmv{m}),
\end{equation}
an identity which trivially follows from the definition of the delta function and the identity matrix $\one$.
By inserting the EOM \eqref{eq:eom_nonint_2p_ddag} for the bare propagator, we find
\begin{align} \label{eq:eom_integral_long}
    &\mcG[\opp_a,\mb{\mcO}^\rmv{m}] \nnnl
    &\quad = \nsint{m'} g^0_{a\bar{a}}(m, m') ( -i\lpartialp_{m} \one - \opH^0 )_{\bar{a}a'} \mcG[\opp_{a'},\mb{\mcO}^\rmv{m}](m', \rmv{m}) \nnnl
    &\quad = \nsint{m'} g^0_{aa'}(m, m') \mc{F}^m[\opp_{a'},\mb{\mcO}^\rmv{m}](m', \rmv{m}).
\end{align}
In the second step, we used integration by parts to convert the partial derivative from acting to the left to acting to the right, i.e., from $-\lpartialp_{m}$ to $\partial'_{m}$, and then used the \lp\ EOM \eqref{eq:eom_int_d}.
Importantly, the boundary term in the intergration by parts can be shown to vanish, see App.~\ref{sec:boundary} or App.~\ref{sec:wo_adia}. 
This procedure is analogous to solving \Eq{eq:2p_eom_matrix} by multiplying $(i\omega - H^0)^{-1}$ on both sides in \Sec{sec:2p_review}, but now in the time domain.

Finally, let us  express \Eq{eq:eom_integral_long} in a concise form hiding orbital indices. To this end, we define
\begin{equation}
\begin{aligned}
    (g^0_m)_{aa'} (m,m')
    &= g^0_{aa'} (m,m'), \\
    \big( \mc{G}[m^{(\dagger)}] \big)_{a} (\cdot)
    &= \mc{G}[\psi^{(\dagger)}_a,\mb{\mc{O}}^\rmv{m}] (\cdot), \\
    \ \big( \mc{F}[m^{(\dagger)}] \big)_{a'} (m', \rmv{m})
    &= \mc{F}^m[\psi^{(\dagger)}_{a'},\mb{\mc{O}}^\rmv{m}] (m', \rmv{m}).
\end{aligned}
\label{eq:eom_implicit_a}
\end{equation}
Viewing $g_m^0$ as an $N_{\rm tot} \times N_{\rm tot}$ matrix and $\mcG[m^{(\dagger)}]$ and $\mc{F}[m^{(\dagger)}]$ as vectors of length $N_{\rm tot}$  w.r.t.\ to their orbital
indices,
the implicit sum $\sum_{a'}$ in \Eq{eq:eom_integral_long} amounts to matrix-vector multiplication.
We thus obtain
\begin{subequations}
\label{eq:eom_integral}
\begin{align}
    \label{eq:eom_integral_d}
    \mcG[m](\mb{z})
    &= \nsint{m'} g^0_m(m, m') \mc{F}[m](m', \rmv{m}), \\
    \label{eq:eom_integral_ddag}
    \mcG[m^\dagger](\mb{z})
    &= -\nsint{m'} \mc{F}[m^\dagger](m', \rmv{m}) g^0_m(m', m) . 
\end{align}
\end{subequations}
\Equ{eq:eom_integral_ddag} follows similarly from \Eq{eq:eom_int_ddag}, where $\mcO^m = \oppd_a$.
The extra minus sign reflects the sign difference in the $\opH^0$ terms in \Eqs{eq:eom_int_d} and \eqref{eq:eom_int_ddag}.

Figure~\ref{fig:diag_eom_bare}(b) diagrammatically depicts the EOM \eqref{eq:eom_integral_d} for $\mc{G}[m]$ with $O^m=d_1$, for the case that the interaction is quartic. The differentiation $i\partial_1$ generates two types of diagrams, both involving a bare propagator: 
it is either connected to the bare interaction vertex associated with 
a composite operator $q_1=[\psi_1,H_\mr{int}]$ or to the ``(anti)commutator leg'' representing $[\psi_1, O^n]_\zeta$.
If $O^n$ equals the single-particle operator $\psi_a$, the (anti)commutator reduces to $\one_{1a}$, thus disconnecting the bare propagator, as exemplified in Figs.~\ref{fig:diag_eom_bare}(e,i,k).

The main upshot of this section is as follows: 
those external legs of a correlator $\mc{G}$ that represent
full single-particle propagators $g$ can be converted, via EOMs, 
to bare single-particle propagators $g_0$ connected to 
various other correlators (schematically, $\mc{G} = g^0 \mc{F}$). This sets the stage for Sec.~\ref{sec:estimator}. 
There, we will remove bare propagators through multiplication by $(g^0)^{-1} = g^{-1} + \Sigma$ (schematically, $\mc{F} = (g^0)^{-1} \mc{G}$, hence 
$g^{-1} \mc{G} = \mc{F} - \Sigma \mc{G}$). In this way, we arrive
at a strategy for amputating legs (computing $g^{-1} \mc{G}$) without explicitly dividing by $g$.

% ======================================================

\subsection{Some remarks on quantum impurity models} \label{sec:eom_impurity}
We briefly pause the development of our general formalism to make some remarks about quantum impurity models.
There, an interacting impurity is coupled to a \textit{non}interacting bath.
Typically, the correlators of interest are ``impurity correlators'', involving only impurity operators.
Here, we show that impurity correlators satisfy a suitably modified version of EOM~\eqref{eq:eom_integral}, involving only impurity operators and indices.

Let us consider a quantum impurity model where the noninteracting Hamiltonian consists of both impurity operators $\opd_i, \opdd_i$ ($i = 1, \ccdots, N_{\rm imp}$) and bath operators $\opc_b, \opcd_b$ ($b = 1, \ccdots, N_{\rm bath}$), while only impurity operators appear in the interacting Hamiltonian:
\begin{equation}
    \opH_{\rm int} = \opH_{\rm int}[\opd_i, \opdd_{i'}].
\end{equation}
The subscripts enumerate both the spin and orbital indices.
We let $\opp_a$ ($a = 1, \ccdots, N_{\rm tot} = N_{\rm imp} + N_{\rm bath}$) enumerate all annihilation operators:
\begin{align}
    \opp_a = \begin{cases}
        d_i & \text{for $i = a = 1, \ccdots, N_{\rm imp}$} \\
        c_b & \text{for $b = a - N_{\rm imp} = 1, \ccdots, N_{\rm bath}$}.
    \end{cases}
\end{align}
It will henceforth be understood that the indices $i$ and $b$ are used exclusively for impurity or bath operators, respectively, while $a$ encompasses both.

The bare propagator $g^{0}_{aa'} = 
g^{0}[\opp_a,\oppd_{a'}]$ of \Eq{eq:define-bare-propagator-general}, 
defined as the propagator with $H_\mr{int}=0$, can be 
obtained by solving the bare EOMs \eqref{eq:eom_nonint_2p}
(e.g., by transforming to the Fourier domain). 
The $aa'= ii'$ components of the resulting bare propagator, $g^{0}_{ii'} = g^{0}[\opd_i,\opdd_{i'}]$, comprise
the ``bare impurity propagator''. It encodes information about the bath via the hybridization function
[see, e.g., \Eq{eq:2p_green_hyb}, or the
$H_\mr{int}=0$ version of Eqs.~\eqref{eq:2p_eom_matrix}--\eqref{eq:2p_eom_matrix_result}].
Together with $H_{\rm int}$, it \emph{fully} specifies the impurity dynamics.
In this sense, once $g^0_{ii'}$ has been found, the bath has in effect been integrated out and needs no further consideration.

The inverse of the bare impurity propagator, $(g^{0})^{-1}_{ii'}$,  is defined as the inverse of the $N_{\rm imp} \!\times\! N_{\rm imp}$ matrix  $g^{0}_{ii'}$, \textit{not}
the $N_{\rm imp} \!\times\! N_{\rm imp}$ block of the inverse of
the $N_{\rm tot} \!\times\! N_{\rm tot}$ matrix $g^{0}_{aa'}$.
Thus, in the Fourier domain we have 
\begin{equation}
    (g^0)^{-1}_{ii '} g^0_{i' i''} = \one_{ii''}.
\end{equation}

Now, we turn our attention to full correlators of $\ell$ impurity operators.
We assume that at least one, say $\mcO^m = d^{(\dagger)}_i$, is a single-particle operator; all others may be  general (single-particle or composite) operators, $\mcO^{n (\neq m)} = \mcO^n[\opd_i, \opdd_{i'}]$.
The EOM for this correlator has the form of the general EOM \eqref{eq:eom_integral_long}, but now containing 
only impurity operators on both sides, either elementary or composite ones.
To see this, notice that
$\mc{F}^m[\opc_b,\mb{\mcO}^\rmv{m}] = \mc{F}^m[\opcd_b,\mb{\mcO}^\rmv{m}] = 0$, 
since the bath operators (anti)commute with $\opH_{\rm int}$ and $\mb{\mcO}^\rmv{m}$.
Therefore, the dummy orbital index $a'$ in \Eq{eq:eom_integral_long} can be limited to impurity orbitals.
The same is true for all implicit orbital indices in  \Eq{eq:eom_integral},
where we now have
\begin{equation}
\begin{aligned}
    (g^0_m)_{ii'} (m,m')
    &= g^0_{ii'} (m,m'), \\
    \big( \mc{G}[m^{(\dagger)}] \big)_{i} (\cdot)
    &= \mc{G}[d^{(\dagger)}_i,\mb{\mc{O}}^\rmv{m}] (\cdot), \\
    \ \big( \mc{F}[m^{(\dagger)}] \big)_{i'} (m,\rmv{m})
    &= \mc{F}^m[d^{(\dagger)}_{i'},\mb{\mc{O}}^\rmv{m}] (m,\rmv{m}).
\end{aligned}
\label{eq:eom_implicit_i}
\end{equation}
The EOMs of impurity correlators are again represented by the diagrams of Fig.~\ref{fig:diag_eom_bare}, with $g^0$ now representing the bare impurity propagator in the presence of a bath.

In what follows, we will keep the discussion general, mostly writing $d_i$. 
The EOM and improved estimators for a generic many-body Hamiltonian without a bath [\Eq{eq:Hamiltonian_general}] can be simply obtained by setting $d_i = \psi_a$ and $N_{\rm bath} = 0$.

% ======================================================
\subsection{EOM in the frequency domain} \label{sec:eom_freq}

Next, we Fourier transform the EOM \eqref{eq:eom_integral} from the time domain to the frequency domain for each of the three formalisms (MF, ZF, KF).

\subsubsection{MF} \label{sec:eom_matsubara}
In the MF, the Fourier transform of the bare-propagator EOM \eqref{eq:eom_nonint_2p_d} reads
\begin{equation} \label{eq:eom_MF_bare_g}
    ( i\omega \one - \opH^0 )_{a\bar{a}} g^{0}_{\bar{a}a'}(i\omega)
    = \one_{aa'}.
\end{equation}
Similarly, the Fourier transform of \Eq{eq:eom_F_def} for $\mcF^m$ reads
\begin{equation}
\begin{aligned}
    F_{\rm M}^m[\mb{\mcO}](i\mb{\omega})
    &= G_{\rm M}\bigl[\comm{\mcO^m}{H_{\rm int}},\mb{\mcO}^\rmv{m}\bigr](i\mb{\omega}) \\
    &\quad + \rmvsum{n}{m} G_{\rm M} \bigl[ \mb{\mcO}^{\rmv{m}n} \bigr](i\mb{\omega}^{\rmv{m}n}) , \nonumber
\end{aligned}
\label{eq:eom_MF_F}
\end{equation}
where $\mb{\omega}^{\rmv{m}n}$ is defined in \Eq{eq:eom_omega_mn_def}.

By Fourier transforming the \lp\  EOM \eqref{eq:eom_integral}, one finds
\begin{equation} \label{eq:eom_MF_impurity_long}
\begin{aligned}
    G_{\rm M}[m](i\mb{\omega})
    &= g^0_{{\rm M},m}(i\omega_m) F_{\rm M}[m](i\mb{\omega}), 
    \\
    G_{\rm M}[m^\dag](i\mb{\omega})
    &= -F_{\rm M}[m^\dag](i\mb{\omega}) g^0_{{\rm M},m}(-i\omega_m),
\end{aligned}
\end{equation}
or, more compactly, with implicit frequency arguments,
\begin{equation} \label{eq:eom_MF_impurity}
\begin{aligned}
    G_{\rm M}[m]
    &= g^0_{{\rm M},m} F_{\rm M}[m], \\
    G_{\rm M}[m^\dagger]
    &= -F_{\rm M}[m^\dagger] g^0_{{\rm M},m}.
\end{aligned}
\end{equation}
The implicit frequency argument of $g^0_{{\rm M},m}$ can be inferred to have $i\omega_m$ in the outer frequency argument.
For the left-multiplication (right-multiplication) of $g^0_{{\rm M},m}$, we use $g^0_{{\rm M},m}(i\omega_m) = g^0_{{\rm M},m}(i\omega_m, -i\omega_m)$ ($g^0_{{\rm M},m}(-i\omega_m) = g^0_{{\rm M},m}(-i\omega_m, i\omega_m)$) [\Eq{eq:green_2p_with_1_frequency}], so that the outer, left (right) frequency argument is $i \omega_m$.
For $\ell=2$, $m=1$, and $[\opd_i, \mcO^2] = [\opd_1, \opdd_2]$, \Eq{eq:eom_MF_F} gives
\begin{equation}
\label{eq:exampleF1MGM}
    F^1_{\rm M}[\opd_1, \opdd_2] = G_{\rm M}[\opq_1, \opdd_2] + \one_{12}.
\end{equation}
Substituting this into the first relation of \Eq{eq:eom_MF_impurity}, one recovers the 2p EOM \eqref{eq:2p_eom_G} derived in \Sec{sec:2p_review}.

% ===================================================
\subsubsection{ZF}  \label{sec:eom_zf}
In the ZF, the Fourier transforms of \Eqs{eq:eom_nonint_2p_d}, \eqref{eq:eom_F_def}, and \eqref{eq:eom_integral} read
\begin{align} \label{eq:eom_ZF_impurity}
    ( \omega \one - \opH^0 )_{a\bar{a}} g^{0}_{\bar{a}a'}(\omega)
    &= \one_{aa'}\,,
    \nnnl
    F_{\rm Z}^{m}[\mb{\mcO}](\mb{\omega})
    & = G_{\rm Z}\bigl[\comm{\mcO^m}{H_{\rm int}},\mb{\mcO}^\rmv{m} \bigr](\mb{\omega}) \nnnl
    & \quad + \rmvsum{n}{m} 
    G_{\rm Z} \bigl[ \mb{\mcO}^{\rmv{m}n} \bigr](\mb{\omega}^{\rmv{m}n}) ,
    \\
    G_{\rm Z}[m](\mb{\omega})
    &= g^{0}_{{\rm Z},m}(\omega_m) F_{\rm Z}[m](\mb{\omega}),
    \nnnl
    G_{\rm Z}[m^\dagger](\mb{\omega})
    &= - F_{\rm Z}[m^\dagger](\mb{\omega}) g^{0}_{{\rm Z},m}(-\omega_m). \nonumber
\end{align}
They have the same structure as the  MF \Eqs{eq:eom_MF_bare_g}, \eqref{eq:eom_MF_F}, and \eqref{eq:eom_MF_impurity}.

% ===================================================
\subsubsection{KF}  \label{sec:eom_keldysh}
Finally, let us derive the EOM in the KF.
We will first do so in the contour basis and subsequently transform the result to the Keldysh basis.
The time-domain EOM for the bare propagator [\Eq{eq:eom_nonint_2p_d}] reads
\begin{equation} \label{eq:eom_kf_2p_example_c}
    ( i\partial_m \one - \opH^0 )_{a\bar{a}} g^{0\,cc'}_{\bar{a}a'}(t, t')
    = Z^{cc'} \delta(t -t') \one_{aa'}, 
\end{equation}
where  we used $\delta(m, n)
= Z^{c_m c_n}\delta(t_m-t_n)$, with  $Z = \begin{psmallmatrix} 1 & \phantom{-} 0 \\ 0 & -1 \end{psmallmatrix}$ the Pauli $z$  matrix defined
in Table~\ref{t:zVariables}.
Fourier transforming this EOM gives
\begin{equation} \label{eq:eom_kf_2p_c_omega}
    ( \omega \one - \opH^0 )_{a\bar{a}} g^{0\,cc'}_{\bar{a}a'}(\omega)
    = Z^{cc'} \one_{aa'} . 
\end{equation}
The Fourier transformation of \Eq{eq:eom_F_def} for $\mcF^m$ reads
\begin{equation}
\begin{aligned}
    F_{\rm K}^{m,\mb{c}}[\mb{\mcO}](\mb{\omega})
    &= G_{\rm K}^\mb{c}\bigl[\comm{\mcO^m}{H_{\rm int}},\mb{\mcO}^\rmv{m} \bigr](\mb{\omega}) \\
    &\quad + \rmvsum{n}{m} Z^{c_m c_n}
    G_{\rm K}^{\mb{c}^{\rmv{m}}} \bigl[ \mb{\mcO}^{\rmv{m}n} \bigr](\mb{\omega}^{\rmv{m}n}).
\end{aligned}
\label{eq:eom_KF_F_def}
\end{equation}
Here, $\mb{c}^{\rmv{m}} = (\ccdots, c_{m-1}, c_{m+1}, \ccdots)$ is defined as in \Eq{eq:list_slashed}, and, in the last term,
$Z^{c_m c_n}$ comes from the KF version of
$\delta(m, n)$ in \Eq{eq:eom_F_def}.
[Note that in \Eq{eq:eom_KF_F_def} all $c$ indices, including $c_m$ and $c_n$, are fixed by the left side.]
Similarly, the Fourier-transformed EOMs \eqref{eq:eom_integral} for $\mc{G}[m^{(\dagger)}]$ read
\begin{equation} \label{eq:eom_KF_impurity_contour}
\begin{aligned}
    G_{\rm K}^\mb{c}[m](\mb{\omega})
    &= g^{0\,c_m c''_m}_{{\rm K},m}(\omega_m) Z^{c''_m c_m'} F_{\rm K}^{\mb{c}^{\prime m}}[m](\mb{\omega}),
    \\
    G_{\rm K}^\mb{c}[m^\dagger](\mb{\omega})
    &= - F_{\rm K}^{\mb{c}^{\prime m}}[m^\dagger](\mb{\omega}) Z^{c_m' c''_m} g^{0\,c''_m c_m}_{{\rm K},m}(-\omega_m),
\end{aligned}
\end{equation}
where $\mb{c}^{\prime m} = (c'_m,\mb{c}^\rmv{m}) = ( \ccdots, c_{m-1}, c_m', c_{m+1}, \ccdots)$
is defined as in \Eq{eq:list_slashed}, and summations $\sum_{c'_m, c''_m}$ are implied.

It is often useful to transform KF correlators from the contour basis $c \in \{-,+\}$ to the Keldysh basis $k \in \{ 1, 2 \}$,
by applying the orthogonal transformation
\begin{equation}
    D^{kc} = \medfrac{1}{\sqrt{2}}
    \begin{pmatrix}
    1 & -1 \\ 1 & \phantom{-}1
    \end{pmatrix}_{kc}
    = \medfrac{1}{\sqrt{2}}(-1)^{k \cdot \delta_{c,+}} 
\end{equation}
to each contour index:
\begin{align}
\label{eq:eom_KF_G_Keldysh}
    G_{\rm K}^{\mb{k}}[\mb{\mcO}](\mb{\omega})
    = \sum_{c_1, \ccdots, c_\ell} 
    \prod_{p=1}^\ell D^{k_p c_p} 
    G_{\rm K}^{\mb{c}}[\mb{\mcO}](\mb{\omega}).
\end{align}
Then, the bare-propagator EOM \eqref{eq:eom_kf_2p_c_omega} becomes
\begin{equation} \label{eq:eom_kf_2p_example_k}
    ( \omega \one - \opH^0 )_{a\bar{a}} g^{0\,kk'}_{\bar{a}a'}(\omega)
    = X^{kk'} \one_{aa'} . 
\end{equation}
Here, we have used the identity
\begin{equation} \label{eq:eom_D_Z_to_X}
    D^{kc} D^{k'c'} Z^{cc'} = (D Z D^{-1})^{kk'} = X^{kk'} ,
\end{equation}
which transforms the Pauli $z$ matrix in the contour basis to the Pauli $x$ matrix,
$X = \begin{psmallmatrix} 0 & 1 \\ 1 & 0 \end{psmallmatrix}$, in the Keldysh basis.

For later use, we also define a rank-$\ell$ tensor 
\begin{align} \label{eq:not_P_def}
    P^{k_1 \ccdots k_\ell}
    & = \!\!\! \sum_{c, c_1, \ccdots,  c_\ell}
    \prod_{n=1}^{\ell} \! Z^{c c_n} D^{k_n c_n}
    = \sum_{c} (-1)^{\ell \cdot \delta_{c,+}} \prod_{n=1}^{\ell} \!D^{k_n c}
    \nnnl
    &= \frac{1 + (-1)^\ell (-1)^{k_1 + \ccdots + k_\ell}}{\sqrt{2^\ell}} \nnnl
    &= \begin{cases}
        \frac{1}{\sqrt{2^{\ell-2}}} & \text{if $k_1 + \ccdots + k_\ell + \ell$ is even,} \\
        0 & \text{if $k_1 + \ccdots + k_\ell + \ell$ is odd}.
    \end{cases}
\end{align}
The rightmost expression in the first line follows
since the $Z$ matrices yield a nonzero result
only if all their indices are equal, $c = c_1 = \ccdots = c_\ell$.
For example, $\ell=3$ gives
\begin{equation} \label{eq:not_P_def_3}
    P^{k_1 k_2 k_3}
    = \begin{cases}
        \frac{1}{\sqrt{2}} & \text{if $k_1 + k_2 + k_3$ is odd,} \\
        0 & \text{if $k_1 + k_2 + k_3$ is even}.
    \end{cases}
\end{equation}
This tensor appears when transforming EOMs from the contour to the Keldysh basis.
It satisfies the identities (sums over repeated indices are implied)
\begin{subequations}
\begin{gather}
    \label{eq:2p_XPX_P}
    X^{k_1 k_1'} X^{k_2 k_2'} P^{k_1' k_2' k_3} 
    = P^{k_1 k_2 k_3}, \\
    \label{eq:3p_PP_P}
    P^{k_1 k_2 k_3} P^{k_3 k_4 k_5}
    = P^{k_1 k_2 k_4 k_5},
\end{gather}
\end{subequations}
which follow directly from its definition.

Next, we transform \Eqs{eq:eom_KF_F_def} and \eqref{eq:eom_KF_impurity_contour} to the Keldysh basis.
First, we multiply \Eq{eq:eom_KF_F_def} by $\prod_{p=1}^{\ell} D^{k_p c_p}$ and sum over the contour indices.
The last term becomes 
\begin{align}
    & \sum_{c_1 , \ccdots , c_\ell}
    \big( \prod_{p=1}^{\ell} D^{k_p c_p} \big)
    Z^{c_m c_n} G_{\rm K}^{\mb{c}^{\rmv{m}}} \nnnl
    &\quad = \sum_{c_m, c_n} D^{k_m c_m} D^{k_n c_n} Z^{c_m c_n} (D^{-1})^{c_n k_{mn}}
     G_{\rm K}^{\mb{k}^{\rmv{m}n}} \nnnl
    &\quad = \sum_{c_n} (-1)^{3 \cdot \delta_{c_n,+}} D^{k_m c_n} D^{k_n c_n} D^{k_{mn} c_n} G_{\rm K}^{\mb{k}^{\rmv{m}n}} \nnnl
    &\quad = P^{k_m k_n k_{mn}} G_{\rm K}^{\mb{k}^{\rmv{m}n}}.
    \label{eq:contour-Keldysh-transformation-of-Z-term}
\end{align}
Here, $\mb{k}^{\rmv{m}n} = (\ccdots, \cancel{k_m}, \ccdots, k_{n-1}, k_{mn}, k_{n+1}, \ccdots)$,
defined as in \Eq{eq:eom_omega_mn_def}, is obtained from $\mb{k}$ by removing $k_m$ from the list and replacing $k_n$ by a new dummy index $k_{mn}$, summation over which is implied.
To arrive at the third line,
we used $Z^{c_m c_n} = \delta_{c_m c_n} (-1)^{\delta_{c_n,+}}
= \delta_{c_m c_n} (-1)^{3 \delta_{c_n,+}}$.
Then, \Eq{eq:eom_KF_F_def} transforms to
\begin{align} \label{eq:eom_KF_F_Keldysh}
    F_{\rm K}^{m,\mb{k}}[\mb{\mcO}](\mb{\omega})
    & = G_{\rm K}^\mb{k}\bigl[\comm{\mcO^m}{H_{\rm int}},\mb{\mcO}^\rmv{m} \bigr](\mb{\omega}) \\
    & \quad + \rmvsum{n}{m} P^{k_m k_n k_{mn}}
    G_{\rm K}^{\mb{k}^{\rmv{m}n}}[\mb{\mcO}^{\rmv{m}n}](\mb{\omega}^{\rmv{m}n}). \nonumber
\end{align}
The $P$ tensor maps the $(\ell\!-\!1)$-element list of Keldysh indices, $\mb{k}^{\rmv{m}n}$, to the original $\ell$-element list $\mb{k}$.
Similarly, by transforming \Eq{eq:eom_KF_impurity_contour} to the Keldysh basis, we find
\begin{subequations} \label{eq:eom_KF_impurity}
\begin{align}
    \label{eq:eom_KF_impurity_d}
    G_{\rm K}[m]
    &= g^0_{{\rm K},m} X_m F_{\rm K}[m], \\
    \label{eq:eom_KF_impurity_ddag}
    G_{\rm K}[m^\dagger]
    &= -F_{\rm K}[m^\dagger] X_m \, g^0_{{\rm K},m}.
\end{align}
\end{subequations}
Here, the subscript on $X_m$ indicates that it acts on 
the Keldysh index $k_m$ of $F_{\rm{K}}[m^{(\dagger)}]$.
We extended \Eq{eq:eom_implicit_i} and defined $g^0_{{\rm K},m}$, $G_{\rm K}[m^{(\dagger)}]$, and $F_{{\rm K}}[m^{(\dagger)}]$ as a matrix and vectors in the basis of the orbital and Keldysh indices:
\begin{equation}
\begin{aligned}
    (g^0_{{\rm K},m})_{ii'}^{kk'} (\omega)
    &= g^{0,kk'}_{{\rm K},ii'} (\omega), \\
    \big( G_{\rm K}[m^{(\dagger)}] \big)_{i}^{k_m} (\mb{\omega})    
    &= G_{\rm K}^{\mb{k}}[d^{(\dagger)}_i,\mb{\mc{O}}^\rmv{m}] (\mb{\omega}), \\
    \big( F_{\rm K}[m^{(\dagger)}] \big)_{i}^{k_m} (\mb{\omega})
    &= F_{\rm K}^{m,\mb{k}}[d^{(\dagger)}_i,\mb{\mc{O}}^\rmv{m}] (\mb{\omega}).
\end{aligned}
\label{eq:eom_implicit_ik}
\end{equation}
For the KF, we use this compact notation only in the Keldysh basis, not in the contour basis, because the $X$ matrix changes to the $Z$ matrix when transformed to the contour basis.

The KF EOMs \eqref{eq:eom_KF_impurity} have the same structure as the MF EOMs \eqref{eq:eom_MF_impurity} and ZF EOMs \eqref{eq:eom_ZF_impurity}, except for the factor $X_m$ acting on the Keldysh indices.
In the rest of this paper, we write formulas only in the KF, and drop the subscripts on $G_\mr{K}$ and $F_\mr{K}$.
The corresponding MF and ZF formulas can be obtained by dropping the Keldysh indices and replacing $X$ and $P$ by unity.

We conclude this subsection by giving, for future reference, the Fourier-transformed KF versions of 
the first lines of \Eqs{eq:eom_F_example_2p} and \eqref{eq:eom_F_example_4p}:
\begin{align}
&
F^{1,\,kk'}[\opd_1, \opdd_2](\omega) = G^{kk'}[\opq_1, \opdd_2](\omega) + X^{kk'} \one_{12}\,, 
\label{eq:eom_KF_example_F_2p}
\\
&
\begin{aligned}
    &F^{1,\,\mb{k}}[\opd_1,\opdd_2,\opd_3,\opdd_4](\mb{\omega})
    = G^\mb{k}[\opq_1, \opdd_2, \opd_3, \opdd_4](\mb{\omega}) \\
    &\qquad - 2\pi i \delta(\omega_{12}) X^{k_1 k_2} \one_{12} G^{k_3 k_4}[\opd_3, \opdd_4](\omega_3, \omega_4) \\
    &\qquad - 2\pi i \delta(\omega_{14}) X^{k_1 k_4} \one_{14} G^{k_2 k_3}[\opdd_2, \opd_3](\omega_2, \omega_3).
\end{aligned}
\label{eq:eom_KF_example_F_4p}
\end{align}
To arrive at these equations from \Eq{eq:eom_KF_F_Keldysh}, we used
\begin{equation}
    \sum_{k_{12}} P^{k_1 k_2 k_{12}} G^{k_{12}}[\one_{12}]
    = \sum_{k_{12}} P^{k_1 k_2 k_{12}} \sqrt{2} \delta_{k_{12}, 2}
    = X^{k_1 k_2}
\end{equation}
for \Eq{eq:eom_KF_example_F_2p}, and, for \Eq{eq:eom_KF_example_F_4p}, we used an analogous equation for $G^{k_{12}k_3 k_4}[\one_{12}, \opd_3, \opdd_4]$.

\subsection{EOM for connected correlators} \label{sec:eom_connected}

\begin{figure}[t]
    \centering
    \includegraphics[width=1.0\columnwidth]{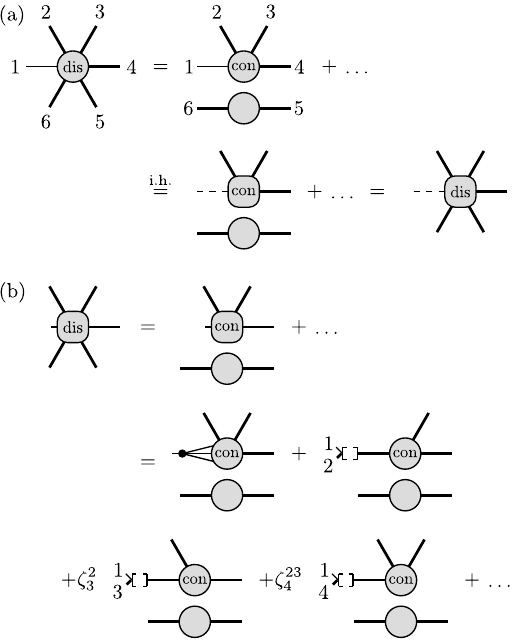}
    \caption{%
     (a) Diagrammatic representation of the inductive proof, starting from \Eq{eq:eom_dis_G}, for a disconnected 6p correlator $G_\mr{dis}[1]$, for which $\mcO^1$ is a single-particle operator.
     We only display diagrams corresponding to $S = \{ 2,3,4 \}$; the dots represent contributions from all other disconnected parts.
     In the second step, we used the induction hypothesis (i.h.) to obtain \Eq{eq:eom_dis_G_2}.
     (b) Corresponding diagrammatic representation of $F_\mr{dis}[1]$, see \Eq{eq:eom_dis_F}.
     In the second equality, we expand $F^1_{\rm con}$ using the definition of $\mc{F}$ [\Eq{eq:eom_F_def}; \Fig{fig:diag_eom_bare}(a)].
    }
    \label{fig:diag_disconnected}
\end{figure}

A vertex is defined in terms of the connected part of the corresponding correlator, i.e., the part that cannot be expressed through products of lower-point correlators.
It is therefore desirable to have an EOM directly applicable to connected correlators.
Here, we show that the EOM \eqref{eq:eom_KF_impurity} holds also if one evaluates only the connected (con) part or only the disconnected (dis) part of the correlators from both sides:
\begin{equation} \label{eq:eom_imp_conn}
\begin{aligned}
    G_{\rm con/dis}[m] &= g^0_m X_m F_{\rm con/dis}[m],
    \\
    G_{\rm con/dis}[m^\dagger] &= -F_{\rm con/dis}[m^\dagger] X_m g^0_m.
\end{aligned}
\end{equation}
When distinguishing between the connected and disconnected parts, we treat a composite operator as a whole, with the single-particle operators comprising it considered to be mutually connected.
For instance, in the expansion of $\mcG[\opd_1\opdd_2\opd_3, \opdd_4](t, 0)$ based on Wick's theorem, the term $\mcG[\opd_1,\opdd_4](t, 0) \mcG[\opdd_2, \opd_3](t, t)$ appears.
Even though the two propagators appear disconnected, this term is classified as connected when concerning the correlator of $\opd_1\opdd_2\opd_3$ and $\opdd_4$, because the two operators $\opdd_2$ and $\opd_3$ from the second correlator are connected to 
the operator $\opd_1$ from the first.

Considering the connected part vastly simplifies the EOM for $\ell \geq 3$.
For example, the connected part of the $F^1$ correlator in \Eq{eq:eom_KF_example_F_4p} simply reads
\begin{equation} \label{eq:eom_conn_example_F_4p}
    F^{1,\,\mb{k}}_{\rm con}[\opd_1,\opdd_2,\opd_3,\opdd_4](\mb{\omega})
    = G^\mb{k}_{\rm con}[\opq_1, \opdd_2, \opd_3, \opdd_4](\mb{\omega}) .
\end{equation}
For 2p correlators, the EOM for the connected part has the same form as the total EOM because the disconnected part is zero for a 1p correlator.

\Equ{eq:eom_imp_conn} can be understood inductively.
Let us assume that the EOM holds for the connected $\ell'$p correlators for all $\ell' < \ell$.
Disconnected \lp\ correlators involve sums over products of connected and disconnected lower-point correlators.
According to the inductive assumption, the connected factors already satisfy the EOMs, while the disconnected factors are spectators regarding the manipulations performed when applying the EOMs.
Therefore, their product also satisfies the EOMs. 
This idea is schematically illustrated in Fig.~\ref{fig:diag_disconnected} for $\ell=6$.

We now develop this idea into a formal proof.
For $\ell=1$, the disconnected part is zero, so the EOM trivially holds for both parts.
Now, we assume that the EOM holds for connected $\ell'$p correlators with all $\ell' < \ell$ and show that the EOM holds for the disconnected $\ell$p correlator.
Without loss of generality, we set $m=1$, as the EOMs for $m \neq 1$ follow from the former by permuting the operators.
The disconnected part of the \lp\ correlator can be expressed as
\begin{equation} \label{eq:eom_dis_G}
    G_{\rm dis}[\opd_1, \mb{\mc{O}}^\rmv{1}]
    = \sum_{\mc{S}} \zeta^\mc{S} G[\mb{\mc{O}}^{\mc{S}'}] G_{\rm con}[\opd_1, \mb{\mc{O}}^\mc{S}]
\end{equation}
where $\mb{\mcO}^\mc{S}$ and $\mb{\mcO}^{\mc{S}'}$ are sublists of $\mb{\mcO}$ listing the operators connected or not connected to $\opd_1$, respectively, indexed by sets $\mc{S}$ and $\mc{S}'$ with $\mc{S} \cup \mc{S}' = \{ 2,\ccdots,\ell\}$.
$\zeta^\mc{S}$ is a sign factor and the sum $\sum_\mc{S}$ enumerates all disconnected contributions.
The diagram for $\ell=6$ and $S=\{2,3,4 \}$ is shown in the first line of Fig.~\ref{fig:diag_disconnected}(a).

By the induction hypothesis, $G_{\rm con}[\opd_1, \mb{\mcO}^\mc{S}]$ satisfies \Eq{eq:eom_imp_conn}. Hence, we can write \Eq{eq:eom_dis_G} as
\begin{align} \label{eq:eom_dis_G_2}
    G_{\rm dis}[\opd_1, \mb{\mc{O}}^\rmv{1}]
    &= \sum_{\mc{S}} \zeta^\mc{S} G[\mb{\mc{O}}^{\mc{S}'}] g_1^0 X_1 F_{\rm con}[\opd_1, \mb{\mc{O}}^\mc{S}] \nnnl
    &= g^0_1 X_1 F_{\rm dis}[\opd_1, \mb{\mc{O}}^\rmv{1}]
    ,
\end{align}
where, in the last step, we identified
\begin{equation} \label{eq:eom_dis_F}
    F^1_{\rm dis}[\mb{\mc{O}}]
    = \sum_{\mc{S}} \zeta^\mc{S} G[\mb{\mc{O}}^{\mc{S}'}] F^1_{\rm con}[\mcO^1, \mb{\mc{O}}^\mc{S}].
\end{equation}
Figure~\ref{fig:diag_disconnected}(b) shows the corresponding diagrams for $\ell=6$ operators and $S=\{2,3,4 \}$.
\Equ{eq:eom_dis_G_2} is the desired \lp\ EOM of the disconnected part.
The \lp\ EOM holds also for the connected part, since $G_{\rm con} = G - G_{\rm dis}$, concluding the proof.

% ======================================================
\subsection{EOM with full propagators}  \label{sec:eom_interacting}

The EOM derived in the previous sections involves full correlators and bare propagators.
For methods where bare and full propagators stem from different numerical settings, it is desirable, as argued before, to exclusively use fully renormalized objects~\cite{1998BullaEstimator,2001BullaNRG,2021LeePRX,2022KuglerEstimator}. 
Here, we derive such an EOM.
The idea, inspired by Ref.~\cite{2022KuglerEstimator}, is to express 
the bare propagator $g^0$ through the full propagator $g$ and the self-energy $\Sigma$ using $(g^0)^{-1} = g^{-1} + \Sigma$.
Applying this manipulation to the EOM [\Eq{eq:eom_imp_conn}] yields
\begin{subequations}\label{eq:eom_imp_final}
\begin{align}
     \label{eq:eom_imp_final_d}
    g^{-1}_m G_{\rm con}[m]
    &= \phantom{-} X_m F_{\rm con}[m] - \Sigma_m G_{\rm con}[m],
    \\
    \label{eq:eom_imp_final_ddag}
    G_{\rm con}[m^\dagger] g^{-1}_m
    &= -F_{\rm con}[m^\dagger] X_m - G_{\rm con}[m^\dagger] \Sigma_m.
\end{align}
\end{subequations}
\Equs{eq:eom_imp_final} are the first main result of this paper.
They generalize the 4p MF EOM of \citet{2012HafermannEstimator} for $m=1$ to an arbitrary \lp\ correlator, to any index $m$, and to the KF and ZF.
The $g^{-1}$ term on the left-hand side amputates one external leg of the correlator; the terms on the right achieve this amputation in a manner that conveniently avoids division by $g$.
Repeated use of such manipulations will amputate the connected correlator and thus yield the vertex without the need to explicitly divide out the propagators.
Thereby, we obtain improved estimators for multipoint vertices.

\begin{figure}[t]
    \centering
    \includegraphics[width=1.0\columnwidth]{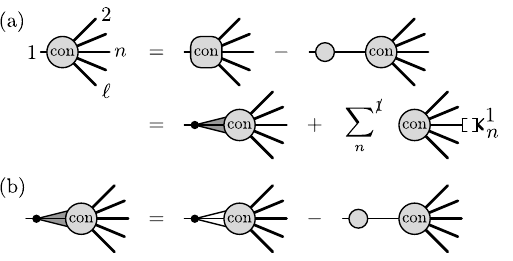}
    \caption{%
    (a) Depiction of the EOMs \eqref{eq:eom_imp_final} (first line) and \eqref{eq:eom_bullet} (second line) for the connected part of a full propagator, $G_{\rm con}[1]$, for a quartic interaction.
    We used \Eq{eq:F_m_bracket_def} to represent $F^{[\ ]}_{\rm con}$ in terms of correlators.
    (b) Depiction of \Eq{eq:G_bullet_def} for $G_{\rm con}[1_\bullet]$ (indicated by dark shading between internal legs), involving the subtraction of one-particle-reducible contributions.
    The short length of the leg labeled 1 indicates that the corresponding external leg has been amputated.
    }
    \label{fig:diag_eom_interacting}
\end{figure}

The second terms on the right of \Eqs{eq:eom_imp_final} subtract one-particle-reducible (1PR) contributions from the first terms.
To make this explicit, we express $F_{\rm con}[m^{(\dagger)}]$ as
\begin{equation} \label{eq:F_con_1PR}
    F_{\rm con}[m^{(\dagger)}]
    = (-) G_{\rm con}[q_m^{(\dagger)}, \mb{\mcO}^\rmv{m}]
    + F^{[\ ]}_{\rm con}[m^{(\dagger)}],
\end{equation}
where we define
\begin{equation} \label{eq:F_m_bracket_def}
    F^{[\ ]}_{\rm con}[m^{(\dagger)}]
    = \rmvsum{n}{m} P^{k_m k_n k_{mn}}
    G_{\rm K,\,con}^{\mb{k}^{\rmv{m}n}}[\mb{\mcO}^{\rmv{m}n}](\mb{\omega}^{\rmv{m}n}) \Bigr|_{\mcO^m = \psi^{(\dagger)}_m}.
\end{equation}
By definition, the first and second terms on the right of \Eq{eq:F_con_1PR} are obtained from those of \Eq{eq:eom_KF_F_Keldysh} by replacing $\mcO^m$ there 
by $\opp_m$ (or $\oppd_m$).
The superscript on $F^{[\ ]}_{\rm con}$
indicates that its operator argument $\mb{\mcO}^{\rmv{m}n}$
involves an (anti)commutator.
The $(-)$ sign before $G_{\rm con}$, applicable for $F_{\rm con}[m^\dagger]$ but not for
$F_{\rm con}[m]$, reflects the  sign difference in definitions \eqref{eq:q_def_general} for $\opqd_m$ and $\opq_m$.
Then, \Eqs{eq:eom_imp_final} can be expressed as
\begin{subequations}\label{eq:eom_bullet}
\begin{align}
     \label{eq:eom_bullet_d}
    g^{-1}_m G_{\rm con}[m]
    &= G_{\rm con}[m_\bullet] + X_m  F^{[\ ]}_{\rm con}[m],
    \\
    \label{eq:eom_bullet_ddag}
    G_{\rm con}[m^\dagger] g^{-1}_m
    &= G_{\rm con}[m^\dagger_\bullet] - F^{[\ ]}_{\rm con}[m^{\dagger}] X_m.
\end{align}
\end{subequations}
where we defined
\begin{subequations}\label{eq:G_bullet_def}
\begin{align}
     \label{eq:G_bullet_def_d}
    G_{\rm con}[m_\bullet]
    &= X_m G_{\rm con}[\opq_m, \mb{\mcO}^\rmv{m}]
    - \Sigma_m G_{\rm con}[m],
    \\
    \label{eq:G_bullet_def_ddag}
    G_{\rm con}[m^\dagger_\bullet]
    &= G_{\rm con}[\opqd_m, \mb{\mcO}^\rmv{m}] X_m
    - G_{\rm con}[m^\dagger] \Sigma_m.
\end{align}
\end{subequations}
The second terms on the right of \Eqs{eq:G_bullet_def} subtract the 1PR contributions from the first terms, completing the amputation of the $m$-th leg.
Figure~\ref{fig:diag_eom_interacting} gives a diagrammatic depiction of \Eqs{eq:eom_imp_final} and \eqref{eq:G_bullet_def}.

% =======================================================================================
\section{Symmetric improved estimators} \label{sec:estimator}

In this section, we use the EOM to derive improved estimators for the self-energy, 3p vertex, and 4p vertex.
Although we write the formula in the KF, let us emphasize that all results of this section apply also to the MF and ZF when all the Keldysh indices are dropped and the coefficients $X^{kk'}$ and $P^{k_1\ccdots k_n}$ are replaced by unity.

We confine ourselves to Hamiltonians with a quartic interaction:
\begin{equation} \label{eq:Hint_quartic}
    \opH_{\rm int} = \sum_{i,i',j,j'} U_{ii'jj'} \opdd_i \opd_{i'} \opdd_j \opd_{j'}.
\end{equation}
Due to the sum over orbital indices, 
different choices of the $U$ tensor can describe the same interaction.
The symmetrized interaction tensor
\begin{equation} \label{eq:Usym}
    U^{\rm sym}_{1234} = U_{1234} + \zeta U_{3214} + \zeta U_{1432} + U_{3412},
\end{equation}
is unique for a given interaction.
For the single-orbital AIM [\Eq{eq:2p_AIM}], one may choose $U_{\uparrow\uparrow\downarrow\downarrow} = U_0$ and let all other components be zero to get
$U^{\rm sym}_{\sigma \sigma \sigma' \sigma'} = \zeta U^{\rm sym}_{\sigma \sigma' \sigma' \sigma} = U_0 \delta_{\bar{\sigma}, \sigma'}$,
where $\bar{\sigma}$ denotes the opposite spin to $\sigma$.
Later, we show that $U^{\rm sym}$ equals the bare vertex [see \Eqs{eq:Gamma_bare_zf} and \eqref{eq:Gamma_bare_kf}].

For notational convenience, we henceforth focus on \lp\ correlators with $\ell \le 4$
(though the strategy presented below can readily be generalized).
We write the 4p connected correlator as
\begin{equation} \label{eq:est_G4_def}
    G_{1234}^{(\cdot,\cdot,\cdot,\cdot)\mb{k}}(\mb{\omega}) = G^{\mb{k}}_{\rm con}[\opd_1, \opdd_2, \opd_3, \opdd_4](\mb{\omega}),
\end{equation}
using odd (even) indices for annihilation (creation) operators.
The superscript $(\cdot,\cdot,\cdot,\cdot)$ indicates that this correlator is a 4p object and will later serve as a ``parent correlator'' for the definition of various auxiliary correlators.
Hereafter, we omit the subscript $1234$ and superscript $\mb{k}$ for 4p correlators.
We primarily work with the connected correlators, 
as vertices are defined from these by amputating their external legs.
Working with the connected part also simplifies the equations after repeated applications of the EOM.
The 4p vertex is defined as
\begin{equation} \label{eq:est_Gamma_def}
    \Gamma(\mb{\omega}) = g^{-1}_1(\omega_1) g^{-1}_3(\omega_3) G^{(\cdot,\cdot,\cdot,\cdot)}(\mb{\omega}) g^{-1}_2(-\omega_2) g^{-1}_4(-\omega_4),
\end{equation}
where $g^{-1}_m$ is matrix-multipied to the $m$-th orbital and Keldysh indices of the 4p correlator.
This 4p vertex is called $F$ in Refs.~\cite{2017GallerDVA,2018RohringerRMP,2019KaufmannEstimator,2020WentzellPRB,2021KuglerPRX,2021LeePRX} and $\gamma$ in Ref.~\cite{2012HafermannEstimator}.

% ================================================
\subsection{Auxiliary correlators}
\label{sec:AuxilaryCorrelators}

As for $G^{2\mr{p}}$, the estimators for  $G^{4\mr{p}}$ will be obtained through multiple applications of EOMs. These will generate various auxiliary correlators, all derived from 
the same ``parent correlator''  $G^{4\mr{p}}$, 
containing not only the single-particle operators $\opd_i$ and $\opdd_i$, 
but also the composite operators $\opq_i = \big[ \opd_i, \opH_{\rm int} \big]$
and $\opqd_i = \big[ \opH_{\rm int}, \opdd_i \big]$, 
and (possibly nested) (anti)commutators of all of these, arising from the 
$[\mc{O}^n,H_\mr{int}]$ and $\delta(m,n)$ terms in the EOMs, respectively. 
For (anti)commutators involving composite operators, we recursively introduce the following compact notation [generalizing \Eq{eq:q2_def}]:
\begin{equation}
\begin{aligned}
    \opq_{i_1 i_2} &= [q_{i_1}^{(\dagger)}, d_{i_2}^{(\dagger)}]_{\zeta^1} , \\
    \opq_{i_1 i_2  i_3} &= [q_{i_1 i_2}, d_{i_3}^{(\dagger)}]_{\zeta^2} , \\
    \opq_{i_1 \ccdots i_n}
    &= [\opq_{i_1 \ccdots i_{n-1}}, d^{(\dagger)}_{i_n}]_{\zeta^{n-1}} .
\end{aligned}
\end{equation}
An anticommutator is taken if both operators are fermionic; a commutator is taken otherwise.
Daggers are used if the corresponding operators in the parent correlator [\Eq{eq:est_G4_def}] have daggers.
Such indices will be labelled with an even integer, 2 or 4.
For example, for a fermionic system, the composite operators derived from \Eq{eq:est_G4_def}
include the following:
\begin{gather} \label{eq:q_def_n}
    \opq_{12} = \{\opq_1, \opdd_2 \}, \
    \opq_{23} = \{\opqd_2, \opd_3 \}, \\
    \opq_{123} = [ \opq_{12}, \opd_3 ], \
    \opq_{234} = [ \opq_{23}, \opdd_4 ], \
    \opq_{1234} = \{\opq_{123}, \opdd_4 \}. \nonumber
\end{gather}

The KF versions of the auxiliary correlators of \Eqs{eq:2p_clamped} are defined as
\begin{equation}
\begin{aligned}
    G^{(\cdot,\cdot)\mb{k}}_{12}(\omega) &= G^{\mb{k}}[\opd_1, \opdd_2](\omega) = g_{12}^{\mb{k}}(\omega), \\
    G^{(1,\cdot)\mb{k}}_{12}(\omega) &= G^{\mb{k}}[\opq_1, \opdd_2](\omega),\\
    G^{(\cdot,2)\mb{k}}_{12}(\omega) &= G^{\mb{k}}[\opd_1, \opqd_2](\omega),\\
    G^{(1,2)\mb{k}}_{12}(\omega) &= G^{\mb{k}}[\opq_1, \opqd_2](\omega),\\
    G^{(12)\mb{k}}_{12} &= P^{k_1 k_2 k_{12}} G^{k_{12}}\bigl[[ \opq_1, \opdd_2 ]_{\zeta}\bigr].
\end{aligned}
\label{eq:est_clamped_2p}
\end{equation}
Hereafter, for 2p auxiliary correlators, we omit the subscript $12$ for orbital indices and the superscript $\mb{k}$ for Keldysh indices.
The diagrammatic representations are given in Fig.~\ref{fig:def_clamp_2p}.
As in the MF case, $G^{(12)}$ is the Hartree self-energy:
\begin{align} \label{eq:hartree}
    G^{(12)}
    = P^{k_1 k_2 k_{12}} \bigl\langle[\opq_1, \opdd_{2}]_{\zeta}\bigr \rangle \sqrt{2} \delta_{k_{12}, 2}
    = \Sigma^{\rm H}_{12} X^{k_1 k_2}.
\end{align}
Here, the $\sqrt{2} \delta_{k_{12}, 2}$ term comes from the transformation of $G^c[\opq_{12}] = \expval{q_{12}}$ to the Keldysh basis as $\sum_c D^{kc} = \sqrt{2} \delta_{k,2}$.
The factor $P^{k_1 k_2 2} = \tfrac{1}{\sqrt{2}} X^{k_1 k_2}$ [\Eq{eq:not_P_def_3}] 
maps the single Keldysh index of $G^{k_{12}}$ via a summation on $k_{12}$ to a two-fold Keldysh index $k_1 k_2$.

Next, we consider connected auxiliary correlators derived from the parent $G^{4 \rm{p}}$ of \Eq{eq:est_G4_def}. We illustrate our notational conventions, described below, with some examples,
assuming all $d_i^{(\dagger)}$ to be fermionic:
\begin{align} \label{eq:est_clamped_4p}
    G^{(\cdot,\cdot,\cdot,\cdot)}
    &= G^{\mb{k}}_{\rm con}[\opd_1, \opdd_2, \opd_3, \opdd_4],
    \nnnl
    G^{(1,2,\cdot,\cdot)}
    &= G^{\mb{k}}_{\rm con}[\opq_1, \opqd_2, \opd_3, \opdd_4],
    \nnnl
    G^{(1,\cdot,3,\cdot)}
    &= G^{\mb{k}}_{\rm con}[\opq_1, \opdd_2, \opq_3, \opdd_4],
    \nnnl
    G^{(12,\cdot,\cdot)}
    &= P^{k_1 k_2 k_{12}} G^{k_{12}k_3k_4}_{\rm con}[\{ \opq_1, \opdd_2 \}, \opd_3, \opdd_4],
    \nnnl
    G^{(23,\cdot,\cdot)}
    &= P^{k_2 k_3 k_{23}} G^{k_{23} k_1 k_4}_{\rm con}[\{ \opqd_2, \opd_3 \}, \opd_1, \opdd_4],
    \\
    G^{(12,3,\cdot)}
    &= P^{k_1 k_2 k_{12}} G^{k_{12}k_3k_4}_{\rm con}[\{ \opq_1, \opdd_2 \}, \opq_3, \opdd_4],
    \nnnl
    G^{(124,3)}
    &= P^{k_1  k_2 k_4 k_{124}} G^{k_{124}k_3}_{\rm con} [[ \{ \opq_1, \opdd_2 \}, \opdd_4], \opq_3],
    \nnnl
    G^{(12,34)}
    &= P^{k_1 k_2 k_{12}} P^{k_3 k_4 k_{34}} G^{k_{12} k_{34}}_{\rm con} [\{ \opq_1, \opdd_2 \}, \{ \opq_3, \opdd_4 \}], 
    \nnnl
    G^{(1234)}
    &= P^{k_1 k_2 k_3 k_4 k_{1234}} G^{k_{1234}}_{\rm con} [\{ [ \{ \opq_1, \opdd_2 \}, \opd_3], \opdd_4 \}]. \nonumber
\end{align}
By definition, all correlators $G^{(\ccdots)}$ carrying superscripts 
in round brackets are \textit{connected} correlators.
Depending on the number and type of composite operators involved, they may be  4p, 3p, 2p, and 1p correlators; correspondingly, the superscripts contain 4, 3, 2, or 1 comma-separated arguments.
As before, `$\cdot$' is a placeholder for $d_i^{(\dagger)}$, a solitary numeral $i$ signals its replacement by $q_i^{(\dagger)}$, and $i_1\ccdots i_n$ denotes the replacement of the corresponding operators by the composite operator $\opq_{i_1 \ccdots i_n}$.

All such auxiliary correlators depend on the same number of indices and frequency arguments: 4 orbital indices, 4 Keldysh indices, and 4 frequency arguments.
These are inherited from those of parent $G^{4\mr{p}}$,
either directly for single-particle operators $d_i^{(\dagger)}$ and $q_i^{(\dagger)}$, 
or indirectly for composite operators, according to the following
rules: to $\opq_{i_1\ccdots i_n}$, assign the frequency $\omega_{i_1 \ccdots i_n} = \omega_{i_1} + \ccdots + \omega_{i_n}$ and the dummy Keldysh index $k_{i_1 \ccdots i_n}$, then map the latter to an $n$-fold index $k_{i_1} \ccdots k_{i_n}$ through multiplication by the rank-$(n+1)$ tensor $P^{k_{i_1} \ccdots k_{i_n} k_{i_1 \ccdots i_n}}$~[\Eq{eq:not_P_def}] and summation over the dummy $k_{i_1 \ccdots i_n}$. 

Finally, when defining auxiliary correlators, we order the operator arguments according to the following conventions: (i) operators with higher nesting come first, and (ii) non-nested operators ($\opq$, $\opd$, $\opqd$, and $\opdd$) are ordered by their subscripts in increasing order.
In \Eq{eq:est_clamped_4p}, we suppressed frequency arguments, since they can be inferred from the structure of the superscripts.
For example, the superscripts $(12,\cdot,\cdot)$ and $k_{12} k_3 k_4$ both indicate a frequency argument $(\omega_{12},\omega_3,\omega_4)$, while superscripts $(124,3)$ and $k_{124} k_3$ both indicate $(\omega_{124}, \omega_3)$, etc.

\begin{figure}[t]
\centering
\includegraphics[width=1.0\columnwidth]{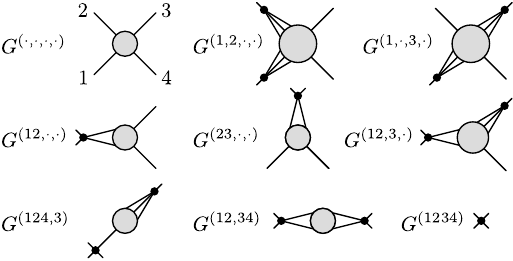}
\caption{Diagrammatic representations of the 4p auxiliary correlators of \Eq{eq:est_clamped_4p} for a quartic interaction.
Only the connected diagrams are evaluated.
 }
\label{fig:def_clamp_4p}
\end{figure}

Figure~\ref{fig:def_clamp_4p} is a diagrammatic representation of the 4p auxiliary correlators listed in \Eqs{eq:est_clamped_4p}.
Some auxiliary correlators, such as $G^{(12,\cdot,\cdot)}$, $G^{(12,34)}$, and $G^{(1234)}$, contain bosonic operators.
Diagrams in which the latter
are disconnected from all the other operators should also be subtracted to obtain the connected part.
As mentioned before, for composite operators, all the external legs of the constituent single-particle operators are regarded as being connected to each other.
As reflected in the diagram, the 1p correlator $G^{(1234)}$ equals the bare vertex up to a sign factor:
\begin{subequations}
\begin{flalign} \label{eq:Gamma_bare}
    -G^{(1234)} &= \Gamma_{\rm bare}, \\
    \label{eq:Gamma_bare_zf}
    \Gamma_{\rm bare} &= U^{\rm sym}_{1234} \qquad \text{(MF, ZF)} ,  \\
    \label{eq:Gamma_bare_kf}
    \Gamma_{\rm bare}^\mb{k}
    & = \begin{cases}
        \frac{1}{2} U^{\rm sym}_{1234} & \text{if $k_1\!+\!k_2\!+\!k_3\!+\!k_4$ is odd} \\
        0 & \text{otherwise}
    \end{cases}  \hspace{-1cm} &
\end{flalign}
\end{subequations}
with $U^{\rm sym}$ defined in \Eq{eq:Usym}.

We also define auxiliary correlators where some operators are replaced by $q_i^{(\dagger)}$, and the corresponding 1PR contributions are subtracted as in \Eq{eq:G_bullet_def}.
We denote such correlators using bullets (`$\bullet$') instead of dots (`$\cdot$') in the superscript and define them as
\begin{equation}
\begin{aligned}
    G^{(\bullet, x_2)}
    &= X_1 G^{(1,x_2)} - \Sigma_1 G^{(\cdot,x_2)}\,,
    \\
    G^{(x_1, \bullet)}
    &= G^{(x_1,2)} X_2 - G^{(x_1, \cdot)} \Sigma_2\,,
    \\
    G^{(12,\bullet, x_4)}
    &= X_3 G^{(12,3,x_4)} - \Sigma_3 G^{(12,\cdot,x_4)}\,,
    \\
    G^{(12,x_3, \bullet)}
    &= G^{(12,x_3,4)} X_4 - G^{(12,x_3,\cdot)} \Sigma_4\,,
    \\
    G^{(\bullet,x_2, x_3, x_4)}
    &= X_1 G^{(1,x_2, x_3, x_4)} - \Sigma_1 G^{(\cdot,x_2, x_3, x_4)}\,,
    \\
    G^{(x_1, \bullet, x_3, x_4)}
    &= G^{(x_1, 2, x_3, x_4)} X_2 - G^{(x_1, \cdot, x_3, x_4)} \Sigma_2 \,,
    \\
    G^{(x_1, x_2, \bullet, x_4)}
    &= X_3 G^{(x_1, x_2, 3, x_4)} - \Sigma_3 G^{(x_1, x_2, \cdot, x_4)}\,,
    \\
    G^{(x_1, x_2, x_3, \bullet)}
    &= G^{(x_1, x_2, x_3, 4)} X_4 - G^{(x_1, x_2, x_3, \cdot)} \Sigma_4 \,,
\end{aligned}
\label{eq:G_aux_bullet_def}
\end{equation}
where $x_n \in \{ \bullet, \cdot, n\}$.
$X_n$ and $\Sigma_n$ are left (right) multiplied for odd (even) $n$, as they correspond to an annihilation (creation) operator in the original 4p correlator [\Eq{eq:est_G4_def}].
This definition will be used repeatedly in the derivation of improved estimators.
Note that $X_n$ and $\Sigma_n$ are left (right) multiplied for 
odd (even) indices, reflecting the absence or presence of a dagger in the corresponding operator of the parent correlator [\Eq{eq:est_G4_def}].
One can apply this definition recursively to evaluate auxiliary correlators with multiple bullets in the superscript, e.g.,
\begin{align}
G^{(\bullet,\bullet,\cdot,\cdot)}
&= G^{(\bullet,2,\cdot,\cdot)} X_2 - G^{(\bullet,\cdot,\cdot,\cdot)} \Sigma_2 \nnnl
&= X_1 G^{(1,2,\cdot,\cdot)} X_2 - \Sigma_1 G^{(\cdot,2,\cdot,\cdot)} X_2 \nnnl
&\quad - X_1 G^{(1,\cdot,\cdot,\cdot)} \Sigma_2
+ \Sigma_1 G^{(\cdot,\cdot,\cdot,\cdot)} \Sigma_2
.
\end{align}

% ================================================
\subsection{Self-energy estimators} \label{sec:est_selfen}

We now derive the sIEs, starting with the self-energy.
We will reproduce the result of Ref.~\cite{2022KuglerEstimator} but will take a slightly different path.
Instead of using the EOM with the bare propagator \eqref{eq:eom_KF_impurity},
we apply the EOM with the full propagator \eqref{eq:eom_imp_final} to $G^{(\cdot, \cdot)}$ twice, once for each external leg.
This amputates the legs and yields the self-energy.
The same procedure will be used to derive the multipoint vertex estimators.

First, using $g = G^{(\cdot,\cdot)} = G[\opd_1, \opdd_2]$ in \Eqs{eq:eom_imp_final}, we find
\begin{subequations}
\begin{align} 
\label{eq:2p_eom_1st-a}    
    g^{-1} g & = X G^{(1,\cdot)}  - \Sigma g + \one , 
  \\ 
\label{eq:2p_eom_1st-b}
g g^{-1} &   = G^{(\cdot,2)} X  - g \Sigma + \one .
\end{align}
\end{subequations}
\begin{subequations}
Solving for $\Sigma$, we find two aIEs for the self-energy, distinguished here by superscripts and illustrated in Fig.~\ref{fig:derivation_symm_Sigma_estimator}(c):
    \label{eq:2p_Sigma_1st}
\begin{align} 
\label{eq:2p_Sigma_1st-a}
    \Sigma^{\rm L}
    & = X G^{(1,\cdot)} g^{-1} ,  \\
\label{eq:2p_Sigma_1st-b}    
    \Sigma^{\rm R} & = g^{-1} G^{(\cdot,2)} X .
\end{align}
\end{subequations}

Next, we employ EOMs for the auxiliary correlators that appear in the aIE by using $G^{(1,\cdot)} = G[\opq_1, \opdd_2]$ in \Eq{eq:eom_imp_final_ddag} or $G^{(\cdot,2)} = G[\opd_1, \opqd_2]$ in \Eq{eq:eom_imp_final_d}:
\begin{subequations} \label{eq:2p_eom_12}
\begin{align} 
    \label{eq:2p_eom_12-a}
    G^{(1,\cdot)} g^{-1} &= \big( G^{(1,2)} + G^{(12)} \big) X - G^{(1, \cdot)} \Sigma^{\rm R} \, ,
    \\
    \label{eq:2p_eom_12-b}
    g^{-1} G^{(\cdot,2)} &= X \big( G^{(1,2)} + G^{(12)} \big) - \Sigma^{\rm L} G^{(\cdot, 2)}.
\end{align}
\end{subequations}
On the right, we used $\Sigma^{\rm R}$ (or $\Sigma^{\rm L}$) in terms containing the self-energy as a factor on the right (or left), because this choice leads to the third, symmetric self-energy estimator discussed in \ref{sec:est_selfen}.
It is obtained by substituting \Eqs{eq:2p_eom_12} into the aIEs of \Eqs{eq:2p_Sigma_1st}.
The two expressions obtained this way,
\begin{subequations}  \label{eq:2p_Sigma_2ndLR}
\begin{align} \label{eq:2p_Sigma_2nd-LRa}
    \Sigma^{\rm S}
    &= X \big[ \big( G^{(1,2)} + G^{(12)} \big) X - G^{(1, \cdot)} \Sigma^{\rm R} \big] , \\ 
    \label{eq:2p_Sigma_2nd-LRb}
    \Sigma^{\rm S} &=  \big[ X \big( G^{(1,2)} + G^{(12)} \big) - \Sigma^{\rm L} G^{(\cdot, 2)}  \big] X ,
\end{align}
\end{subequations}
are equal (hence we denote both by $\Sigma^{\rm S}$), as can be seen by inserting the aIEs of \Eq{eq:2p_Sigma_1st} on the right (we also use $X G^{(12)} X = \Sigma^{\rm H} X$ [\Eq{eq:hartree}]):
\begin{align} \label{eq:2p_Sigma_2nd}
    \Sigma^{\rm S}&= X G^{(1,2)} X + \Sigma^{\rm H} X - X G^{(1,\cdot)} g^{-1} G^{(\cdot,2)} X.
\end{align}
\Equ{eq:2p_Sigma_2nd} is the Keldysh version of the sIE for the self-energy illustrated in \Fig{fig:derivation_symm_Sigma_estimator}(e).
\begin{figure}[!tbp]
    \centering
    \includegraphics[width=1.0\columnwidth]{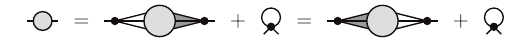}
    \caption{Depiction of \Eq{eq:SigmaSubtractionBulletNotation} for the symmetric self-energy estimator $\Sigma^{\rm S}$, expressed through 1PR-subtracted correlators [cf.~\Fig{fig:diag_eom_interacting}(b)].
     }
    \label{fig:self_energy_shaded}
\end{figure}
The term involving $G^{(1,\cdot)} g^{-1} G^{(\cdot,2)}$ subtracts all 1PR diagrams from $G^{(1,2)}$ (cf.\ Fig.~\ref{fig:pert_exp}).
Using \Eqs{eq:2p_Sigma_2ndLR}, this 1PR subtraction can also be expressed as 
\begin{align}
\label{eq:SigmaSubtractionBulletNotation}
    \Sigma^{\rm S} =  X G^{(1, \bullet)}  + X \Sigma^{\rm H} 
    = G^{(\bullet,2)} X + \Sigma^{\rm H} X ,  
\end{align}
where we employed notation analogous to that of \Eqs{eq:G_aux_bullet_def}. 
Figure~\ref{fig:self_energy_shaded} illustrates \Eq{eq:SigmaSubtractionBulletNotation}
using the shaded vertex of \Fig{fig:diag_eom_interacting}(b).
The choice of using $\Sigma^{\rm L}$ or $\Sigma^{\rm R}$ 
in subtraction terms containing the self-energy as left or right factors will also be used in later sections when evaluating the multipoint estimators.

% ===============================================
\subsection{3p vertex estimators}
The sIE for the 4p vertex turns out to depend, among others, on a number of 3p vertices.
In this section, we therefore explain how to obtain sIEs for these.
To be concrete, we consider the 3p vertex for the auxiliary correlator $G^{(ab,\cdot,\cdot)}$, defined by amputating the external legs that correspond to $\opd$ or $\opdd$.
For example, the vertex for $G^{(12,\cdot,\cdot)}$, called $\Gamma^{(12,\cdot,\cdot)}$, or $\Gamma^{(12)}$ for short, is
\begin{align} \label{eq:3p_G12_est_def}
    &\Gamma^{(12,\cdot,\cdot)} = g^{-1}_3 G^{(12,\cdot,\cdot)} g^{-1}_4
    =\includegraphics[valign=c]{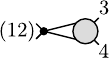}\,.
\end{align}
Here, $g^{-1}_3$ and $g^{-1}_4$ matrix-multiply the third and fourth orbital and Keldysh indices of $G^{(12,\cdot,\cdot)}$.

For fermionic systems, these 3p vertices are fermion-boson vertices which are related to the Hedin vertex~\cite{1965Hedin}.
They are an important ingredients of, e.g., diagrammatic extensions of dynamical mean-field theory and the calculation of response properties~\cite{2009Katanin,2012Rubtsov,2015Ayral,2017GallerDVA,2018vanLoon,2019KrienSBE1,2019KrienSBE2,2020KrienSBE,2021KrienSBE}.
We note that although $\Gamma^{(12)}$ has four Keldysh indices, one can easily convert it to have 3 Keldysh indices to more clearly reveal its 3p nature:
\begin{equation}
    \sum_{k_1, k_2} P^{k_1 k_2 k_{12}} \Gamma^{(12)\,\mb{k}}
    = \Gamma^{(12)\,k_{12}k_3k_4}.
\end{equation}

Our goal is to derive an estimator for $\Gamma^{(12)}$, symmetric with respect to legs 3 and 4, by using EOMs \eqref{eq:eom_imp_final} to amputate the external legs of $G^{(12,\cdot,\cdot)}$.
We first find the EOM w.r.t.\ $\omega_3$ for $G^{(12,\cdot,\cdot)} = G_{\rm con}[\opq_{12},\opd_3,\opdd_4](\omega_{12},\omega_3,\omega_4)$ by using \Eq{eq:eom_imp_final_d} with $m=2$:
\begin{align} \label{eq:3p_G12_derivation}
    &\quad
    X_3 [g^{-1}_{3} + \Sigma_{3}] G^{(12,\cdot,\cdot)}(\mb{\omega}) \nnnl
    &=  P^{k_1 k_2 k_{12}} F_{\rm con}^{2, k_{12} k_3 k_4}[\opq_{12},\opd_3,\opdd_4](\omega_{12},\omega_3,\omega_4) \nnnl
    &=
    P^{k_1 k_2 k_{12}}
    \Bigg[ G_{\rm con}^{k_{12} k_3 k_4}[\opq_{12},\opq_3,\opdd_4](\omega_{12},\omega_3,\omega_4) \nnnl
    &\quad + P^{k_{12} k_3 k_{123}} G_{\rm con}^{k_{123} k_4} \left[ [\opd_3, \opq_{12} ],\opdd_4 \right](\omega_{123},\omega_4) \nnnl
    &\quad + P^{k_3 k_4 k_{34}} G_{\rm con}^{k_{12} k_{34}}\left[ \opq_{12}, [\opd_3, \opdd_4]_{\zeta} \right](\omega_{12},\omega_{34}) \Bigg].
\end{align}
The first term in the square bracket gives $G^{(12,3,\cdot)}(\mb{\omega})$ [\Eq{eq:est_clamped_4p}].
The second term gives $-G^{(123,\cdot)}(\mb{\omega})$ when \Eq{eq:3p_PP_P}, the identity $P \cdot P = P$, is used; the minus sign comes from $[\opd_3, \opq_{12}] = -\opq_{123}$ [\Eq{eq:q_def_n}].
The third term vanishes because
the connected part of $G\big[ \opq_{12}, [\opd_3, \opdd_4]_{\zeta} \big] = G[ q_{12}, \one_{34} ]$ is zero: the identity operator does not have any external leg and thus cannot be connected with other operators.
Therefore, \Eq{eq:3p_G12_derivation} becomes
\begin{equation}
    X_3 \bigl[ g^{-1}_{3} + \Sigma_{3} \bigr] G^{(12,\cdot,\cdot)} = G^{(12,3,\cdot)} - G^{(123,\cdot)}.
\end{equation}
Writing this equation in terms of the 1PR-subtracted auxiliary correlators  [\Eq{eq:G_aux_bullet_def}], we find
\begin{equation}
    g^{-1}_{3} G^{(12,\cdot,\cdot)} = G^{(12,\bullet,\cdot)} - X_3 G^{(123,\cdot)},
\end{equation}
Amputating the remaining fourth leg from $g^{-1}_{3} G^{(12,\cdot,\cdot)}$, one finds the aIE for the 3p vertex:
\begin{equation} \label{eq:3p_G12_1st}
    \Gamma^{(12)}
    = \bigl[ G^{(12,\bullet,\cdot)} - X_3 G^{(123,\cdot)} \bigr] g^{-1}_4.
\end{equation}

Now, consider the EOM w.r.t.\ $\omega_4$ for each of the auxiliary correlators on the right of \Eq{eq:3p_G12_1st}:
\begin{align} \label{eq:3p_G12_eom}
    G^{(12,\cdot,\cdot)} g^{-1}_{4} &= G^{(12,\cdot,\bullet)} + \zeta G^{(124,\cdot)} X_4 , \nnnl
    G^{(12,3,\cdot)} g^{-1}_{4} &= G^{(12,3,\bullet)} + \bigl[ G^{(12,34)} + \zeta G^{(124,3)} \bigr] X_4 , \nnnl
    G^{(123,\cdot)} g^{-1}_{4} &= G^{(123,\bullet)} + G^{(1234)} X_4.
\end{align}
By substituting these equations to \Eq{eq:3p_G12_1st}, one obtains the sIE for the 3p vertex:
\begin{align} \label{eq:3p_G12_2nd_der}
    &\quad \Gamma^{(12)} \nnnl
    &= G^{(12,\bullet,\bullet)}
    + X_3 \bigl[ G^{(12,34)} + \zeta G^{(124,3)} - G^{(1234)} \bigr] X_4  \nnnl
    &\quad - \zeta \Sigma_{3} G^{(124,\cdot)} X_4 - X_3 G^{(123,\bullet)} \nnnl
    &= \mcK^{(12,\cdot,\cdot)}
    + X_3 \bigl[ G^{(12,34)} - G^{(1234)} \bigr] X_4  \nnnl
    &\quad + \zeta G^{(124,\bullet)} X_4 - X_3 G^{(123,\bullet)}.
\end{align}
Here, $\mcK^{(12,\cdot,\cdot)}$, or $\mcK^{(12)}$ for short, is defined as
\begin{align} \label{eq:3p_G12_U3}
    \mcK^{(12,\cdot,\cdot)}
    &= G^{(12,\bullet,\bullet)} \nnnl
    &= X_3 G^{(12,3,4)} X_4 - X_3 G^{(12,3,\cdot)} \Sigma_{4} \nnnl
    &\quad - \Sigma_{3} G^{(12,\cdot,4)} X_4 + \Sigma_{3} G^{(12,\cdot,\cdot)} \Sigma_{4}
    .
\end{align}
$\mcK^{(12)}$ is one-particle irreducible (1PI) in the third and fourth legs thanks to the 1PR subtraction shown in Fig.~\ref{fig:diag_eom_interacting}(b).
It is a sum of four terms, obtained by performing one of the two operations for both $n = 3$ and $n=4$: (i) multiply by $-\Sigma_n$ ($x_n = \cdot$) or (ii) insert $n$ into the superscript for the auxiliary correlator and multiply by $X_n$ ($x_n = n$).
The symbol $\mcK$ is used as this term is identical (up to a sign) to the $\mcK_2$ asymptotic class of the 4p vertex~\cite{2020WentzellPRB} (see \Sec{sec:asymptotic}).

\begin{figure}[t]
    \centering
    \includegraphics[width=1.0\columnwidth]{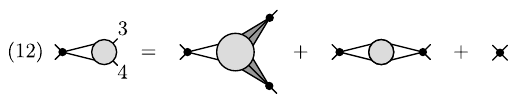}
    \caption{Diagrammatic representation of the sIE for the 3p vertex $\Gamma^{(12)}$ [\Eq{eq:3p_symmetric}], comprising only connected diagrams.
    }
    \label{fig:diag_3p}
\end{figure}

Next, we note that the last line of \Eq{eq:3p_G12_2nd_der} vanishes.
Since $\opH_{\rm int}$ is a 4p interaction, $\bigl[ \{ q_a^{(\dagger)}, d_b^{(\dagger)} \}, d_c^{(\dagger)} \bigr] = x^{(abc)}_{abcd} d^{(\dagger)}_d$ holds for a constant factor $x^{(abc)}_{abcd}$, yielding
\begin{equation}
\begin{aligned}
    G^{(123,\bullet)}
    &= x^{(123)}_{123d} (G^{(\cdot,2)}_{d4} X_4 - G^{(\cdot,\cdot)}_{d4} \Sigma_{4})
    = 0, \\
    G^{(124,\bullet)}
    &= x^{(124)}_{124d} (X_3 G^{(1,\cdot)}_{3d} - \Sigma_{3} G^{(\cdot,\cdot)}_{3d})
    = 0,
\end{aligned}
\label{eq:3p_G12_der_use}
\end{equation}
where the cancellations follow from \Eq{eq:2p_Sigma_1st}.
Using $X_3 G^{(12,34)} X_4 = G^{(12,34)}$ [via \Eq{eq:2p_XPX_P}] and $G^{(1234)} = -\Gamma_{\rm bare}$ [\Eq{eq:Gamma_bare}], we find the following compact sIE for $\Gamma^{(12)}$, represented diagrammatically in \Fig{fig:diag_3p}:
\begin{equation}
\begin{aligned} 
    \Gamma^{(12)} &= \mcK^{(12)} + G^{(12,34)} + \Gamma_{\rm bare} , \\
    \Gamma^{(13)} &= \mcK^{(13)} + G^{(13,24)} - \zeta \Gamma_{\rm bare} , \\
    \Gamma^{(14)} &= \mcK^{(14)} - G^{(14,23)} + \Gamma_{\rm bare} , \\
    \Gamma^{(23)} &= \mcK^{(23)} + G^{(14,23)} - \Gamma_{\rm bare} , \\
    \Gamma^{(24)} &= \mcK^{(24)} - G^{(13,24)} + \zeta \Gamma_{\rm bare}  , \\
    \Gamma^{(34)} &= \mcK^{(34)} + G^{(12,34)} + \Gamma_{\rm bare}
    .
\end{aligned}
\label{eq:3p_symmetric}
\end{equation}
We also listed analogous sIEs for the other 3p vertices, which can be derived similarly.

% =====================================================================
\subsection{4p vertex estimators} \label{sec:4p}
Finally, we derive a sIE for the 4p vertex $\Gamma$ [\Eq{eq:est_Gamma_def}].
We use the same strategy of repeatedly applying the EOM \eqref{eq:eom_imp_final} to 4p auxiliary correlators.
At the $m$-th order, we use the EOM w.r.t.\ $\omega_m$ as well as the lower-order estimators.

The EOM of the 4p connected correlator $G^{(\cdot,\cdot,\cdot,\cdot)}$ w.r.t.\ $\omega_1$ is
\begin{equation}
     g^{-1}_1 G^{(\cdot,\cdot,\cdot,\cdot)}= G^{(\bullet,\cdot,\cdot,\cdot)}.
\end{equation}
By amputating the remaining external legs, we find a first-order 4p aIE:
\begin{equation} \label{eq:4p_1st}
    \Gamma = g^{-1}_3 G^{(\bullet,\cdot,\cdot,\cdot)} g^{-1}_2 g^{-1}_4.
\end{equation}
This equation is the 4p aIE used in Eq.~(84) of Ref.~\cite{2021LeePRX}.
The same formula holds if one takes the correlators themselves instead of their connected parts as the disconnected parts on both sides cancel via the 2p EOM. Then, \Eq{eq:4p_1st} becomes Eq.~(26) of Ref.~\cite{2012HafermannEstimator}.

The relevant EOMs w.r.t.\ $\omega_2$ are
\begin{equation}
\begin{aligned}
    G^{(\cdot,\cdot,\cdot,\cdot)} g^{-1}_2 &= G^{(\cdot, \bullet,\cdot,\cdot)}, \\
    G^{(1,\cdot,\cdot,\cdot)} g^{-1}_2 &= G^{(1,\bullet,\cdot,\cdot)} + G^{(12,\cdot,\cdot)} X_2\,.
\end{aligned}
\label{eq:4p_eom_2}
\end{equation}
By inserting \Eq{eq:4p_eom_2} into \Eq{eq:4p_1st} and using the definition \eqref{eq:G_aux_bullet_def}, we find a second-order 4p aIE:
\begin{equation} \label{eq:4p_2nd}
    \Gamma
    = g^{-1}_3 \Big[ G^{(\bullet,\bullet,\cdot,\cdot)} + G^{(12,\cdot,\cdot)} \Big] g^{-1}_4\,.
\end{equation}

Inserting the EOMs
\begin{align} \label{eq:4p_eom_3}
     g^{-1}_3 G^{(\cdot,\cdot,\cdot,\cdot)}&= G^{(\cdot,\cdot,\bullet,\cdot)} \nnnl
     g^{-1}_3 G^{(1,\cdot,\cdot,\cdot)} &= G^{(1,\cdot,\bullet,\cdot)} - X_3 \zeta G^{(13,\cdot,\cdot)} \nnnl
     g^{-1}_3 G^{(\cdot,2,\cdot,\cdot)} &= G^{(\cdot,2,\bullet,\cdot)} - X_3 G^{(23,\cdot,\cdot)} \nnnl
     g^{-1}_3 G^{(1,2,\cdot,\cdot)} &= G^{(1,2,\bullet,\cdot)} - X_3 \bigl[ \zeta G^{(13,2,\cdot)} +  G^{(23,1,\cdot)} \bigr ],
\end{align}
into the second-order aIE [\Eq{eq:4p_2nd}], we obtain a third-order 4p aIE:
\begin{align} \label{eq:4p_3rd}
    \Gamma
    &= \Big[ G^{(\bullet,\bullet,\bullet,\cdot)} + g^{-1}_3 G^{(12,\cdot,\cdot)}+ \zeta G^{(13,\cdot,\cdot)} \Sigma_2 \nnnl
    &\quad - \zeta G^{(13,2,\cdot)} X_2 + \Sigma_1 G^{(23,\cdot,\cdot)} - X_1 G^{(23,1,\cdot)} \Big] g^{-1}_4 \nnnl
    &= \Big[ G^{(\bullet,\bullet,\bullet,\cdot)} + g^{-1}_3 G^{(12,\cdot,\cdot)} - \zeta G^{(13,\bullet,\cdot)} - G^{(23,\bullet,\cdot)} \Big] g^{-1}_4 .
\end{align}
The second and third rows of this formula can be simplified using the 3p EOMs,
\begin{equation}
\begin{aligned}
    G^{(13,\cdot,\cdot)} g^{-1}_2
    &= G^{(13,\bullet,\cdot)} + \zeta G^{(123,\cdot)} X_2, \\
    g^{-1}_1 G^{(23,\cdot,\cdot)}
    &= G^{(23,\bullet,\cdot)} + X_1 G^{(123,\cdot)}.
\end{aligned}
\label{eq:4p_3rd_deriv}
\end{equation}
Further simplification is possible using
\begin{equation} \label{eq:G123_to_bare}
    X_1 G^{(123,\cdot)} g_4^{-1} = G^{(123,\cdot)} X_2 g_4^{-1} = -\Gamma_{\rm bare},
\end{equation}
which is derived by substituting \Eqs{eq:3p_G12_der_use} and \eqref{eq:Gamma_bare} into the third equation of \Eq{eq:3p_G12_eom}.
Here, the $X_1$ and $X_2$ terms cancel with the $X_4$ term in \Eq{eq:3p_G12_der_use} due to the structure of the KF bare vertex [\Eq{eq:Gamma_bare_kf}].
Substituting \Eqs{eq:4p_3rd_deriv} and \eqref{eq:G123_to_bare} into \Eq{eq:4p_3rd} yields
\begin{align}
    \Gamma
    &= G^{(\bullet,\bullet,\bullet,\cdot)} g^{-1}_4
    + \Gamma^{(12)} - \zeta \Gamma^{(13)} - \Gamma^{(23)}
    - 2\Gamma_{\rm bare}.
\end{align}

\begin{figure}[t]
    \centering
    \includegraphics[width=1.0\columnwidth]{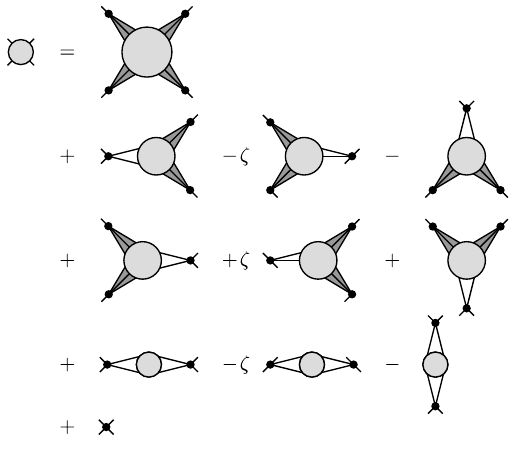}
    \caption{Diagrammatic representation of the 4p sIE of \Eq{eq:4p_4th} (listing the contributions in the same order as there).
    Only connected diagrams are involved.
    Indices of the amputated external legs are indicated by their orientation.
    }
    \label{fig:diag_4p}
\end{figure}

Finally, using the EOMs w.r.t.\ $\omega_4$,
\begin{align} \label{eq:4p_eom_4}
G^{(\cdot,\cdot,\cdot,\cdot)} g^{-1}_4 &= G^{(\cdot,\cdot,\cdot,\bullet)} 
, \nnnl
G^{(1,\cdot,\cdot,\cdot)} g^{-1}_4 &= G^{(1,\cdot,\cdot,\bullet)} + G^{(14,\cdot,\cdot)} X_4 
, \nnnl
G^{(\cdot,2,\cdot,\cdot)} g^{-1}_4 &= G^{(\cdot,2,\cdot,\bullet)} + \zeta G^{(24,\cdot,\cdot)} X_4 
\, \nnnl
G^{(\cdot,\cdot,3,\cdot)} g^{-1}_4 &= G^{(\cdot,\cdot,3,\bullet)} + G^{(34,\cdot,\cdot)} X_4 
, \nnnl
G^{(1,2,\cdot,\cdot)} g^{-1}_4 &= G^{(1,2,\cdot,\bullet)} + \bigl[ G^{(14,2,\cdot)} + \zeta G^{(24,1,\cdot)} \bigr] X_4 
, \nnnl
G^{(1,\cdot,3,\cdot)} g^{-1}_4 &= G^{(1,\cdot,3,\bullet)} + \bigl[ G^{(14,\cdot,3)} + G^{(34,1,\cdot)} \bigr] X_4 
, \nnnl
G^{(\cdot,2,3,\cdot)} g^{-1}_4 &= G^{(\cdot,2,3,\bullet)} + \bigl[ \zeta G^{(24,\cdot,3)} + G^{(34,\cdot,2)} \bigr] X_4 
, \nnnl
G^{(1,2,3,\cdot)} g^{-1}_4 &= G^{(1,2,3,\bullet)} + \bigl[ G^{(14,2,3)} + \zeta G^{(24,1,3)} 
\nnnl
&\qquad\qquad\qquad + G^{(34,1,2)} \bigr] X_4
,
\end{align}
we find the desired fourth-order 4p sIE,
\begin{align} \label{eq:4p_4th}
\Gamma 
& = 
\Gamma_{\rm core} \nnnl
& \ + \mcK^{(12)} - \zeta \mcK^{(13)} - \mcK^{(23)}
+ \mcK^{(34)} + \zeta \mcK^{(24)} + \mcK^{(14)} \nnnl
& \ + G^{(12,34)} - \zeta G^{(13,24)} - G^{(14,23)} \nnnl
& \ + \Gamma_{\rm bare},
\end{align}
fully symmetric in all four frequencies.
Here, we used
\begin{equation} \label{eq:4p_U4}
    \Gamma_{\rm core}
    = G^{(\bullet,\bullet,\bullet,\bullet)}.
\end{equation}
This term is defined recursively in \Eq{eq:G_aux_bullet_def} and contains $2^4=16$ terms that can be evaluated
using the same rule as the 3p case [\Eq{eq:3p_G12_U3}]: either (i) multiply $-\Sigma_n$ ($x_n = n$) or (ii) add $n$ in the superscript of the auxiliary correlator and multiply $X_n$ ($x_n = \cdot$), 
for $n = 1, 2, 3, 4$.
We also used the definition of $\mcK^{(ab)}$ [\Eq{eq:3p_G12_U3}] to isolate  the bosonic 2p correlators $G^{(12,34)}$, $G^{(13,24)}$, and $G^{(14,23)}$, and the bare vertex $G^{(1234)}$. From the top to bottom of \Eq{eq:4p_4th}, the rows contain 4p, 3p, 2p, and 1p correlators.
\Equ{eq:4p_4th}, giving a sIE for the 4p vertex, is our second main result. 
Figure~\ref{fig:diag_4p} depicts it diagrammatically.

% ===============================================

\subsection{Perturbative behavior of the estimators} \label{sec:perturbative}
One regime where the robustness of sIEs
against numerical errors becomes particularly evident is the weak-interaction limit.
In this regime, when using diagonalization-based methods like NRG without improved estimators, numerical artifacts may dominate the signal due to the small magnitude of the vertex~\cite{2021LeePRX}.
We hence discuss the perturbative limit explicitly in the following.
However, note that the improved estimators are of course formally exact at all interaction strengths.

First, let us consider the self-energy.
In the limit of small $U$, directly calculating the self-energy 
from the Dyson equation [\Eq{eq:2p_selfen_dyson}] can lead to an error of order $\mcO(U^0)$ due to an imperfect cancellation between the bare and full propagators.
In the aIE [\Eq{eq:2p_Sigma_1st}], the leading error is $\mcO(U^1)$ as the auxiliary correlator $G^{(1,\cdot)}$ contains $H_{\rm int}$.
Analogously, with the sIE [\Eq{eq:2p_Sigma_2nd}], the error in the frequency-dependent part is $\mcO(U^2)$ because all terms in the estimator (except the Hartree self-energy, which is computed directly via the equilibrium density matrix, using \Eq{eq:hartree}) include $H_{\rm int}$ at least twice.

Next, for the 3p sIEs in \Eq{eq:3p_symmetric} and Fig.~\ref{fig:diag_3p}, a perturbative expansion of the three terms gives contributions of order $\mcO(U^3)$, $\mcO(U^2)$, and $\mcO(U^1)$, respectively.
The third, $\mcO(U^1)$ term is the exact bare vertex.
The second, $\mcO(U^2)$ term is a bosonic 2p correlator which can be computed much more accurately than the 3p correlators.
Only the first, $\mcO(U^3)$ term $\mcK^{(12)}$ involves 3p correlators.
Finally, the 4p sIE also contains the exact bare vertex at $\mcO(U)$ and involves only bosonic 2p correlators up to $\mcO(U^2)$:
\begin{equation} \label{eq:4p_perturbative}
    \Gamma = \Gamma_{\rm bare} + G^{(12,34)} - \zeta G^{(13,24)} - G^{(14,23)} + \mcO(U^3).
\end{equation}
Thus, the 4p sIE is highly accurate in the weak-coupling limit, with errors from multipoint calculations entering only at $\mcO(U^3)$.
In contrast, a direct amputation of the correlator [\Eqs{eq:3p_G12_est_def} and \eqref{eq:est_Gamma_def}] or the use of first-order aIEs [\Eq{eq:3p_G12_1st} and \eqref{eq:4p_1st}] introduces errors three and two orders (two and one orders) earlier, respectively, than for the 4p (3p) sIE.

Our estimators for the 3p and 4p vertex are invariant under a shift of $\opH_{\rm int}$ by a quadratic term:
\begin{equation}
    \opH_{0} \to \opH_{0} + \lambda_{ij} \opdd_i \opd_j, \quad
    \opH_{\rm int} \to \opH_{\rm int} - \lambda_{ij} \opdd_i \opd_j.
\end{equation}
The self-energy estimators transform as
\begin{equation}
    \Sigma \to \Sigma - \lambda.
\end{equation}
Since NRG is linear in each argument of the correlator, the choice of $\lambda$ does not affect the numerical results calculated with the estimators (for a given $z$ shift)~\cite{2022KuglerEstimator}.
An interesting choice of the shift is $\lambda = \Sigma^{\rm H}$.
In this case, $\Sigma$ scales as $\mcO(U^2)$ in the small-$U$ limit.
Then, the first and second terms of the 1PR-subtracted vertex [Fig.~\ref{fig:diag_eom_interacting}(b)] scale as $\mcO(U^1)$, and $\mcO(U^2)$, respectively.
Hence, $\mcK^{(12)}$ [\Eq{eq:3p_G12_U3}] and $\Gamma_{\rm core}$ [\Eq{eq:4p_U4}] can be decomposed into terms that enter at different orders in the perturbative expansion according to the number of occurrences of $\Sigma$.
For example, in \Eq{eq:3p_G12_U3}, we have a $\mcO(U^3)$ term ($X_3 G^{(12,3,4)} X_4$),
two $\mcO(U^4)$ terms ($\Sigma_3 G^{(12,\cdot,4)} X_4$ and $X_3 G^{(12,3,\cdot)} \Sigma_4$),
and a $\mcO(U^5)$ term ($\Sigma_3 G^{(12,\cdot,\cdot)} \Sigma_4$).
Similarly, for $\Gamma_{\rm core}$, the perturbative order of each term can be classified into orders ranging from $\mcO(U^4)$ to $\mcO(U^8)$.

% ===============================================
\subsection{Relation to the vertex asymptotics} \label{sec:asymptotic}
A similar numerical advantage of the sIEs is also expected in the large frequency limit. When the input frequencies are much larger than any intrinsic energy scales of the system, the propagator $g$ is inversely proportional to the frequency. At high frequencies, numerical results for $g$ become noisy due to the vanishing magnitude and a small signal-to-noise ratio.
Direct amputation, i.e.\ division by $g$, introduces a large error in the vertex.
The sIEs for $\Gamma^{\rm 3p}$ and $\Gamma^{\rm 4p}$ are free from this error because they do not require any amputation.

The 4p sIE of \Eq{eq:4p_eom_4} bears a close connection to the asymptotic behaviors of the 4p vertex.
If any of the external frequency arguments is taken to infinity, a diagram carrying this frequency in a (non-amputated) line vanishes~\cite{2017KaufmannAsymptotics,2020WentzellPRB}.
We now use this property to connect the 4p sIE [\Eq{eq:4p_4th}] and its diagrammatic representation (Fig.~\ref{fig:diag_4p}) to the asymptotic classes of the 4p vertex.

If all four frequency arguments are taken to infinity without any particular constraint except $\omega_{1234} = 0$, the 4p vertex reduces to the bare interaction.
The last term of the 4p sIE \eqref{eq:4p_4th} is this bare interaction [\Eq{eq:Gamma_bare}]:
\begin{equation}
    \lim_{\abs{\omega_1}, \ccdots, \abs{\omega_4} \to \infty} \Gamma(\mb{\omega}) = \Gamma_{\rm bare}.
\end{equation}

Nontrivial asymptotic classes are defined by the limits of some or all frequencies going to infinity while keeping the sum of two frequencies to a fixed, finite value.
Concretely, we have~\cite{2020WentzellPRB}
\begin{align} \label{eq:4p_asymp_def}
    \lim_{\abs{\nu_r} \to \infty} \lim_{\abs{\nu_r'} \to \infty} \Gamma(\mb{\omega_r})
    &= \Gamma_{\rm bare} + \mcK_{1}^{r}(\omega_r), \nnnl
    \lim_{\abs{\nu'_r} \to \infty} \Gamma(\mb{\omega_r})
    &= \Gamma_{\rm bare} + \mcK_{1}^{r}(\omega_r) + \mcK_{2}^{r}(\nu_r, \omega_r), \nnnl
    \lim_{\abs{\nu_r} \to \infty} \Gamma(\mb{\omega_r})
    &= \Gamma_{\rm bare} + \mcK_{1}^{r}(\omega_r) + \mcK_{2'}^{r}(\nu_r', \omega_r) ,
\end{align}
where we parametrize the frequencies as
\begin{equation} \label{eq:freq_parametrization}
    \mb{\omega_r} = \begin{cases}
        (\nu_r, -\nu_r - \omega_r, \nu_r' + \omega_r, -\nu_r') & \text{for $r=t$ (ph)}, \\
        (\nu_r, -\nu_r', -\nu_r - \omega_r, \nu_r' + \omega_r) & \text{for $r=p$ (pp)} \\
        (\nu_r, -\nu_r', \nu_r' + \omega_r, -\nu_r - \omega_r) & \text{for $r=a$ ($\overline{\rm ph}$)}.
    \end{cases}
\end{equation}
Here, $t$, $p$, and $a$ denote the transverse, parallel, and antiparallel channels, according to, e.g., the conventions of Ref.~\cite{Kugler2018b,Kugler2018c,Gievers2022}.

The first asymptotic class $\mcK_1$ corresponds to the bosonic 2p correlators~\cite{2020WentzellPRB} in the third line of \Eq{eq:4p_4th}:
\begin{equation}
\begin{aligned}
    \mcK^t_{1}(\omega_t) &= G^{(12,34)}(-\omega_t, \omega_t), \nnnl
    \mcK^p_{1}(\omega_p) &= -\zeta G^{(13,24)}(-\omega_p, \omega_p), \nnnl
    \mcK^a_{1}(\omega_a) &= -G^{(14,23)}(-\omega_a, \omega_a).
\end{aligned}
\label{eq:4p_asymp_K1}
\end{equation}
The second asymptotic class, involving  $\mcK_{2}$ and $\mcK_{2'}$, comes from the $\mcK^{(ab)}$ terms:
\begin{align} \label{eq:4p_asymp_K2}
    \mcK^{t}_{2}(\nu_t, \omega_t)   &=   \mcK^{(34)}(\omega_t, \nu_t, -\nu_t - \omega_t), \nnnl
    \mcK^{t}_{2'}(\nu_t', \omega_t) &=   \mcK^{(12)}(-\omega_t, \nu_t' + \omega_t, -\nu_t'), \nnnl
    \mcK^{p}_{2}(\nu_p, \omega_p)   &= \zeta \mcK^{(24)}(\omega_p, \nu_p, -\nu_p - \omega_p), \nnnl
    \mcK^{p}_{2'}(\nu_p', \omega_p) &= - \zeta \mcK^{(13)}(-\omega_p, -\nu_p', \nu_p' + \omega_p), \nnnl
    \mcK^{a}_{2}(\nu_a, \omega_a)   &= - \mcK^{(23)}(\omega_a, \nu_a, -\nu_a - \omega_a), \nnnl
    \mcK^{a}_{2'}(\nu_a', \omega_a) &=   \mcK^{(14)}(-\omega_a, -\nu_a', \nu_a' + \omega_a).
\end{align}
The remaining core contribution, which does not contribute to the asymptotics, is \Eq{eq:4p_U4}:
\begin{equation} \label{eq:4p_asymp_rest}
    \Gamma - \Big[ \Gamma_{\rm bare} + \sum_{r} \left( \mcK^{r}_{1} + \mcK^{r}_{2} + \mcK^{r}_{2'}\right) \Big] = \Gamma_{\rm core}.
\end{equation}

For computational schemes built on the asymptotics-based parametrization of the 4p vertex, using the 4p sIE is highly advantageous because each asymptotic class is calculated separately.
The core contribution, in particular, which decays in all high-frequency limits, can be calculated using \Eq{eq:4p_U4}. This is expected to be much more accurate than subtracting terms that belong to different asymptotic classes from the full vertex.

\begin{figure*}[t]
    \centering
    \includegraphics[width=1.0\textwidth]{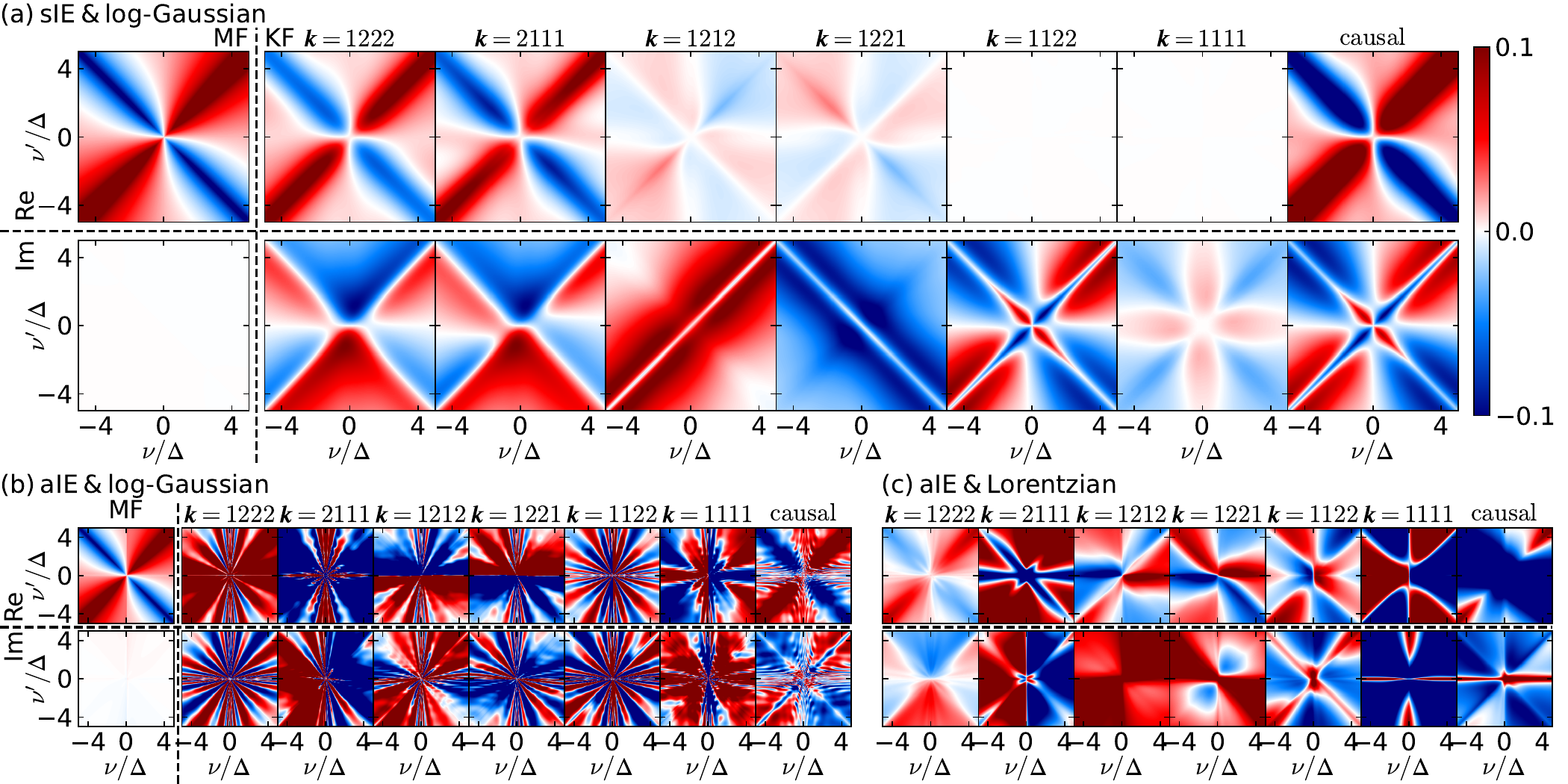}
    \caption{%
    MF and KF 4p vertices $(\Gamma_{\uparrow\downarrow} - \Gamma_{{\rm bare};\uparrow\downarrow}) / U$ in the AIM at weak interaction ($\Delta/D = 0.1$, $U / \Delta = 0.5$, and $T / \Delta = 0.01$) at $\omega=0$.
    They were computed using (a) the sIE of \Eq{eq:4p_4th}, and (b,c) the aIE of \Eq{eq:4p_1st}, with (a,b) log-Gaussian broadening or (c) Lorentzian broadening (see App.~\ref{subsec:NRG-broadening} for details).
    Only the KF vertices are shown for (c) as broadening does not affect MF vertices.
    For all three panels (a,b,c), the upper (lower) row shows the real (imaginary) part.
    For the sIE, the MF vertex is purely real, and the KF vertex at $\mb{k}=1122$ and $\mb{k}=1111$ are purely imaginary.
    Keldysh components not plotted are related to those plotted by crossing symmetry, complex conjugation, or both (see App.~\ref{sec:symmetry}).
    The aIE vertex breaks these symmetries.
    In panels (b) and (c), the KF data is clearly inferior (more noisy, showing spurious features, etc.) than in (a), illustrating that aIEs are not suitable for computing all components of the KF vertex.
    }
    \label{fig:weak_vertex_overview}
\end{figure*}

% ========================================================
\subsection{Subtracting the disconnected contributions} \label{eq:est_total}
So far, we presented 3p and 4p sIEs involving \textit{connected} auxiliary correlators.
Such estimators are suitable for NRG, where correlators are computed using spectral representations which offer a natural way for obtaining connected correlators by subtracting disconnected parts on the level of partial spectral functions~\cite{2021KuglerPRX,2021LeePRX}.
Yet, other methods, like QMC, only have access to the total correlator.
It is then useful to have total correlators instead of their connected parts in the vertex estimators.

Leaving the derivation to App.~\ref{sec:disconnected_estimator}, we here present 
a KF 4p sIE involving only total correlators:
\begin{equation} \label{eq:tot_4p_est}
    \Gamma
    = \Gamma_{\rm tot}
    - 2\pi\delta(\omega_{12}) \Sigma_{12}^{k_1 k_2} \Sigma_{34}^{k_3 k_4}
    - 2\pi\delta(\omega_{14}) \zeta \Sigma_{14}^{k_1 k_4} \Sigma_{32}^{k_3 k_2}.
\end{equation}
Here, the subscript `tot' indicates that the connected auxiliary correlators in \Eq{eq:4p_4th} are replaced by total correlators, i.e., the sum of the connected and disconnected parts.
The additional self-energy terms cancel the disconnected diagrams in the total correlator.
In the MF, the Dirac delta function $2\pi\delta(\omega)$ is replaced by the Kroneker delta $\beta \delta_{\omega, 0}$.

The additional disconnected terms involve the self-energy; hence, they vanish in the noninteracting case, as well as in the perturbative limit up to $\mcO(U^1)$ [or $\mcO(U^2)$ if $H_{\rm int}$ is shifted to give $\Sigma^{\rm H}=0$, cf.\ \Sec{sec:perturbative}].
Moreover, they are much smaller than those obtained by direct amputation, where disconnected terms involve the square of the inverse propagator.
When using sIEs, significant cancellations occur between the disconnected parts of the various auxiliary 4p correlators involved; thus, the disconnected parts surviving these cancellations are much smaller.

In App.~\ref{sec:disconnected_estimator}, we also show that the estimators using total correlators share the same perturbative and asymptotic properties as the original estimators expressed through connected correlators discussed in the previous sections.

One remaining choice to be made is which self-energy estimators to use when evaluating the vertex estimators.
Possible choices include the aIEs $\Sigma^{\rm L}$ [\Eq{eq:2p_Sigma_1st-a}] and $\Sigma^{\rm R}$ [\Eq{eq:2p_Sigma_1st-b}], and the sIE $\Sigma^{\rm S}$ [\Eq{eq:2p_Sigma_2nd}].
Although these estimators are all equivalent analytically, this choice may affect the results in the presence of numerical noise.

We propose to use the aIE $\Sigma^{\rm L}$ ($\Sigma^{\rm R}$) for the self-energies $\Sigma_1$ and $\Sigma_3$ ($\Sigma_2$ and $\Sigma_4$) which left- and right-multiply the auxiliary correlators.
This choice maximizes the cancellation of disconnected diagrams: e.g., the disconnected term in the 3p sIE for $\Gamma^{(34)}$ is proportional to
\begin{align*}
    &\Sigma^{\rm L} G^{(\cdot,\cdot)} \Sigma^{\rm R} \!-\! X G^{(1,\cdot)} \Sigma^{\rm R}
    \!-\! \Sigma^{\rm L} G^{(\cdot,2)} X \!+\! X G^{(1,2)} X \!+\! \Sigma^{\rm H} X .
\end{align*} 
If \Eqs{eq:2p_Sigma_1st} are used, the first three terms are all equal up to signs.
Thus, two of them mutually cancel (even if $\Sigma^{\rm L}$ and $\Sigma^{\rm R}$ differ due to numerical noise), so that the expression
simplifies to
\begin{align}
    - X G^{(1,\cdot)} g^{-1} G^{(\cdot,2)} X + X G^{(1,2)} X + \Sigma^{\rm H} X  &= \Sigma^{\rm S}.
\end{align}
Then, by using $\Sigma^{\rm S}$ in \Eq{eq:tot_4p_est} to remove the remaining disconnected terms, the cancellation is made exact.
Such a cancellation may be particularly beneficial for QMC where the total correlators are computed.
For NRG, the disconnected parts are already subtracted on the level of partial spectral functions.
Still, we use $\Sigma^{\rm L}$ ($\Sigma^{\rm R}$) for left (right) multiplications in \Eq{eq:4p_4th}, expecting that this helps with the cancellation of any remnant disconnected terms that might have survived as numerical artifacts.

% =============================================================
\section{Numerical results} \label{sec:results}

\begin{figure*}[tb]
    \centering
    \includegraphics[width=1.0\textwidth]{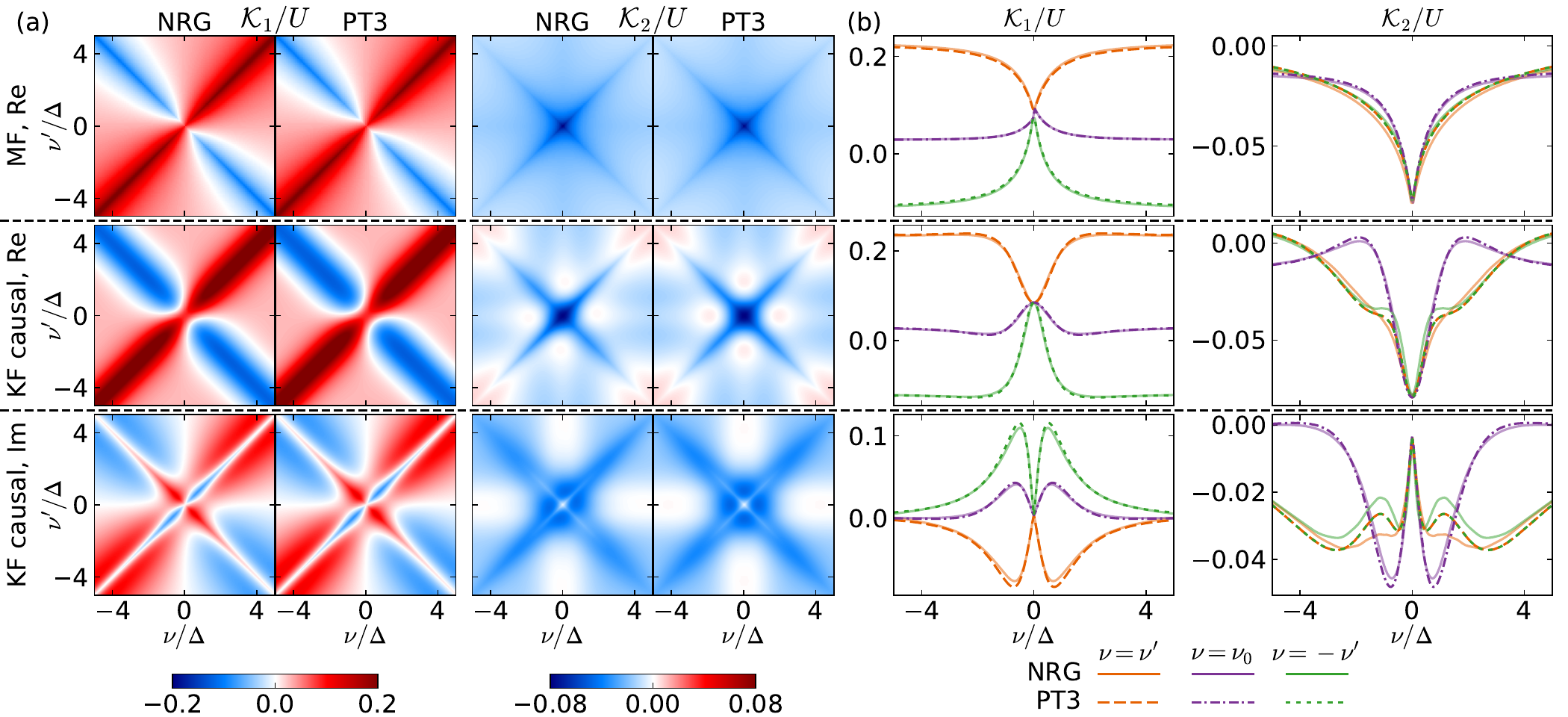}
    \caption{%
    (a) MF and KF 4p vertices in the first and second asymptotic classes, $\mcK_1 = \sum_{r=a,p,t} \mcK_1^r$ and $\mcK_2 = \sum_{r=a,p,t} \mcK_2^r + \mcK_{2'}^r$, normalized by $U$, in the AIM at weak interaction (parameters: same as for Fig.~\ref{fig:weak_vertex_overview}).
    The first, second, and third rows show the MF vertex (which is purely real), and the real and imaginary parts of the causal component of the KF vertex, respectively, comparing results from NRG (first, third columns) and
    third-order perturbation theory (PT3) (second, fourth columns).
    (b) Line cuts of the vertices in (a) along the diagonal (orange), $y$ axis (purple), and anti-diagonal (green). $\nu_0$ is $\pi/\beta$ for the MF and $0$ for the KF.
    Overall, the agreement is excellent both qualitatively, see (a), and quantitatively, see (b).
    For $\mcK_2$ in the right column of (b), the PT3 and NRG results do not fully agree, because PT3 
    yields a symmetry between diagonal and anti-diagonal cuts which is lifted by terms beyond third order (for orange and green, dashed PT3 lines coincide, solid NRG ones do not).
    }
    \label{fig:weak_K12_NRG_vs_PT3}
\end{figure*}

\begin{figure}[!htbp]
\centering
\includegraphics[width=1.0\columnwidth]{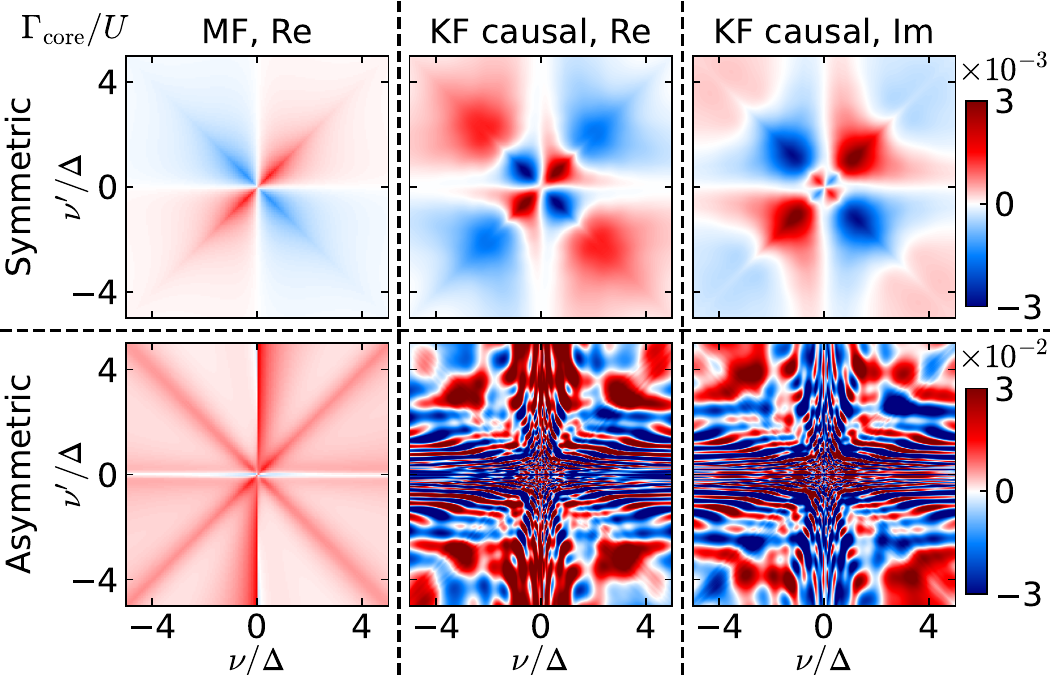}
\caption{
Core part of the MF and KF 4p vertices $\Gamma_{\rm core} / U$, in the AIM at weak interaction (parameters: same as for Fig.~\ref{fig:weak_vertex_overview}).
The first, second, and third columns show the MF vertex (which is purely real), and the real part and imaginary parts of the causal component of the KF vertex, respectively, computed using sIEs (top row) or aIEs (bottom row). The latter are dominated by numerical noise, the former not.
}
\label{fig:weak_core_asym_vs_sym}
\end{figure}

\begin{figure*}[!htbp]
    \centering
    \includegraphics[width=1.0\textwidth]{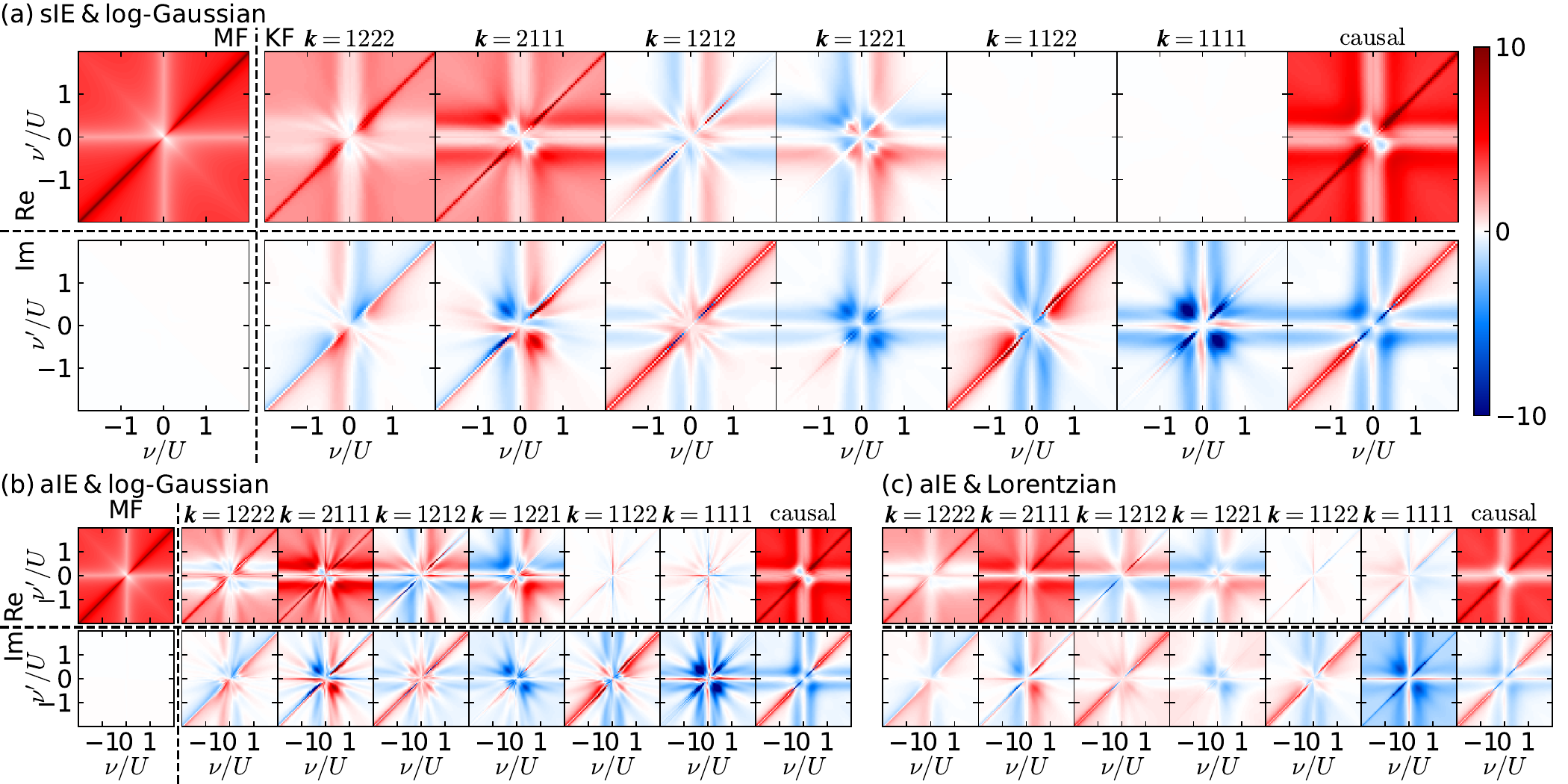}
    \caption{%
    MF and KF 4p vertices $(\Gamma_{\uparrow\downarrow} - \Gamma_{{\rm bare};\uparrow\downarrow}) / U$, for the AIM, analogous to \Fig{fig:weak_vertex_overview}, but at a much stronger interaction ($\Delta/D = 0.04$, $U / \Delta = 5$, and $T / \Delta = 0.0025$), with $\omega =0$.
    }
    \label{fig:strong_vertex_overview}
\end{figure*}

In this section, we demonstrate the advantages of sIEs over aIEs for NRG computations of the 4p vertex~\cite{2021KuglerPRX,2021LeePRX}.
To this end, we consider the AIM and compare results from NRG to those of third-order perturbation theory (PT3) and renormalized perturbation theory (RPT)~\cite{2004HewsonRPT,2018OguriRPTprl,2018OguriRPT,2022OguriRPT}.
For our purposes, NRG, PT3, and RPT may all be viewed as black-box methods for computing MF and KF vertices, where PT3 and RPT are restricted to weak interaction and asymptotically low energies, respectively.
(Reference \cite{2021LeePRX} describes the inner workings of NRG vertex computations and App.~\ref{app:NRG} some further refinements needed for present purposes.)
We also refrain from discussing the physics of the AIM or analyzing
the physical information encoded in its 4p vertex. Instead, we focus on the advantages of sIEs over aIEs.

The Hamiltonian of the AIM was already given in \Eq{eq:2p_AIM}. 
We here take a rectangular hybridization function $-\mathrm{Im}\Delta^R(\nu) = \pi \sum_b \abs{V_b}^2 \delta(\nu - \veps_b) = \Delta \Theta(D - \abs{\nu})$ with half-bandwidth $D$ and hybridization strength $\Delta$.
We focus on the particle-hole symmetric case, $\veps_d = -U/2$.

In the following, we represent the 4p vertex in the $t$-channel parametrization [\Eq{eq:freq_parametrization}] at vanishing transfer frequency,
\begin{align} \label{eq:ph_representation}
\Gamma_{\sigma\sigma'}(\nu, \nu')
& = 
\Gamma_{\sigma\sigma'}(\nu, \nu',\omega=0)
\nonumber \\
& = 
\Gamma[\opd_\sigma,\opdd_\sigma, \opd_{\sigma'}, \opdd_{\sigma'}](\nu, -\nu, \nu', -\nu')
.
\end{align}
Thus, $\Gamma$ describes the effective interaction of an electron with energy $\nu$ and spin $\sigma$ and an electron with energy $\nu'$ and spin $\sigma'$.
We will analyze $\Gamma$ in the MF and the KF. In the KF, we will consider its components in the Keldysh basis as well as the causal component in the contour basis (corresponding to $\mb{c}=\mmmm$).
The latter is a particularly sensitive probe to the numerical accuracy as it involves a sum over all components in the Keldysh basis:
\begin{equation}
\textstyle \Gamma^{\mathrm{causal}} = \frac{1}{4} \sum_{k_1, \ccdots, k_4} \Gamma^{k_1 k_2 k_3 k_4}.
\end{equation}
By crossing and complex conjugation symmetries, one has
(see App.~\ref{sec:symmetry})
\begin{subequations}
\begin{align}
\mathrm{Re}\, \Gamma^{\mathrm{causal}}
& =
\tfrac{1}{2} \mathrm{Re}\,
( \Gamma^{1222} + \Gamma^{2111} )
+ (\nu \leftrightarrow \nu')
,
\\
\mathrm{Im}\, \Gamma^{\mathrm{causal}}
& =
\tfrac{1}{4} \mathrm{Im}\,
( \Gamma^{1212} + \Gamma^{1221} + \Gamma^{1122} + \tfrac{1}{2} \Gamma^{1111} )
\nonumber \\
& \ + (\nu \leftrightarrow \nu')
.
\end{align}
\end{subequations}

\begin{figure*}[!htbp]
    \centering
    \includegraphics[width=1.0\textwidth]{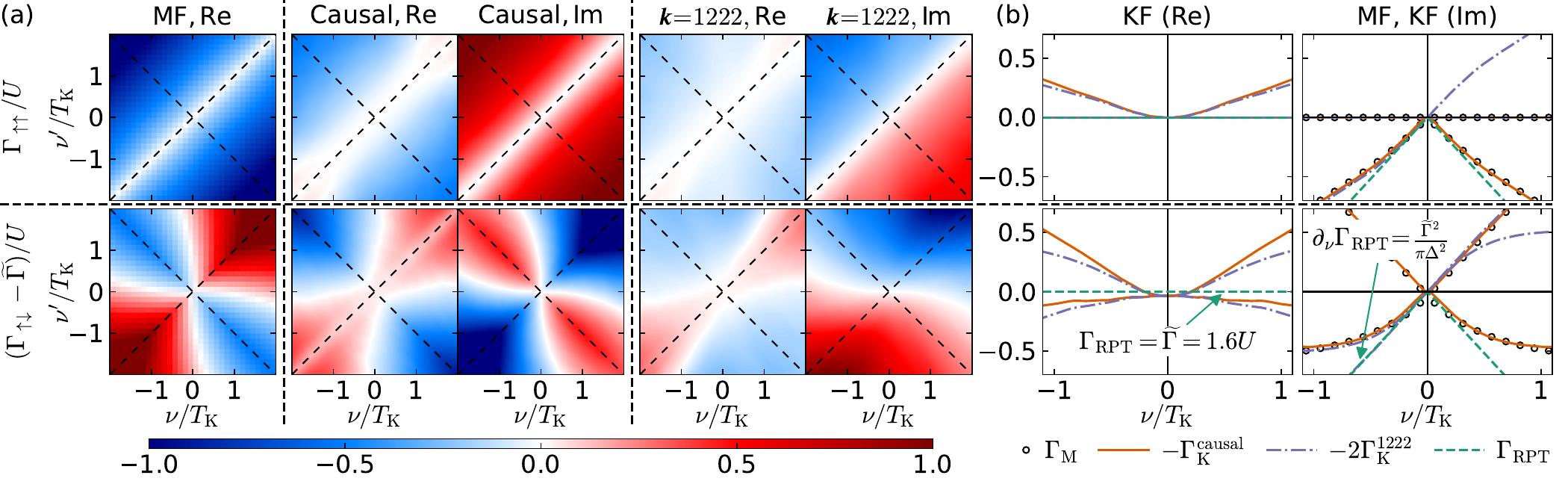}
    \caption{%
    (a) Low-energy part of the MF and KF 4p vertices $(\Gamma_{\sigma\sigma'} - \widetilde{\Gamma}_{\sigma\sigma'}) / U$ in the AIM at strong interaction
    (parameters: same as for Fig.~\ref{fig:strong_vertex_overview}), computed with the sIE.
    The vertices are an order of magnitude smaller than those shown in \Fig{fig:strong_vertex_overview}.
    (b) One-dimensional cuts of the vertices along $\nu = \nu'$ and $\nu = -\nu'$ (dashed lines in (a)).
    The Fermi-liquid predictions for the vertex, obtained by RPT [\Eq{eq:rpt}], are shown using green dashed lines.
    In the regime of validity of RPT, $|\nu|,|\nu'|,T \ll T_{\rm K}$, the agreement between NRG and RPT is remarkably good.
    The RPT parameters are $\widetilde{U} = 0.20U$, $Z = 0.36$, and $\widetilde{\Gamma} = \widetilde{U} / Z^2 = 1.6U$.}
    \label{fig:strong_RPT}
\end{figure*}

\begin{figure}[!htbp]
    \centering
    \includegraphics[width=1.0\columnwidth]{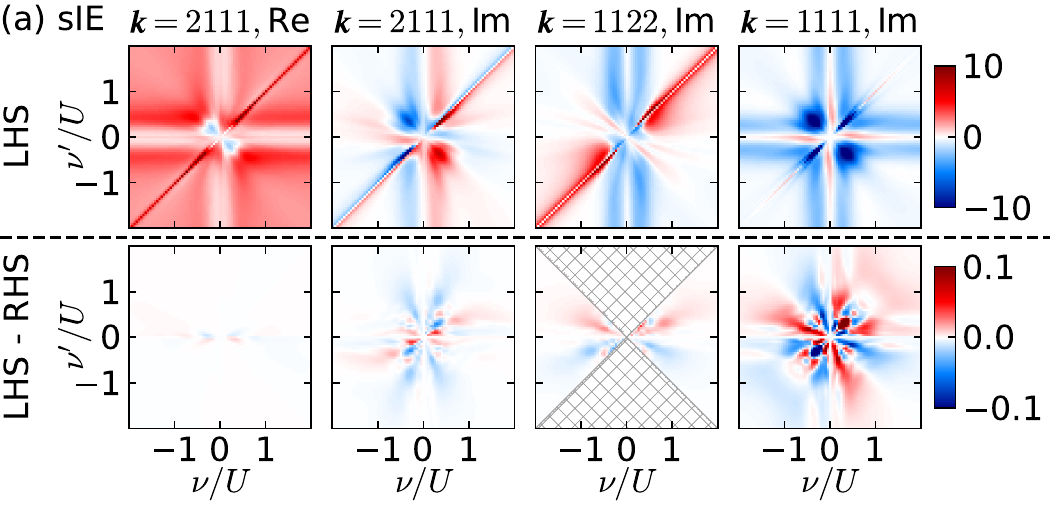}
    \includegraphics[width=1.0\columnwidth]{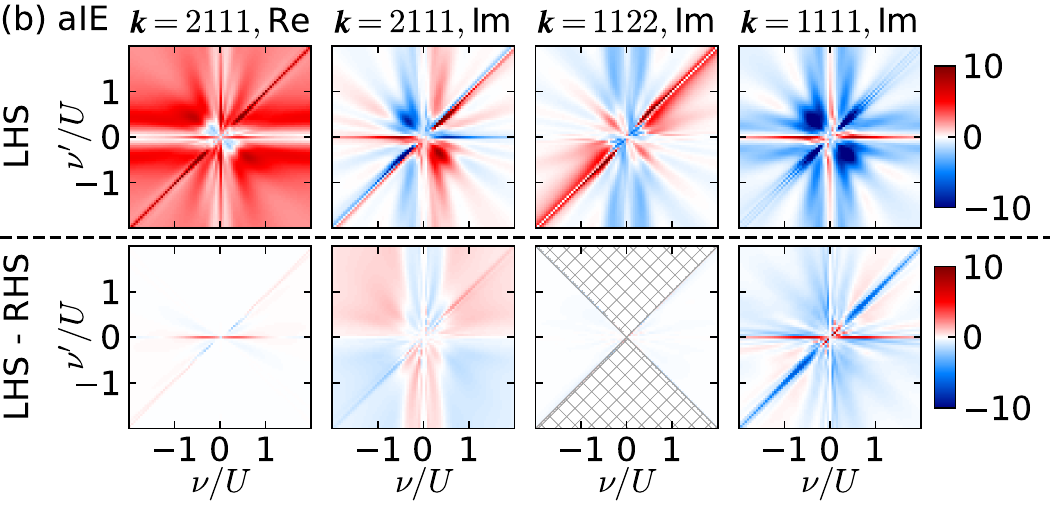}
    \caption{FDRs of the KF 4p vertices in the AIM at a strong interaction (parameters: same as for Fig.~\ref{fig:strong_vertex_overview}), computed using (a) the sIE and (b) an aIE.
    The four columns show the real and imaginary parts of \Eq{eq:4p_vertex_fdt_1}, \Eq{eq:4p_vertex_fdt_2}, and \Eq{eq:4p_vertex_fdt_3}, respectively, normalized by $U$.
    Their left sides are shown in the upper rows of (a,b), the difference between the left and right sides (violation of the FDR) in the bottom rows.
    Remarkably, for the sIE the violations are two orders of magnitude smaller than the vertices, and our 4p sIE NRG results thus satisfy vertex FDRs to within a few percent.
    By contrast, for the aIE, the violations are not much smaller than the vertices themselves, implying sizeable violations of the FDRs.
    In the bottom row for $\mb{k}=1122$, we do not plot results in the hatched region where $\abs{\nu'} > \abs{\nu} + 4T$, as the right-hand side of \Eq{eq:4p_vertex_fdt_2} becomes numerically unstable due to the $\cosh  (\nu'/2T) / \cosh (\nu/2T)$ term, which increases exponentially with $\abs{\nu'} - \abs{\nu}$.}
    \label{fig:strong_FDR}
\end{figure}

% =====================================
\subsection{Weak interaction}

As the first benchmark, we study the AIM at weak interaction, $\Delta/D = 0.1$, $U / \Delta = 0.5$, and $T / \Delta = 0.01$, 
and compare our NRG results against those from PT3.
In the weak coupling regime, defined by $U/(\pi \Delta) \ll 1$ \cite{Zlatic1983,GeRitz2023}, PT3 yields fairly accurate results and thus serves as a useful benchmark. For NRG, being a diagonalization-based method, the accuracy of the results does not depend on the strength of the coupling, i.e., weak coupling is as non-trivial a challenge as strong coupling.

Figures~\ref{fig:weak_vertex_overview}(a) and ~\ref{fig:weak_vertex_overview}(b) compare the 4p vertex obtained using the sIE [\Eq{eq:4p_4th}] and the aIE [\Eq{eq:4p_1st}], respectively, both broadened the same way, using a narrow (log-Gaussian) broadening kernel (see App.~\ref{subsec:NRG-broadening} for details).
The aIE results are completely dominated by fan-shaped noise, an artifact of NRG discretization.
By contrast, the sIE results are almost completely free from such artifacts.
This clearly illustrates the advantage of the sIE over the aIE.
For comparison, 
Fig.~\ref{fig:weak_vertex_overview}(c) shows aIE results broadened with a much broader (Lorentzian) kernel (in the same way as for the aIE results of Ref.~\cite{2021KuglerPRX}, Fig.~12). 
This hides the discretization artifacts by smearing them out, at the cost of strongly over-broadening the physically meaningful features seen in the sIE results of Fig.~\ref{fig:weak_vertex_overview}(a).
In addition, Keldysh components of the aIE vertex other than the fully retarded $\mb{k}=1222$ component strongly deviate from the corresponding sIE result.

Figure~\ref{fig:weak_K12_NRG_vs_PT3} displays the contributions from the $\mcK_1$ and $\mcK_2$ asymptotic classes obtained with NRG and PT3.
We find an excellent agreement for all asymptotic terms, in the MF and the KF.
The fact that even the line cuts match perfectly is testament to the accurate broadening of the NRG data.

Next, we focus on the core part, which is very challenging to compute from perturbation theory due to the leading-order contribution of the envelope diagram.
Figure~\ref{fig:weak_core_asym_vs_sym} compares the core vertex obtained using 
the sIE [\Eq{eq:4p_U4}] (top row) and an aIE (bottom row).
Since the aIE [\Eq{eq:4p_1st}] does not contain a decomposition into asymptotic classes, one must subtract the asymptotic contributions from the full vertex to get the core part [\Eq{eq:4p_asymp_rest}].
For the AIM parameters used here, the core vertex is around two orders of magnitude smaller than the full vertex.
Hence, the subtraction entails a large numerical error.
Indeed, the data in the bottom row of Fig.~\ref{fig:weak_core_asym_vs_sym} is completely dominated by numerical noise, ten times larger than the true core vertex (upper row), in both the MF and the KF. By contrast, using the sIE, the core vertex is determined from its own estimator~\eqref{eq:4p_U4}, which involves no subtraction of terms with different asymptotics or perturbative order.
Thereby, the sIE is much less susceptible to numerical errors than the aIE.

% =====================================================
\subsection{Strong interaction}

We now turn to the nonperturbative regime with a stronger interaction $\Delta/D = 0.04$, $U / \Delta = 5$, and $T / \Delta = 0.0025$.
Figure~\ref{fig:strong_vertex_overview} compares sIE and aIE results for the 4p vertex.
The MF and KF vertices differ significantly from the weak-coupling case (\Fig{fig:weak_vertex_overview}).
The discretization artifacts observed with the aIE in \Fig{fig:strong_vertex_overview}(b) are less prominent at strong coupling than at weak coupling, but are still noticeable.
Yet, being asymmetric, the aIE breaks several symmetries of the vertex, having, e.g., a nonzero real part in the $\mb{k}=1122$ and $1111$ components.

In the nonperturbative regime, accurate reference results for the entire 4p vertex are not available.
However, in the low-energy Fermi-liquid regime, where the temperature and all frequencies are much lower than the Kondo temperature $T_{\rm K}$ (here, $T \simeq 0.02 T_{\rm K}$~\cite{2021KuglerPRX}), RPT predicts a specific behavior of the vertex~\cite{2004HewsonRPT,2018OguriRPTprl,2018OguriRPT,2022OguriRPT}.
For an SU(2)-symmetric single-orbital AIM at half filling, the MF, causal KF, and fully retarded KF vertices in the low-temperature, low-frequency limit have the following form~\cite{2018OguriRPTprl,2018OguriRPT,2022OguriRPT},
\begin{subequations} \label{eq:rpt}
\begin{align}
    \Gamma_{{\rm M},\,\sigma \sigma'}(i\nu, i\nu')
    &= \widetilde{\Gamma} \delta_{\sigma\bar{\sigma}'} - \frac{\widetilde{\Gamma}^2}{\pi \Delta^2} (\abs{\nu \!-\! \nu'} - \delta_{\sigma\bar{\sigma}'}\abs{\nu \!+\! \nu'}),
    \\
    \Gamma_{{\rm K},\,\sigma \sigma'}^{\mathrm{causal}}(\nu, \nu')
    &= \widetilde{\Gamma} \delta_{\sigma\bar{\sigma}'} + i \frac{\widetilde{\Gamma}^2}{\pi \Delta^2} (\abs{\nu \!-\! \nu'} - \delta_{\sigma\bar{\sigma}'}\abs{\nu \!+\! \nu'}),
    \\
    2\Gamma_{{\rm K},\,\sigma \sigma'}^{1222}(\nu, \nu')
    &= \widetilde{\Gamma} \delta_{\sigma\bar{\sigma}'} + i \frac{\widetilde{\Gamma}^2}{\pi \Delta^2} [(\nu \!-\! \nu') - \delta_{\sigma\bar{\sigma}'}(\nu \!+\! \nu')],
\end{align}
\end{subequations}
where $\bar{\sigma} = -\sigma$.
The last equation for the fully retarded KF vertex is derived using the analytic continuation of the absolute value $f(i\nu) = \abs{\nu}$ to $f^{R/A} = -i(\nu \pm i0^+)$.
The effective static interaction $\widetilde{\Gamma}$ is given by $\widetilde{\Gamma}_{\sigma\sigma'} = \delta_{\sigma \bar{\sigma}'}\widetilde{U} / Z^2 $, where $\widetilde U$ is the quasiparticle interaction and $Z$ the
quasiparticle weight.
These can be directly extracted from the low-energy eigenspectrum spectrum of NRG~\cite{Hewson1993RPT,2004HewsonRPT,Pandis2015RPT,Kugler2020SRO,2021KuglerPRX}; for our strong-coupling parameters, we find $\widetilde{U}=0.20U$, $Z=0.36$, and $\widetilde{\Gamma} = \widetilde{U} / Z^2 = 1.6U$.
These are the same values as in Ref.~\cite{2021KuglerPRX}. While, there, the agreement with RPT in the limit $\nu,\nu' \to 0$ was checked for the MF and fully retarded KF vertices, we here significantly extend this comparison by including the linear order in $\nu$ and $\nu'$ and all Keldysh components.

Figure~\ref{fig:strong_RPT}(a) shows the low-energy part of the sIE vertex, with $\abs{\nu}, \abs{\nu'} \lesssim T_{\rm K}$.
Figure~\ref{fig:strong_RPT}(b) compares NRG and RPT results for line cuts of the vertex, showing remarkably good agreement in the low-energy regime $\abs{\nu'} \ll T_{\rm K}$, for both the MF and KF.
This provides strong confirmation of the accuracy of the imaginary- and real-frequency vertices computed from NRG using the sIE.
We note that the small undershooting of $\Gamma_{\uparrow\downarrow}(0, 0, 0)$ can be systematically improved by using a denser grid for binning and a smaller broadening parameter, at the expense of increased computational costs (see App.~\ref{app:NRG}).

As a final test, we check how well the NRG results for $\Gamma^{\mb{k}}$
satisfy generalized fluctuation-dissipation relations (FDRs). These FDRs are known from the literature~\cite{Wang2002FDR,2020GeThesis,Ge2023} and take a particularly simple form at $\omega=0$:
\begin{subequations}
\begin{flalign}
\label{eq:4p_vertex_fdt_1}
\Gamma^{2111} &=
\Gamma^{2122} + t_{\nu'} (\Gamma^{2112} - \Gamma^{2121}) 
&
\nnnl
& - 2i t_\nu t_{\nu'} \Im \Gamma^{2212} + it_{\nu} \Im \Gamma^{2211},
&
\\
\label{eq:4p_vertex_fdt_2}
\Im \Gamma^{1122} &= 
c_{\nu'}^2 / c_{\nu}^2 \Im \Gamma^{2211} 
&
\nnnl
&+ 2t_\nu \Im \Gamma^{1222}
- 2t_{\nu'} c_{\nu'}^2 / c_{\nu}^2 \Im \Gamma^{2212},
&
\\
\label{eq:4p_vertex_fdt_3}
\Im \Gamma^{1111} &=
[ 2 t_{\nu} (1 \!+\! t_{\nu'}^2) \Im \Gamma^{1222}
- t_{\nu'}^2 \Im \Gamma^{1122} 
&
\nnnl
& + t_{\nu} t_{\nu'} (\Im \Gamma^{2121} \!-\! \Im \Gamma^{2112}) ] + (\nu \!\leftrightarrow\! \nu'). 
\hspace{-0.5cm} &
\end{flalign}
\end{subequations}
Here, we used $c_\nu = \cosh \tfrac{\nu}{2T}$ and $t_\nu = \tanh \tfrac{\nu}{2T}$ (same for $\nu'$) for short and omitted the frequency argument $(\nu, \nu', \omega=0)$ for the vertices.
The FDRs for the other Keldysh components with one $2$ follow from \Eq{eq:4p_vertex_fdt_1} by crossing symmetry or complex conjugation (cf.\ App.~\ref{sec:symmetry}).
Figure~\ref{fig:strong_FDR} shows that the FDRs are all satisfied remarkably well for the sIE vertex [\Fig{fig:strong_FDR}(a)], with errors two orders of magnitude smaller than the signal. By contrast, the FDRs are strongly violated for the aIE vertex [\Fig{fig:strong_FDR}(b)].

% =============================================================
\section{Summary and outlook} \label{sec:conclusion}
We presented a new estimator for the 4p vertex which is symmetric in all indices and involves only full (interacting) correlators.
This sIE achieves the amputation of external legs via EOMs, without dividing the correlators by propagators, and also maximizes the cancellation of the disconnected parts between multipoint objects.
The asymptotic decomposition of the vertex naturally arises from the sIE, ensuring the accuracy of every term via a separate estimator for each, without any large-frequency limits or numerically unstable subtractions.
We demonstrate the utility of the sIE by calculating the 4p vertex of the AIM at weak coupling and strong coupling using multipoint NRG.
Both the imaginary-frequency MF and real-frequency KF vertices agree very well with known limits, namely weak-coupling perturbation theory and low-energy Fermi-liquid theory, and the latter accurately satisfies the generalized fluctuation-dissipation relations.
We expect that the sIE may also be useful for other computational methods such as QMC.
For NRG, it provides a robust way of computing the real-frequency Keldysh vertex, opening up the possiblility for studying real-frequency nonlocal correlations via diagrammatic extensions of DMFT~\cite{2018RohringerRMP}.

% =============================================================
\begin{acknowledgments}
We thank Andreas Gleis for useful suggestions on the NRG methodology.
The authors gratefully acknowledge the \href{www.gauss-centre.eu}{Gauss Centre for Supercomputing e.V.} for funding this project by providing computing time on the GCS Supercomputer SuperMUC-NG at \href{www.lrz.de}{Leibniz Supercomputing Centre}.
This research was supported in part by the National Science Foundation under PHY-1748958.
JML was supported by the Creative-Pioneering Research Program through Seoul National University, Korean NRF No-2023R1A2C1007297, and the Institute for Basic Science (No.~IBSR009-D1).
JH, JS, and JvD were supported by the Deutsche Forschungsgemeinschaft (DFG) under Germany’s Excellence Strategy EXC-2111 (Project No.~390814868), and the Munich Quantum Valley, supported by the Bavarian state government with funds from the Hightech Agenda Bayern Plus. 
JH acknowledges support by the International Max Planck Research School for Quantum Science and Technology (IMPRS-QST).
JS acknowledges the DFG Grant No.~LE3883/2-2 (Project No.~403832751).
SSBL is supported by the New Faculty Startup Fund from Seoul National University, and also by the National Research Foundation of Korea (NRF) grant funded by the Korean government (MSIT) (No.~RS-2023-00214464).
The Flatiron Institute is a division of the Simons Foundation.
\end{acknowledgments}

% ======================================================================================================
\appendix
\section{Boundary conditions of correlators} \label{sec:boundary}
In this Appendix, we show that the boundary term that arises when integrating by parts in \Eq{eq:eom_integral_long} vanishes.
In the MF, this can be easily seen using the boundary condition of the imaginary-time correlator on the contour of Fig.~\ref{fig:contours_simple}(a).
However, in the ZF and the KF, correlators defined by \Eq{eq:green_def} and the contours of Figs.~\ref{fig:contours_simple}(b) and \ref{fig:contours_simple}(c) do not have simple boundary conditions.
We can nevertheless show that the boundary term vanishes, using (i) correlators ordered on a modified (L-shaped) contour (Fig.~\ref{fig:contours})~\cite{StefanucciBook}
and (ii) the adiabatic assumption, which is widely adopted (also in the original work of Keldysh~\cite{1965Keldysh}) as it simplifies the derivation.

It may be surprising that the adiabatic assumption is evoked in the ZF and KF but not in the MF.
After all, in thermal equilibrium (which we assume in this work), the entire information is encoded in MF correlators.
Indeed, one can obtain the retarded components of the KF by analytic continuation~\cite{2021KuglerPRX,2020GeThesis,Ge2023} and all other components by further accounting for the discontinuities of the MF correlator in the complex frequency plane~\cite{Ge2023}.
We resolve this issue in App.~\ref{sec:wo_adia} by showing that it is indeed possible (albeit more tedious) to derive the KF EOMs in thermal equilibrium without the adiabatic assumption.

\subsection{Contour formalism for MF, ZF, KF correlators}
Using the notation for \lp\ correlators from \Sec{sec:eom}, we define a correlator of $\ell$ operators $\mb{\mcO}$ at times $\mb{z}$ on a contour with a (possibly) time-dependent Hamiltonian $\opH(z)$ as
\begin{flalign} 
\label{eq:G_def}
\mcG[\mb{\mcO}](\mb{z})
& = (-i)^{\ell-1} \frac{\Tr \left[ \mcT e^{-i \nint d\bar{z} H(\bar{z})} \mcO^1_{\rm S}(1) \ccdots \mcO^\ell_{\rm S}(\ell) \right] }{\Tr \left[ \mcT e^{-i \nint d\bar{z} \opH(\bar{z})} \right] },
\hspace{-0.5cm} &
\end{flalign}
where $\mcT$ denotes the contour ordering of the operators and $\bar{z}$ denotes a contour integration variable (not the complex conjugate of $z$).
Here, $\mcO^m_{\rm S}(m) = \mcO^m_{\rm S}(z_m)$ denote operators in the Schr\"odinger picture, in contrast to the Heisenberg operators $\mcO^m(m)$ of \Eq{eq:Heisenberg_def}, because time dependence is generated by $e^{-i \nint d\bar{z} H(\bar{z})}$.
The correlator \eqref {eq:G_def} satisfies the Kubo--Martin--Schwinger (KMS) boundary condition
\begin{equation} \label{eq:KMS}
\mcG[\mb{\mcO}](\mb{z}) \big|_{z_m = z_{\rm i}}
    = \zeta_m \mcG[\mb{\mcO}](\mb{z}) \big|_{z_m = z_{\rm f}}\,,
\end{equation}
where $z_{\rm i}$ and $z_{\rm f}$ are the endpoints of $\mcC$, and $\zeta_m$ is $+1$ ($-1$) if $\mcO^m$ is a bosonic (fermionic) operator.
The sign factor arises from commuting $\mcO^m$ past all other operators.
(The correlator is nonzero only for an even number of fermionic operators.
Hence, if $\mcO^m$ is fermionic, the remaining operators include an odd number of fermionic operators, leading to a $-1$ sign factor.)
The KMS boundary condition is easily proven using the cyclicity of the trace~\cite{StefanucciBook}.

Such a simple boundary condition does not hold for the correlators from \Eq{eq:green_contour}, defined as the thermal expectation value of time-ordered operators, because in general $\mcO^m$ does not commute with
the thermal density matrix $\rho = e^{-\beta H} / \Tr e^{-\beta H}$ involved in the thermal average $\expval{\ldots}$.
In the KF (Fig.~\ref{fig:contours_simple}(c)), e.g., we have $z_{\rm i} = -\infty^{-}$ and $z_{\rm f} = -\infty^+$, which leads to
\begin{align} \label{eq:contours_boundary_neq}
    \mcG[\mb{\mcO}](z_{\rm i}, z_2, \ccdots )
    &= \Tr\big[ \rho  \mcO^1(-\infty) \mcO^2(2) \ccdots \big] \nnnl
    \neq \zeta_1 \mcG[\mb{\mcO}](z_{\rm f}, z_2, \ccdots)
    &= \Tr\big[ \rho  \mcO^2(2) \ccdots \mcO^\ell(\ell) \mcO^1(-\infty) \big] \nnnl
    &= \Tr\big[ \mcO^1(-\infty) \rho  \mcO^2(2) \ccdots \big],
\end{align}
if $\mcO^1$ does not commute with $\rho$.

\begin{figure}[t]
    \centering
    \includegraphics[width=1.0\columnwidth]{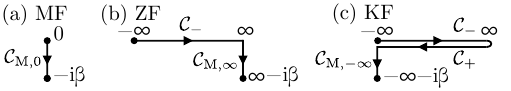}
    \caption{Time contour for each many-body formalism considered similar to \Fig{fig:contours_simple}, but with the vertical imaginary-time branches for (b) and (c).
    }
    \label{fig:contours}
\end{figure}

To connect \Eq{eq:G_def} with the correlators of the MF, ZF, and KF, we choose the contours
\begin{subequations} \label{eq:contour_contours}
\begin{align}
    {\rm MF:} & \; \overline{\mcC}  = \mcC_{\mathrm{M}, 0} = [0, -i\beta], \\
    {\rm ZF:} & \; \overline{\mcC}  = \mcC_- \oplus \mcC_{\mathrm{M},\infty}, \\
    \label{eq:contour_contours_KF}
    {\rm KF:} & \; \overline{\mcC}  = \mcC_- \oplus \mcC_+ \oplus \mcC_{\mathrm{M},-\infty},
\end{align}
\end{subequations}
respectively, as illustrated in Fig.~\ref{fig:contours}, where $\mcC_{\mathrm{M}, a} = [a, a-i\beta]$.
The overline distinguishes these contours from those used in the main text [\Eq{eq:contour_contours_simple} and \Fig{fig:contours_simple}].
In the MF, we set
\begin{subequations} \label{eq:contour_H}
\begin{align}
\label{eq:contour_MF}
{\rm MF:}\ &\opH(z) = \opH.
\end{align}
The time evolution on the vertical branch $\mcC_{\mathrm{M},0}$ then generates $e^{-\beta H}$, the interacting thermal state.
Thus, we readily find that, in the MF, the contour-ordered correlators are identical to the imaginary-time-ordered correlators [\Eq{eq:green_mf_expval}].

In the ZF and KF, we instead use
\begin{align}
\label{eq:contour_ZF}
{\rm ZF:}\ &\opH(z) = \begin{cases}
    \opH_\eta(t) & \text{if $z \in [-\infty, \infty]$} \\
    \opH^0 & \text{if $z \in [\infty, \infty - i\beta]$}
\end{cases},\\
\label{eq:contour_KF}
{\rm KF:}\ &\opH(z) = \begin{cases}
    \opH_\eta(t) & \text{if $z = t^{\pm}, t < 0$} \\
    \opH & \text{if $z = t^{\pm}, t \geq 0$} \\
    \opH^0 & \text{if $z \in [-\infty, -\infty - i\beta]$}
\end{cases},
\end{align}
\end{subequations}
where
$t^-$ ($t^+$) denotes time $t$ on the forward (backward) branch, and
\begin{equation} \label{eq:contour_adia_switching}
    \opH_\eta(t) = \opH^0 + e^{-\eta \abs{t}}\opH_{\rm int}
\end{equation}
describes the adiabatic switching of the interaction with an infinitesimal rate $\eta = 0^+$ on the horizontal branches.
The interaction is fully switched off on the vertical branch.

In the KF, the time evolution on the vertical branch generates $e^{-\beta H^0}$, the noninteracting thermal state.
According to the adiabatic assumption, the adiabatic switching of the interaction on the horizontal branches connects this state to the interacting thermal state~\cite{1965Keldysh}:
\begin{equation} \label{eq:contour_adiabatic_assumption}
    \frac{e^{-\beta \opH}}{\Tr e^{-\beta \opH}} \stackrel{\rm adia.}{=} U_\eta(0, -\infty) \frac{e^{-\beta \opH^0}}{\Tr e^{-\beta \opH^0}} U_\eta(-\infty, 0),
\end{equation}
where $U_\eta$ is the time-evolution operator for $\opH_\eta$.
Under this adiabatic assumption, the contour-ordered correlator is identical to the ordinary Keldysh correlators defined as the equilibrium expectation value of contour-time-ordered operators [\Eq{eq:green_kf_adia_expval}]~\cite{StefanucciBook}:
\begin{equation}
    \mcG_{\rm K}^\mb{c}[\mb{\mcO}](\mb{t})
    \stackrel{\rm adia.}{=} (-i)^{\ell - 1} \Bigexpval{ \mcT [\mb{\mcO}] (\mb{t}^\mb{c})}.
\end{equation}

In the ZF, the adiabatic switching connects the noninteracting ground state $\ket{\Psi^0}$ to the interacting ground state $\ket{\Psi}$ (assuming no level crossing)~\cite{1951GellMannLow}:
\begin{equation} \label{eq:contour_adiabatic_assumption_zf}
    \ket{\Psi} \stackrel{\rm adia.}{=} U_\eta(0, -\infty) \ket{\Psi^0}.
\end{equation}
Then, again, the ZF contour-ordered correlator equals the ground-state expectation value of time-ordered operators [\Eq{eq:green_zf_expval}]:
\begin{equation}
    \mcG_{\rm Z}[\mb{\mcO}](\mb{t})
    \stackrel{{\rm adia.},\,T=0}{=} (-i)^{\ell - 1} \Bigexpval{ \mcT [\mb{\mcO}] (\mb{t}) }.
\end{equation}

\subsection{Derivation of vanishing boundary terms} \label{sec:boundary_adia_der}
Let us now prove \Eq{eq:eom_integral_long}.
The derivation of the EOMs leading up to \Eq{eq:eom_integral_long} [including Eqs.~\eqref{eq:eom_operators}, \eqref{eq:eom_general}, \eqref{eq:eom_nonint_2p}, and \eqref{eq:eom_int}]
holds unchanged for the contour correlators defined by \Eq{eq:G_def}.
The step from \Eq{eq:eom_integral_derivation_start} to \Eq{eq:eom_integral_long}, with the boundary term made explicit, reads
\begin{align} \label{eq:boundary_derivation}
    &\quad \mcG[\opp_a,\mb{\mcO}^\rmv{m}] \nnnl
    &= \nsint{m'} g^0_{a\bar{a}}(m, m') ( -i\lpartialp_{m} \one - \opH^0 )_{\bar{a}a'} \mcG[\opp_{a'},\mb{\mcO}^\rmv{m}](m', \rmv{m}) \nnnl
    &= \nsint{m'} g^0_{a\bar{a}}(m, m') (i \partial'_{m} \one - \opH^0 )_{\bar{a}a'} \mcG[\opp_{a'},\mb{\mcO}^\rmv{m}](m', \rmv{m}) \nnnl
    &-i g^0_{aa'}(m, m') \mcG[\opp_{a'},\mb{\mcO}^\rmv{m}](m', \rmv{m}) \big|_{z_{\rm i}}^{z_{\rm f}}.
\end{align}
Thanks to the KMS boundary condition [\Eq{eq:KMS}], which holds for all three choices of the contour
$\overline{\mcC}$ defined in \Eq{eq:contour_contours}, the last line vanishes:
\begin{equation}
    g^0(m, z_{\rm i}) \mcG(z_{\rm i}, \rmv{m})
    = g^0(m, z_{\rm f}) \mcG(z_{\rm f}, \rmv{m}).
\end{equation}
Here, we omitted the orbital subscript and operator arguments for brevity.
The sign factors coming from $g^0$ and $\mcG$ are both $\zeta_m$ and thus cancel as $\zeta_m^2 = 1$.
We emphasize that this logic cannot be used for the ZF and KF contours \textit{without} vertical branches [\Eq{eq:contour_contours_simple}] as the KMS boundary condition~\eqref{eq:KMS} then does not hold [\Eq{eq:contours_boundary_neq}].

Inserting the EOM \eqref{eq:eom_int_d} into \Eq{eq:boundary_derivation}, we find the analogue of \Eq{eq:eom_integral_long},
\begin{flalign} 
\label{eq:boundary_derivation_2}
\mcG[\opp_a,\mb{\mcO}^\rmv{m}]
& = \nsint{m'} g^0_{aa'}(m, m') \mc{F}^m[\opp_{a'},\mb{\mcO}^\rmv{m}](m', \rmv{m}),
\hspace{-0.5cm} &
\end{flalign}
where the integral over $z_m'$ is performed over $\overline{\mcC}$ [\Eq{eq:contour_contours}].
In the MF, this concludes the proof because $\overline{\mcC}_{\rm M} = \mcC_{\rm M}$.
In the ZF and KF, $\overline{\mcC}$ and $\mcC$ differ by the vertical branches $\mcC_{\mathrm{M},\infty}$ and $\mcC_{\mathrm{M},-\infty}$, respectively.
If all the time arguments $\mb{z}$ are on the horizontal branches $C_\pm$, $z_m'$ on the vertical branch does not contribute to the integral of \Eq{eq:boundary_derivation_2}, because the interaction is zero on the vertical branch [\Eqs{eq:contour_ZF} and \eqref{eq:contour_KF}]:
\begin{equation} \label{eq:contour_F_zero}
    \left. \mc{F}^m(z_m', \mb{z}^\rmv{m}) \right|_{\Im z_m' \neq 0\ \text{and}\ z_{n} \in C_\pm} = 0.
\end{equation}
Note that the adiabatic assumption is crucial here: if the interaction were nonzero on the vertical branch, \Eq{eq:contour_F_zero} would not hold.
With the vertical branch not contributing, the integration domain in \Eq{eq:boundary_derivation_2} becomes $\mcC$, thus concluding the proof of \Eq{eq:eom_integral_long}.

% ==============================================================
\section{EOM derivation without adiabatic assumption} \label{sec:wo_adia}
In this Appendix, we prove the EOM in the integral form [\Eq{eq:eom_integral}] in the KF without resorting to the adiabatic assumption.
Without the adiabatic assumption, one needs to use a contour with the interaction present on both the horizontal and vertical branches.
We will show that the contribution of the vertical branch to the EOM vanishes.
We thus recover the EOM with only the horizontal branchs, as used in the main text.

After introducing the relevant contours in \Sec{sec:wo_adia_corr}, we prove the EOM in \Sec{sec:wo_adia_other} by rewriting the correlators in terms of \lp\ greater correlators [\Eq{eq:wo_adia_greater_def}].
We finish in \Sec{sec:wo_adia_k1_1} by presenting a much simpler proof that applies only to a subset of Keldysh components.

\begin{figure}[t]
    \centering
    \includegraphics[width=1.0\columnwidth]{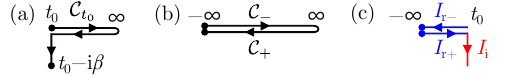}
    \caption{Time contours for KF (a) without or (b) with the adiabatic assumption [\Fig{fig:contours_simple}(c)], and (c) the difference between (a) and (b).
    }
    \label{fig:wo_adia_contour}
\end{figure}

\subsection{Correlators without the adiabatic assumption} \label{sec:wo_adia_corr}
We use the contour
\begin{gather} \label{eq:wo_adia_contour}
    \mcC_{t_0} = \mcC_{-,t_0} \oplus \mcC_{+,t_0} \oplus \mcC_{{\rm M}, t_0}, \\
    \mcC_{-,t_0} = [t_0, \infty],\ \mcC_{+,t_0} = [\infty, t_0],\ \mcC_{{\rm M}, t_0} = [t_0, t_0 - i\beta], \nonumber
\end{gather}
as shown in \Fig{fig:wo_adia_contour}(a), and set
\begin{equation}
    \opH(z) = \opH.
\end{equation}
Since the interaction is present on the vertical branch, the adiabatic assumption [\Eq{eq:contour_adiabatic_assumption}] is \textit{not} needed to equate the contour-ordered correlators with the equilibrium expectation values.
Instead, \Eq{eq:G_def} directly yields
\begin{equation}
    \mcG_{\rm KF,\ w\!/\!o\,adia.}^\mb{c}[\mb{\mcO}](\mb{t})
    = (-i)^{\ell - 1} \Bigexpval{ \mcT \prod_{i=1}^{\ell} [\mb{\mcO}](\mb{t}^\mb{c}) }.
\end{equation}

If all time arguments lie on the horizontal branches, i.e., for real valued $t_1, \ccdots, t_\ell$,
the contour $\mcC_{t_0}$ itself defines a real-time correlator $\mcG^\mb{c}(\mb{t})$ only for $t_i \geq t_0$.
Still, we can extend the definition to $t_i < t_0$ in a manner that yields a time-translation-invariant correlator by construction.
We define a time shift $\Delta t = t_0 - \min(t_0, t_1, \ccdots, t_\ell)$ so that
\begin{equation}
    \mb{t} + \Delta t = \big( t_1 + \Delta t, \ccdots, t_\ell + \Delta t \big)
\end{equation}
are on the contour $\mcC_{t_0}$, i.e. $t_i + \Delta t \geq t_0$.
Then, we define
\begin{equation} \label{eq:wo_adia_G_time_translation}
    \mcG^\mb{c}(\mb{t}) = \mcG^\mb{c}(\mb{t} + \Delta t),
\end{equation}
where the left side is given by the right side.
This extended definition enlarges the domain of $\mcG^\mb{c}(\mb{t})$ to include $\mcC_{-} \oplus \mcC_{+}$ [\Fig{fig:wo_adia_contour}(b)], the domain of the correlator of \Eq{eq:green_kf_adia_expval}, as a subset.
By construction, the resulting $\mcG^\mb{c}(\mb{t})$ satisfies time-translational invariance.
We note that one cannot simply set $t_0 = -\infty$ because this limit is ill-defined for the EOM in the integral form [e.g., \Eq{eq:wo_adia_I_i} evaluates to \Eq{eq:wo_adia_Ii_integral}].

Let us consider the EOM \eqref{eq:eom_integral_d} with $m=1$.
(Other EOMs easily follow by permutation and complex conjugation.)
Our goal is to prove the EOM \eqref{eq:eom_integral_d} with $z_1'$ integrated over $\mcC_{-} \oplus \mcC_{+}$ [\Fig{fig:wo_adia_contour}(b) or \Fig{fig:contours_simple}(c)]:
\begin{equation} \label{eq:wo_adia_eom_1}
    \mcG[1](\mb{z})
    = \nsint{1' \in \mcC_{\pm}} g^0_1(1, 1') \mc{F}[1](1', \rmv{1}).
\end{equation}
A similar equation, but with $z_1'$ integrated over $\mcC_{t_0}$ [\Fig{fig:wo_adia_contour}(a)], can readily be derived as in \Sec{sec:boundary_adia_der}:
\begin{equation} \label{eq:wo_adia_eom_2}
    \mcG[1](\mb{z})
    = \nsint{1' \in \mcC_{t_0}} g^0_1(1, 1') \mc{F}[1](1', \rmv{1}).
\end{equation}
We will now derive \Eq{eq:wo_adia_eom_1} from \Eq{eq:wo_adia_eom_2} by showing that the right sides of the two equations are equal.

The difference between the right sides comes from the integral over the segments shown in \Fig{fig:wo_adia_contour}(c). This is given by $I_{\rm r} + I_{\rm i}$,
where $I_{\rm r} = I_{{\rm r}+} + I_{{\rm r}-}$, and
\begin{subequations}
\begin{align}
    \label{eq:wo_adia_I_r}
    I_{{\rm r}\,\pm}(\mb{t}^\mb{c})
    &= \pm \nbint{-\infty}{t_0} d t' g^0_{aa'}(1, t^{\prime \pm}) \mc{G}[\opq_{a'},\mb{\mcO}^\rmv{1}](t^{\prime \pm}, \rmv{1}),
    \\
    \label{eq:wo_adia_I_i}
    I_{\rm i}(\mb{t}^\mb{c})
    &= - i\nbint{0}{\beta} d \tau g^0_{aa'}(1, t_0-i\tau) \mc{G}[\opq_{a'},\mb{\mcO}^\rmv{1}](t_0 - i\tau, \rmv{1}).
\end{align}
\end{subequations}
The right sides contain $\mcG$, not $\mcF$, because the equal-time commutators in $\mcF$ [the last term of \Eq{eq:eom_F_def}] do not contribute since the domains of $1'$ and $\rmv{1}$ do not overlap.
We used the shorthand $\opq_{a'} = [\opp_{a'},H_{\rm int}]$ and
\begin{equation}
    (1, t') = (t_1^{c_1}, t'),\quad
    (z, \rmv{1}) = (z, t_2^{c_2}, \ccdots, t_{\ell}^{c_\ell})
\end{equation}
as in \Eq{eq:list_slashed}.
Henceforth, we drop the superscript $0$ on $g$ and abbreviate $\mcG[\opq_{a'},\mb{\mcO}^\rmv{1}]$ as $\mcG_{a'}$.
The time arguments $\mb{t}^\mb{c}$ are treated as given, fixed points on the contour $C_{-}^{t_0} \oplus C_{+}^{t_0}$, so that
\begin{equation} \label{eq:wo_adia_time}
t_i \geq t_0
.
\end{equation}

\subsection{General proof of the EOM} \label{sec:wo_adia_other}

To prove $I_{\rm r} + I_{\rm i} = 0$, we begin by defining an \lp\ ``greater correlator''
\begin{equation} \label{eq:wo_adia_greater_def}
\mcG^{>\mb{c}^\rmv{1}}[\mb{\mcO}](z, \mb{t}^\rmv{1})
= (-i)^{\ell - 1} \Bigexpval{ \mcO^1(z) \,
\mcT [\mb{\mcO}^{\rmv{1}}] (\mb{t}^{\rmv{1}\,\mb{c}^\rmv{1}}) }
,
\end{equation}
where the superscript $>$ denotes that $\mcO^1(z)$ does not follow the contour ordering and is always ordered as the last, put at the leftmost position within the thermal expectation value.
This correlator is a generalization of the greater component of the 2p correlator $\mcG^{>}(t, t') = \mcG(t^+, t^{\prime -})$.
We allow $z$ to be any complex number in \Eq{eq:wo_adia_greater_def}.
The Fourier transform of the greater correlator reads
\begin{equation} \label{eq:wo_adia_greater_FT}
    \mcG^{>\mb{c}^\rmv{1}}(t, \mb{t}^\rmv{1})
    = \nint \frac{d^\ell \omega}{(2\pi)^\ell} e^{-i\mb{\omega}\cdot\mb{t}} G^{>\mb{c}^\rmv{1}}(\mb{\omega}) 2\pi\delta(\omega_{1\ccdots\ell}).
\end{equation}

We now rewrite the integrands of $I_{{\rm r}\pm}$ and $I_{\rm i}$ in terms of greater correlators, starting with $I_{\rm i}$ [\Eq{eq:wo_adia_I_i}].
While $t_0 - i\tau$ is the first argument of $\mc{G}[\opq_{a'},\mb{\mcO}^\rmv{1}](t_0 - i\tau, \rmv{1})$ and thus already in the right place for equating this correlator with a greater correlator, this is not the case for $g_{aa'}(1, t_0-i\tau)$.
However, using the simple relation $G[\mcO^1, \mcO^2](z_1, z_2) = \zeta G[\mcO^2, \mcO^1](z_2, z_1)$, which holds irrespective of the contour used in \Eq{eq:G_def}, we can switch the two arguments of $g$.
Thereby, we obtain
\begin{align} \label{eq:wo_adia_greater_1}
    g_{aa'}(1, t_0 -i\tau)
    &= \zeta g^{>}[\oppd_{a'},\opp_a](t_0 -i\tau, 1),
    \nnnl
    \mcG(t_0 -i\tau, \rmv{1})
    &= \mcG^{>}(t_0 - i\tau, \rmv{1}).
\end{align}
Rewriting $I_{\rm i}(\mb{t}^\mb{c})$ in terms of the greater correlators yields
\begin{align} \label{eq:wo_adia_Delta_G_1}
    I_{\rm i}(\mb{t}^\mb{c})
    &= - i\nbint{0}{\beta} d \tau g_{aa'}(1, t_0 - i\tau) \mc{G}_{a'}(t_0 - i\tau, \rmv{1}) \nnnl
    &= - i\nbint{0}{\beta} d \tau
    \zeta g^{>}[\oppd_{a'},\opp_a](t_0 - i\tau, 1)
    \mcG^{>}_{a'}(t_0 - i\tau, \rmv{1}) \nnnl
    &= - i\nbint{0}{\beta} d \tau f(t_0 - i\tau,\mb{t}^\mb{c}).
\end{align}
Here, we defined
\begin{equation} \label{eq:wo_adia_f_def}
    f(z,\mb{t}^\mb{c}) = \zeta g^{>}[\oppd_{a'},\opp_a](z, 1) \mc{G}^{>}_{a'}(z, \rmv{1}),
\end{equation}
which appears with $z = t_0 - i\tau$.
Below, it will reappear with other complex time arguments.

Similarly, for $I_{{\rm r},+}$ [\Eq{eq:wo_adia_I_r}], $t^{\prime+}$ is the last contour argument because $t' \leq t_0$.
Applying the same permutation to the arguments of $g$, we find
\begin{align} \label{eq:wo_adia_greater_3}
    g_{aa'}(1, t^{\prime +})
    &= \zeta g^{>}[\oppd_{a'},\opp_a](t', 1),
    \nnnl
    \mcG_{a'}(t^{\prime+}, \rmv{1})
    &= \mcG^{>}_{a'}(t', \rmv{1}).
\end{align}
Substituting these equations into \Eq{eq:wo_adia_I_r} yields
\begin{align} \label{eq:wo_adia_Delta_G_2}
    I_{{\rm r},+}(\mb{t}^\mb{c})
    &=  \nbint{-\infty}{t_0} d t^{\prime} g_{aa'}(1, t^{\prime+}) \mc{G}_{a'}(t^{\prime+}, \rmv{1}) \nnnl
    &= \nbint{-\infty}{t_0} d t' \zeta g^{>}[\oppd_{a'},\opp_a](t', 1) \mc{G}^{>}_{a'}(t', \rmv{1}) \nnnl
    &= \nbint{-\infty}{t_0} d t' f(t', \mb{t}^\mb{c}).
\end{align}

Finally, for $I_{{\rm r},-}$ [\Eq{eq:wo_adia_I_r}], $t' \leq t_0$ is the smallest time argument but is on the forward branch ($c' = -$).
Yet, we can still relate the correlators to greater correlators by applying time translation by $t_0 - t'$ and using the KMS boundary condition [\Eq{eq:KMS}]:
\begin{align} \label{eq:wo_adia_greater_5}
    g_{aa'}(1, t^{\prime -})
    &= g^{c_1 -}_{aa'}(t_1 - t' + t_0, t_0) \nnnl
    &= \zeta g^{c_1{\rm M}}_{aa'}(t_1 - t' + t_0, t_0 - i\beta) \nnnl
    &= g^{>}[\oppd_{a'},\opp_a](t' - i\beta, 1),
\\
\label{eq:wo_adia_greater_6}
    \mcG_{a'}(t^{\prime -}, \rmv{1})
    &= \mcG^{-\mb{c}^\rmv{1}}_{a'}(t_0, \mb{t}^{\rmv{1}} - t' + t_0) \nnnl
    &= \zeta \mcG^{\mb{c}^\rmv{1}}_{a'}(t_0-i\beta, \mb{t}^{\rmv{1}} - t' + t_0) \nnnl
    &= \zeta \mcG^{>}_{a'}(t'-i\beta, \rmv{1}).
\end{align}
The KMS boundary condition is used in the second equalities of \Eqs{eq:wo_adia_greater_5} and \eqref{eq:wo_adia_greater_6}.
For the third equalities, we used \Eq{eq:wo_adia_greater_1} and the time-translational invariance.
Substituting \Eqs{eq:wo_adia_greater_5} and \eqref{eq:wo_adia_greater_6} into \Eq{eq:wo_adia_I_r} gives
\begin{align} \label{eq:wo_adia_Delta_G_3}
    I_{{\rm r},-}(\mb{t}^\mb{c})
    &= - \nbint{-\infty}{t_0} d t' g_{aa'}(1, t'^-) \mc{G}_{a'}(t'^{-}, \rmv{1}) \nnnl
    &= - \nbint{-\infty}{t_0} d t' \zeta g^{>}[\oppd_{a'},\opp_a](t' - i\beta, 1) \mc{G}^{>}_{a'}(t' - i\beta, \rmv{1}) \nnnl
    &= -\nbint{-\infty}{t_0} d t' f(t' - i\beta, \mb{t}^\mb{c}).
\end{align}

By combining \Eqs{eq:wo_adia_Delta_G_1}, \eqref{eq:wo_adia_Delta_G_2} and \eqref{eq:wo_adia_Delta_G_3}, we find
\begin{subequations}
\label{eq:wo_adia_time_integral}
\begin{align} 
\label{eq:wo_adia_time_integral_1}
I_{\rm i}(\mb{t}^\mb{c})
&= \nbint{t_0}{t_0 - i\beta} d z f(z, \mb{t}^\mb{c}), \\
\label{eq:wo_adia_time_integral_2}
I_{{\rm r}}(\mb{t}^\mb{c})
&= \nbint{-\infty}{t_0}  d t' \bigl[f(t', \mb{t}^\mb{c}) - f(t' - i\beta, \mb{t}^\mb{c}) \bigr]
.
\end{align}
\end{subequations}
Next, we evoke the Fourier representation of $f$,
\begin{align}
\label{eq:f_z_tc_Fourier}
f(z,\mb{t}^\mb{c})
&= \nint \frac{d \Omega}{2\pi}  \frac{d^\ell \omega}{(2\pi)^{\ell}} 
e^{-i\Omega z}
e^{-i\mb{\omega}\cdot\mb{t}} f^\mb{c}(\Omega,\mb{\omega})
.
\end{align}
Although it is not needed in what follows,
one can show
\begin{align} 
f^\mb{c}(\Omega, \mb{\omega}) 
& = 2\pi\delta(\Omega + \omega_{1\ccdots\ell}) f^\mb{c}(\mb{\omega}),
\nnnl
f^\mb{c}(\mb{\omega})
& = 
\zeta g^{>}[\oppd_{a'},\opp_a](-\omega_1, \omega_1)
G^{>\mb{c}^\rmv{1}}_{a'}(-\omega_{2\ccdots\ell}, \mb{\omega}^\rmv{1})
,
\end{align}
since Fourier transforms turn \Eq{eq:wo_adia_f_def} into a convolution, and a convolution of two delta functions is a single delta function, $\delta(\Omega + \omega_{1\ccdots\ell})$.
In evaluating \Eqs{eq:wo_adia_time_integral}
with \Eq{eq:f_z_tc_Fourier},
we switch the $z$ (or $t'$) and $\Omega$ integrals.
For $I_{\rm i}$, the $z$ integral reads
\begin{equation} \label{eq:wo_adia_Ii_integral}
\nbint{t_0}{t_0 - i\beta} d z \, e^{i\Omega z}
=
\frac{-i}{\Omega} e^{i\Omega t_0} \left( e^{\Omega \beta} - 1 \right).
\end{equation}
For $I_{\rm r}$, we use the Fourier transform of the step function,
\begin{align}
\nbint{-\infty}{t_0} d t' e^{i\Omega t'}
& =
\nbint{-\infty}{\infty} d t' \Theta(t_0 - t') e^{i\Omega t'}
=
\frac{-i}{\Omega - i 0^+} e^{i\Omega t_0},
\nnnl
\nbint{-\infty}{t_0} d t' e^{i\Omega (t' - i\beta)}
& =
\frac{-i}{\Omega - i 0^+} e^{i\Omega t_0} e^{\Omega \beta}.
\end{align}
In total, we get
\begin{align} \label{eq:wo_adia_final_time}
I_{\rm i}(\mb{t}^\mb{c}) & + I_{{\rm r}}(\mb{t}^\mb{c}) 
=
i \nint \frac{d \Omega}{2\pi} \frac{d^\ell \omega}{(2\pi)^\ell} e^{i\Omega t_0}(e^{\Omega\beta} - 1)  \nnnl
& \ \times 
\left(\frac{1}{\Omega - i0^+} - \frac{1}{\Omega} \right)  e^{-i\mb{\omega} \cdot \mb{t}} f^\mb{c}(\Omega,\mb{\omega})
\nnnl
& =
\nint \frac{d \Omega}{2\pi} \frac{d^\ell \omega}{(2\pi)^\ell} e^{i\Omega t_0}(1 - e^{\Omega\beta}) 
\pi\delta(\Omega) e^{-i\mb{\omega} \cdot \mb{t}} f^\mb{c}(\Omega,\mb{\omega})
\nnnl
& = 0.
\end{align}
In conclusion, adding the contributions from the horizontal and vertical branches to the difference between the right sides of \Eqs{eq:wo_adia_eom_1} and \eqref{eq:wo_adia_eom_2} yields $(1 - e^{\Omega\beta}) \delta(\Omega) = 0$, thus proving $I_{\rm i} + I_{\rm r} = 0$ and the EOM.

Note that this derivation applies only to the equilibrium KF as we have set $H(z) = H$.
In the non-equilibrium case where $H(z)$ depends on $z$, correlators that mix horizontal and vertical branches need to be taken into account to close the EOM.

We further note that well-known FDRs in the frequency domain can be derived from the KMS boundary condition using identities similar to those derived in this Appendix~\cite{StefanucciBook}.
As an illustration, let us derive the 2p FDR.
We begin by noting that \Eq{eq:wo_adia_greater_6} is valid for all $t_1$ and $t_2$ if $c_1 = +$.
Then, one finds
\begin{equation}
    g^{<}(t_2, t_1) = \zeta g(t_1^+, t_2^-) = \zeta g^{>}(t_2 - i\beta, t_1),
\end{equation}
whose Fourier transformation reads
\begin{equation}
    g^{<}(\omega) = \zeta e^{-\beta\omega} g^{>}(\omega).
\end{equation}
By substituting $g^{>/<} = \frac{1}{2}(\pm g^{R} \mp g^{A} + g^{K})$, we find the familiar 2p FDR
\begin{align}
    g^{K}(\omega)
    &= \frac{1 + \zeta e^{-\beta\omega}}{(1 - \zeta e^{-\beta\omega})} [g^{R}(\omega) - g^{A}(\omega)] \nnnl
    &= [\coth(\beta\omega/2)]^\zeta  [g^{R}(\omega) - g^{A}(\omega)].
\end{align}

\subsection{Derivation of the EOMs for fully retarded correlators} \label{sec:wo_adia_k1_1}

Let us finish this Appendix by presenting two cases where $I_{\rm r}$ and $I_{\rm i}$ are both trivially zero: correlators whose Keldysh components satisfy $k_1 = 1$ or $\mb{k}^\rmv{1} = (1,\ccdots,1)$.
EOMs in these two cases suffice to derive the sIEs for the fully retarded correlators, which have only a single $2$ in the list of Keldysh indices.
For example, for $\mb{k}=(2111)$, we may use the latter case when deriving the EOM for the first operator, and the former otherwise.
The EOMs for these components may as well be derived by the analytic continuation of the MF EOMs.

In the Keldysh basis, the integrals of interest are
\begin{subequations}
\begin{align} \label{eq:wo_adia_I_r_Keldysh}
I^{\mb{k}}_{{\rm r}}(\mb{t})
= - \nbint{-\infty}{t_0} d t' g^{k_1 k'}_{aa'}(t_1, t') X^{k' k''} \mc{G}^{k''\mb{k}^\rmv{1}}(t', \mb{t}^\rmv{1}),
\\
\label{eq:wo_adia_I_i_Keldysh}
I^{\mb{k}}_{\rm i}(\mb{t})
= - i\nbint{0}{\beta} d \tau g^{k_1 {\rm M}}_{aa'}(t_1, t_0-i\tau) \mc{G}^{{\rm M}\mb{k}^\rmv{1}}(t_0 - i\tau, \mb{t}^\rmv{1}),
\end{align}
\end{subequations}
where the superscript M denotes that $t_0 - i\tau$ is on the vertical branch.
For $I_{{\rm r}} = I_{{\rm r},-} + I_{{\rm r},+}$, the sum over the forward and backward branches converts to the sum over dummy Keldysh indices $k'$ and $k''$.
The $(-1)^{\delta_{c,+}}$ sign factor converts to $X^{k' k''}$ via \Eq{eq:eom_D_Z_to_X}.

To show $I_{\rm r} = I_{\rm i} = 0$,
we use a well-known property of the KF correlators: the correlator is zero if the Keldysh index of the largest real-time argument is 1~\cite{2010JakobsReview}:
\begin{equation} \label{eq:wo_adia_keldysh_zero}
    \mcG^\mb{k}(\mb{z}) = 0 \text{ if $k_n = 1$ and $z_n = t_n \geq \Re z_i$ for all $i$}.
\end{equation}
This holds because moving $z_n$ from the forward to the backward branch keeps that operator order and thus the correlator invariant.
A subsequent Keldysh rotation with $D^{k=1,c} \propto (-1)^{\delta_{c,+}}$ then yields zero.

Let us begin with the $k_1 = 1$ components.
$I_{\rm r}$ in the Keldysh basis [\Eq{eq:wo_adia_I_r_Keldysh}] contains the propagator $g^{1k'}(t_1, t')$.
In the integration domain $t' \in [-\infty, t_0]$, $t' \leq t_0 \leq t_1$ holds [\Eq{eq:wo_adia_time}], and $t_1$ is the largest real-time argument.
Since $k_1 = 1$, \Eq{eq:wo_adia_keldysh_zero} yields
\begin{equation}
g^{1k'}(t_1, t') = 0
\quad\Rightarrow\quad
I_{\rm r} = 0
.
\end{equation}
Similarly, $I_{\rm i}$ in the Keldysh basis includes the propagator $g^{1{\rm M}}(t_1, t_0 - i\tau)$.
This term is again zero because $t_1$ is the largest real-time argument:
\begin{equation}
g^{1{\rm M}}(t_1, t_0 - i\tau) = 0
\quad\Rightarrow\quad
I_{\rm i} = 0
.
\end{equation}

Similarly, one can derive $I_{\rm r} = I_{\rm i} = 0$ for $\mb{k}^{\rmv{1}} = (1,\ccdots,1)$, i.e., for a fully retarded correlator.
Let $n$ be the index where $t_n$ is the largest among $t_2, \ccdots, t_\ell$.
Since $t' \in [-\infty, t_0]$, we find $t_n \geq t'$.
Then, \Eq{eq:wo_adia_keldysh_zero} gives (because $k_n = 1$)
\begin{equation}
\mc{G}^{k''\mb{k}^{\rmv{1}}}(t', \mb{t}^\rmv{1}) = 0
\quad\Rightarrow\quad
I_{\rm r} = 0
.
\end{equation}
Analogously, one has
\begin{equation}
\mc{G}^{{\rm M}\mb{k}^\rmv{1}}(t_0 - i\tau, \mb{t}^\rmv{1}) = 0
\quad\Rightarrow\quad
I_{\rm i} = 0
.
\end{equation}

% ===============================================
\section{Symmetries of the 4p vertex} \label{sec:symmetry}
In this Appendix, we present symmetries of the 4p vertex $\Gamma_{\sigma\sigma'}(\nu, \nu')$ in the $t$-channel parametrization at vanishing transfer frequency [\Eq{eq:ph_representation}].
A general 4p vertex $\Gamma_{1234}$ satisfies the crossing symmetry
\begin{align}
\Gamma_{1234} = -\Gamma_{3214} = -\Gamma_{1432} = \Gamma_{3412}
\end{align}
and the complex conjugation symmetry
\begin{align}
    \Gamma_{1234}^*(\mb{\omega}_{1234}) = \zeta_\mb{k}\, \Gamma_{2143}(-\mb{\omega}_{2143}),
\end{align}
where $\zeta_\mb{k} = (-1)^{1+\sum_i k_i}$.
Here, the numeric subscripts refer to all the relevant arguments: Keldysh, spin, orbital, and frequency.

For a single-orbital system at zero transfer frequency, we get a chain of identities:
\begin{subequations}
\begin{flalign}
& \hspace{0.2cm} \phantom{=\;\,}
\Gamma_{\sigma\sigma\sigma'\sigma'}^{k_1 k_2 k_3 k_4}(\nu,\nu')
\hspace{-0.5cm}
&& =
\Gamma_{\sigma'\sigma'\sigma\sigma}^{k_3 k_4 k_1 k_2}(\nu',\nu)
& \\
& \hspace{0.2cm} =
\zeta_{\mb{k}} [\Gamma_{\sigma\sigma\sigma'\sigma'}^{k_2 k_1 k_4 k_3}(\nu,\nu')]^*
\hspace{-0.5cm}
&& =
\zeta_{\mb{k}} [\Gamma_{\sigma'\sigma'\sigma\sigma}^{k_4 k_3 k_2 k_1}(\nu',\nu)]^*
.
&
\end{flalign}
\end{subequations}
Since $\Gamma_{\uparrow\uparrow\downarrow\downarrow} = \Gamma_{\downarrow\downarrow\uparrow\uparrow}$ under SU(2) symmetry, assumed in this work, there is no restriction from spin space.

Thereby, we obtain the symmetries
\begin{subequations}
\begin{flalign}
& \hspace{1cm} \phantom{=\;\,}
\Gamma_{\sigma\sigma'}^{1222}(\nu,\nu')
\hspace{-1cm} && =
\Gamma_{\sigma\sigma'}^{2212}(\nu',\nu)
& \nonumber \\
& \hspace{1cm} =
[\Gamma_{\sigma\sigma'}^{2122}(\nu,\nu')]^*
\hspace{-1cm} && =
[\Gamma_{\sigma\sigma'}^{2221}(\nu',\nu)]^*
,
& \\
& \hspace{1cm} \phantom{=\;\,}
\Gamma_{\sigma\sigma'}^{2111}(\nu,\nu')
\hspace{-1cm} && =
\Gamma_{\sigma\sigma'}^{1121}(\nu',\nu)
& \nonumber \\
& \hspace{1cm} =
[\Gamma_{\sigma\sigma'}^{1211}(\nu,\nu')]^*
\hspace{-1cm} && =
[\Gamma_{\sigma\sigma'}^{1112}(\nu',\nu)]^*
,
& \\
& \hspace{1cm} \phantom{=\;\,}
\Gamma_{\sigma\sigma'}^{1212}(\nu,\nu')
\hspace{-1cm} && =
\Gamma_{\sigma\sigma'}^{1212}(\nu',\nu)
& \nonumber \\
& \hspace{1cm} =
-[\Gamma_{\sigma\sigma'}^{2121}(\nu,\nu')]^*
\hspace{-1cm} && =
-[\Gamma_{\sigma\sigma'}^{2121}(\nu',\nu)]^*
,
& \\
& \hspace{1cm} \phantom{=\;\,}
\Gamma_{\sigma\sigma'}^{1221}(\nu,\nu')
\hspace{-1cm} && =
\Gamma_{\sigma\sigma'}^{2112}(\nu',\nu)
& \nonumber \\
& \hspace{1cm} =
-[\Gamma_{\sigma\sigma'}^{2112}(\nu,\nu')]^*
\hspace{-1cm} && =
-[\Gamma_{\sigma\sigma'}^{1221}(\nu',\nu)]^*
,
& \\
& \hspace{1cm} \phantom{=\;\,}
\Gamma_{\sigma\sigma'}^{1122}(\nu,\nu')
\hspace{-1cm} && =
\Gamma_{\sigma\sigma'}^{2211}(\nu',\nu)
& \nonumber \\
& \hspace{1cm} =
-[\Gamma_{\sigma\sigma'}^{1122}(\nu,\nu')]^*
\hspace{-1cm} && =
-[\Gamma_{\sigma\sigma'}^{2211}(\nu',\nu)]^*
,
& \\
& \hspace{1cm} \phantom{=\;\,}
\Gamma_{\sigma\sigma'}^{1111}(\nu,\nu')
& & =
\Gamma_{\sigma\sigma'}^{1111}(\nu',\nu)
& \nonumber \\
& \hspace{1cm} =
-[\Gamma_{\sigma\sigma'}^{1111}(\nu,\nu')]^*
\hspace{-1cm} && =
-[\Gamma_{\sigma\sigma'}^{1111}(\nu',\nu)]^*
.
& 
\end{flalign}
\end{subequations}
We note that the $\Gamma^{2222}_{\sigma\sigma'}$ component is zero by the property of KF vertices~\cite{2010JakobsReview}.

We conclude that the independent components are
\begin{align}
    & 
    \Gamma_{\sigma\sigma'}^{1222}(\nu,\nu')
    , \quad
    \Gamma_{\sigma\sigma'}^{2111}(\nu,\nu')
    , \quad
    \Gamma_{\sigma\sigma'}^{1122}(\nu,\nu'\!\leq\!\nu)
    , \nonumber \\
    & 
    \Gamma_{\sigma\sigma'}^{1221}(\nu,\nu'\!\leq\!\nu)
    , \quad
    \Im \Gamma_{\sigma\sigma'}^{1212}(\nu,\nu')
    , \quad
    \Im \Gamma_{\sigma\sigma'}^{1111}(\nu,\nu'\!\leq\!\nu)
    .
\end{align}
With the arguments 
$(\nu,\nu'\!\leq\!\nu)$,
we indicate that there is a symmetry in $(\nu \leftrightarrow \nu')$,
so that it in principles suffices to compute only half of the data points in the frequency plane.

By using these symmetries, we find the following identities for the causal vertex:
\begin{subequations} \label{eq:symm_causal}
\begin{align}
\Re \Gamma^{\rm causal}
& =
\tfrac{1}{2} \Re
( \Gamma^{1222} + \Gamma^{2212} + \Gamma^{2111} + \Gamma^{1121} )
\nonumber \\
& =
\tfrac{1}{2} \Re
( \Gamma^{1222} + \Gamma^{2111} )
+ (\nu \!\leftrightarrow\! \nu'),
\\
\Im \Gamma^{\rm causal}
& =
\tfrac{1}{4} \Im
( 2 \Gamma^{1212} + 2 \Gamma^{1221} + \Gamma^{1122} 
\nonumber \\
&
+ \Gamma^{2211} + \Gamma^{1111} )
\nonumber \\
& =
\tfrac{1}{4} \Im
( \Gamma^{1212} + \Gamma^{1221}
+ \Gamma^{1122} + \tfrac{1}{2} \Gamma^{1111} )
\nonumber \\
&
+ (\nu \!\leftrightarrow\! \nu').
\end{align}
\end{subequations}

% ===============================================
\section{Vertex estimators using the total correlator} \label{sec:disconnected_estimator}

In this Appendix, we derive sIEs for 3p and 4p vertices involving only total correlators, useful for methods like QMC, as discussed in \Sec{eq:est_total}.
The main results are \Eqs{eq:disc_2p}, \eqref{eq:disc_3p_est}, and \eqref{eq:disc_4p_est}.
These estimators have a form similar to those using the connected correlators
and the same perturbative and asymptotic properties (see Secs.~\ref{sec:perturbative} and \ref{sec:asymptotic}).
The estimators of this section have additional terms involving the self-energy to cancel disconnected contributions.

Below, we derive the relations for fermionic systems in normal (non-superconducting) phases preserving the $\mr{U}(1)$ charge symmetry,
ruling out terms like $\smallexpval{d_{\sigma}^{(\dagger)}}$ and $\smallexpval{d_{\uparrow} d_{\downarrow}}$ in the disconnected parts.
We add subscripts to correlators to distinguish the connected part (`con'), the disconnected part (`dis'), and the total (`tot') correlator; $G_{\rm tot} = G_{\rm con} + G_{\rm dis}$.
For later use, we define
\begin{equation}
    \widetilde{\Sigma} = \Sigma - \Sigma^{\rm H}X.
\end{equation}
This modified self-energy is $\mcO(U^2)$ in the perturbative limit and decays to zero in the large-frequency limit.

We begin with the 3p vertex estimator [\Eq{eq:3p_symmetric}]
\begin{equation} \label{eq:disc_3p_vertex}
    \Gamma^{(34)} = \mcK^{(34)}_{\rm con} + G^{(12,34)}_{\rm con} + \Gamma_{\rm bare}.
\end{equation}
The second term is an auxiliary correlator defined in terms of a 2p bosonic connected correlator [\Eq{eq:est_clamped_4p}].
The corresponding disconnected part is
\begin{align} \label{eq:disc_2p_dis}
    G^{(12,34)}_{\rm dis}
    &= P^{k_1 k_2 k_{12}} P^{k_3 k_4 k_{34}} G^{k_{12} k_{34}}_{\rm dis} [\opq_{12}, \opq_{34}] \nnnl
    &= 2\pi\delta(\omega_{12}) X^{k_1 k_2} X^{k_3 k_4} \Sigma^{\rm H}_{12} \Sigma^{\rm H}_{34},
\end{align}
where we used $\smallexpval{\opq_{12}} = \Sigma^{\rm H}_{12}$ [\Eq{eq:hartree}].
Note that $\delta(\omega_{12})$ and $\delta(\omega_{34})$ can be used interchangeably because the energy conservation constraint, $\omega_{1234} = 0$, is implicitly understood.
In the MF, the Dirac delta function $2\pi\delta(\omega_{12})$ is replaced by the Kronecker delta $\beta \delta_{\omega_{12}, 0}$.
Similarly deriving the disconnected parts for other bosonic 2p correlators and using $\opq_{23} = -\zeta \opq_{32}$, we find
\begin{align} \label{eq:disc_2p}
    G^{(12,34)}_{\rm con}
    &= G^{(12,34)}_{\rm tot} - 2\pi\delta(\omega_{12}) X^{k_1 k_2} \Sigma^{\rm H}_{12} X^{k_3 k_4} \Sigma^{\rm H}_{34},
    \nnnl
    G^{(13,24)}_{\rm con}
    &= G^{(13,24)}_{\rm tot},
    \nnnl
    G^{(14,23)}_{\rm con}
    &= G^{(14,23)}_{\rm tot} + 2\pi\delta(\omega_{14}) \zeta X^{k_1 k_4} \Sigma^{\rm H}_{14} X^{k_3 k_2} \Sigma^{\rm H}_{32}.
\end{align}

For the first term $\mcK^{34}_{\rm con}$ in \Eq{eq:disc_3p_vertex}, which is defined via \Eq{eq:3p_G12_U3} and is related to the $\mcK_2$ asymptotic class [\Eq{eq:4p_asymp_K2}], the disconnected part is given by
\begin{align} \label{eq:disc_3p}
    \mcK^{(34)}_{\rm dis}
    &= \sum_{\substack{x_n \in \{n, \cdot\} \\ n \in \{1, 2\}}} L_{x_1} G^{(x_1, x_2, 34)}_{\rm dis} L_{x_2} \nnnl
    &= \Sigma_1 G^{(\cdot,\cdot,34)}_{\rm dis} \Sigma_2
    - X_1 G^{(1,\cdot,34)}_{\rm dis} \Sigma_2 \nnnl
    &\quad - \Sigma_1 G^{(\cdot,2,34)}_{\rm dis} X_2 + X_1 G^{(1,2,34)}_{\rm dis} X_2,
\end{align}
where $L_{x_n = \,\cdot} = X_n$ and $L_{x_n = n} = \Sigma_n$.
This result involves diagrams in which the first and second legs are disconnected from the third and fourth legs.
For example, the disconnected part of $G^{(\cdot,\cdot,34)}$ reads
\begin{align}
    G^{(\cdot,\cdot,34)}_{\rm dis}
    &= P^{k_3 k_4 k_{34}} G^{k_1 k_2 k_{34}}_{\rm dis}[\opd_1, \opdd_2, \opq_{34}] \nnnl
    &= 2\pi\delta(\omega_{12}) X^{k_3 k_4} \Sigma^{\rm H}_{34} G^{(\cdot,\cdot)}_{12} ,
\end{align}
where we again used $\smallexpval{\opq_{34}} = \Sigma^{\rm H}_{34}$ [\Eq{eq:hartree}].
By applying the same procedure to all terms in \Eq{eq:disc_3p}, we find
\begin{align} \label{eq:disc_3p_2}
    \mcK^{(34)}_{\rm dis}
    &= 2\pi\delta(\omega_{12}) X^{k_3 k_4} \Sigma^{\rm H}_{34} \big(
    \Sigma G^{(\cdot,\cdot)} \Sigma - X G^{(1,\cdot)} \Sigma \nnnl
    &\qquad - \Sigma G^{(\cdot,2)} X + X G^{(1,2)} X
    \big)_{12}^{k_1 k_2} \nnnl
    &= 2\pi\delta(\omega_{12}) X^{k_3 k_4} \Sigma^{\rm H}_{34} \widetilde{\Sigma}_{12}^{k_1 k_2}.
\end{align}
For the second equality, we employed the self-energy estimators \eqref{eq:2p_Sigma_1st} and \eqref{eq:2p_Sigma_2nd}.
Using \Eqs{eq:disc_2p} and \eqref{eq:disc_3p_2} to convert the right side of \Eq{eq:disc_3p_vertex} to the total correlator, we find the desired 3p vertex estimator.
In the following, we also list the sIEs for other 3p vertices which can be derived similarly,
\begin{equation}
\begin{aligned} 
\mcK^{(12)}_{\rm con} &= \mcK^{(12)}_{\rm tot} - 2\pi\delta(\omega_{12}) \widetilde{\Sigma}_{34}^{k_3 k_4} X^{k_1 k_2} \Sigma^{\rm H}_{12}
, \\
\mcK^{(13)}_{\rm con} &= \mcK^{(13)}_{\rm tot}
, \\
\mcK^{(14)}_{\rm con} &= \mcK^{(14)}_{\rm tot} - 2\pi\delta(\omega_{14}) \zeta \widetilde{\Sigma}_{32}^{k_3 k_2} X^{k_1 k_4} \Sigma^{\rm H}_{14}
, \\
\mcK^{(23)}_{\rm con} &= \mcK^{(23)}_{\rm tot} + 2\pi\delta(\omega_{14}) \zeta \widetilde{\Sigma}_{14}^{k_1 k_4} X^{k_3 k_2} \Sigma^{\rm H}_{32}
, \\
\mcK^{(24)}_{\rm con} &= \mcK^{(24)}_{\rm tot}
, \\
\mcK^{(34)}_{\rm con} &= \mcK^{(34)}_{\rm tot} - 2\pi\delta(\omega_{12}) \widetilde{\Sigma}_{12}^{k_1 k_2} X^{k_3 k_4} \Sigma^{\rm H}_{34}
.
\end{aligned}
\label{eq:disc_3p_est}
\end{equation}
After replacing the connected auxiliary correlators by the total ones, a disconnected term involving self-energies is subtracted.
In \Eq{eq:disc_3p_est}, thanks to $\widetilde{\Sigma}$, the $\mcK^{(34)}$ estimator is still manifestly $\mcO(U^3)$ in the perturbative limit and decays in the $\abs{\omega_1} \to \infty$ or $\abs{\omega_2} \to \infty$ limits.
Note that by shifting the noninteracting and interacting Hamiltonian by $\Sigma^{\rm H}_{ij} \opdd_i \opd_j$ and $-\Sigma^{\rm H}_{ij} \opdd_i \opd_j$, respectively, the Hartree self-energy and the disconnected part of the 3p vertex estimators can be eliminated.

We close with the 4p vertex estimator \eqref{eq:4p_4th}.
As most terms already appear in the 3p estimator, it suffices to consider the core vertex $\Gamma_{\rm core}$ [\Eq{eq:4p_U4}].
In the normal (non-superconducting) phase, a disconnected 4p correlator has legs 1 and 2 disconnected from 3 and 4, or legs 1 and 4 disconnected from 2 and 3.
The first case gives
\begin{align}
    &\Gamma_\mathrm{core,\, dis\textrm{-}12} \nnnl
    &= 2\pi\delta(\omega_{12})
    \!\!\!\! \sum_{\substack{x_n \in \{n, \cdot\} \\ n \in \{1, 2, 3, 4\}}} \!\!\!\!
    L_{x_1} L_{x_3} G^{(x_1,x_2)}_{12} G^{(x_3,x_4)}_{34} L_{x_2} L_{x_4} \nnnl
    &= 2\pi\delta(\omega_{12})
    \! \sum_{\substack{x_n \in \{n, \cdot\} \\ n \in \{1, 2\}}} \! L_{x_1} G^{(x_1,x_2)}_{12} L_{x_2}
    \! \sum_{\substack{x_n \in \{n, \cdot\} \\ n \in \{3, 4\}}} \! L_{x_3} G^{(x_3,x_4)}_{34} L_{x_4} \nnnl
    &= 2\pi\delta(\omega_{12}) \widetilde{\Sigma}_{12}^{k_1 k_2} \widetilde{\Sigma}_{34}^{k_3 k_4},
\end{align}
where $\widetilde{\Sigma}$ appears as in \Eq{eq:disc_3p}.
Similarly, the second term reads
\begin{equation}
    \Gamma_\mathrm{core,\, dis\textrm{-}14}
    = 2\pi\delta(\omega_{14}) \zeta \widetilde{\Sigma}_{12}^{k_1 k_2} \widetilde{\Sigma}_{34}^{k_3 k_4}.
\end{equation}
Thus, the vertex estimator is
\begin{align} \label{eq:disc_4p_est}
    \Gamma_\mathrm{core}
    = \Gamma_\mathrm{core,\, tot}
    &- 2\pi\delta(\omega_{12}) \widetilde{\Sigma}_{12}^{k_1 k_2} \widetilde{\Sigma}_{34}^{k_3 k_4} \nnnl
    &- 2\pi\delta(\omega_{14}) \zeta \widetilde{\Sigma}_{14}^{k_1 k_4} \widetilde{\Sigma}_{32}^{k_3 k_2}.
\end{align}
Again, disconnected terms involving the self-energy are subtracted.
Since $\widetilde{\Sigma}$ is of order $\mcO(U^2)$ in the perturbative limit and decays in the large-frequency limit, the perturbative and asymptotic behavior of $\Gamma_\mathrm{core}$ is preserved.
Substituting \Eqs{eq:disc_2p}, \eqref{eq:disc_3p_est}, and \eqref{eq:disc_4p_est} into \Eq{eq:4p_4th}, we get the estimator
\eqref{eq:tot_4p_est} for the total vertex.

% ========================================
\section{Details of the NRG implementation}
\label{app:NRG}

For quantum impurity models, it is possible to compute
local multipoint correlators with NRG using a strategy described
in detail in Ref.~\cite{2021LeePRX}.
In short, MF or KF multipoint correlators are computed by convolving a set of MF or KF kernels, known analytically, with a set of so-called real-frequency partial spectral functions (PSFs), obtained via NRG~\cite{2021KuglerPRX,2021LeePRX}.
In this appendix, we describe some further refinements of this strategy, needed to obtain the results shown in the present paper.

\subsection{Barycentric binning}
\label{subsec:binning}

In essence, the NRG is a scheme iteratively diagonalizing
a discretized representation of a quantum impurity model \cite{2008BullaRMP}.
The states generated during the iterative diagonalization can be used to construct a complete set of approximate eigenstates for the interacting Wilson chain Hamiltonian, $\mc{H} \ket{\ul{i}} \approx E_{\ul{i}} \ket{\ul{i}}$.
These can be used in Lehmann-type representations for partial spectral functions (PSFs), with matrix elements $\mel{\ul{i}}{\mcO}{\ul{j}} = \mcO_{\ul{i}\ul{j}}$ and energy differences $E_{\ul{i}\ul{1}} = E_{\ul{i}} - E_{\ul{1}}$.
The Lehmann representation for an \lp\ PSF has the form of a sum over the transitions between these eigenstates, containing many discrete delta functions~\cite{2021KuglerPRX,2021LeePRX}:
\begin{equation} \label{eq:NRG_PSF}
    S[\mb{\mc{O}}](\veps_1, \ccdots, \veps_{\ell-1}) = \sum_{\ul{1},\ccdots,\ul{\ell}}
    \rho_{\ul{1}} \prod_{i=1}^{\ell-1} [\mcO^{i}_{\ul{i}\ul{i+1}} \delta(\veps_i - E_{\ul{i}\ul{1}})] \mcO^{\ell}_{\ul{\ell},\ul{1}}.
\end{equation}

These delta functions are collected by ``binning''~\cite{Peters2006NRG,Weichselbaum2007NRG,Weichselbaum2012NRG}, which shifts the energy $E_{\ul{i} \ul{1}}$ to the nearest grid point of a pre-determined grid.
The grid is chosen to have logarithmic spacing to capture the logarithmic discretization of the bath in NRG.
For $\ell > 2$, one must be able to capture the dependence of multiple frequencies of different magnitudes.
To this end, Ref.~\cite{2021LeePRX} introduced a ``slicing'' scheme, which encodes the frequency binning directly in the tensor representation of $\mcO$.
However, the slicing approach yields a factor $(N_{\veps})^{\ell - 2}$ to the computational cost, with $N_{\veps}$ the number of bins.
Thus, it is desirable to use as few bins as possible to reduce the computational cost.
On the other hand, using too coarse a grid leads to large errors from shifting energies for binning.

To minimize the error arising from binning with moderate computational cost, we use a barycentric binning scheme, which attributes the spectral weight to both adjacent bins (instead of the single nearest bin) in order to conserve the barycenter.
Hence, to bin a delta function $\delta(\veps - E_{\ul{2}\ul{1}})$, we first find the grid points $\ul{\veps}_{n}$ and $\ul{\veps}_{n+1}$ just below and above the transition energy: $\ul{\veps}_{n} \leq E_{\ul{2}\ul{1}} \leq \ul{\veps}_{n+1}$.
Then, we split the weight of the delta function to these two bins with weights inversely proportional to the distance in the logarithmic scale:
\begin{gather}
    \delta(\veps - E_{\ul{2}\ul{1}})
    \approx w \delta(\veps - \ul{\veps}_{n})
    + (1-w) \delta(\veps - \ul{\veps}_{n+1}), \nnnl
    w = \frac{\ln\abs{\ul{\veps}_{n+1}} - \ln\abs{E_{\ul{2}\ul{1}}}}{\ln\abs{\ul{\veps}_{n+1}} - \ln\abs{\ul{\veps}_n}}.
\end{gather}
Though one needs to consider two bins for a single spectral contribution, the total number of relevant bins does not double, but merely increase by $\sim 30\%$, since the bins are typically clustered~\cite{Lee2016NRG,2021LeePRX}.
Additionally, with the barycentric scheme, the binning error is largely suppressed, allowing for a much coarser frequency grid (cf.\ \Sec{sec:app_comp_params}), decreasing the overall computational cost.

\subsection{Log-Gaussian broadening}
\label{subsec:NRG-broadening}
To obtain smooth functions, the discrete PSFs obtained from the Lehmann representation need to be broadened.
Conversion from a binned, discrete PSF $S(\mb{\ul{\veps}})$ to a continuous one $\widetilde{S}(\mb{\veps})$ can be written as an integration involving a broadening kernel $B_i$:
\begin{equation}
    \widetilde{S}(\mb{\veps})
    = \nint d^{\ell-1}\ul{\veps} \, S(\mb{\ul{\veps}}) \prod_{i=1}^{\ell-1} B_i(\veps_i, \ul{\veps}_i').
\end{equation}

For 2p calculations, the usual choice of the broadening kernel  is a convolution of a log-Gaussian and a Fermi-Dirac function~\cite{2008BullaRMP,Weichselbaum2007NRG,Lee2016NRG,2021LeePRX}:
\begin{align} \label{eq:brodening_log_gaussian}
    B_{\rm LG+F}(\veps, \ul{\veps}) &= \nint d\veps' \delta_{\rm F}(\veps, \veps') \delta_{\rm LG}(\veps', \ul{\veps}),
    \nnnl
    \delta_{\rm sLG}(\veps', \ul{\veps})
    &= \frac{\Theta(\veps' \ul{\veps})}{\sqrt{\pi} \sigma_{\rm LG} \abs{\ul{\veps}}} \exp\left[ -\left( \frac{\ln\abs{\ul{\veps} / \veps'}}{\sigma_{\rm LG}} - \frac{\sigma_{\rm LG}}{4} \right)^2 \right],
    \nnnl
    \delta_{\rm cLG}(\veps', \ul{\veps})
    &= \frac{\Theta(\veps' \ul{\veps})}{\sqrt{\pi} \sigma_{\rm LG} \abs{\ul{\veps}}} \exp\left( -\frac{\ln^2\abs{\ul{\veps} / \veps'}}{\sigma_{\rm LG}^2} - \frac{\sigma_{\rm LG}^2}{4} \right),
    \nnnl
    \delta_{\rm F}(\veps, \veps') &= \frac{1}{2\gamma_{\rm F}} \left( 1 + \cosh \frac{\veps - \veps'}{\gamma_{\rm F}} \right)^{-1}.
\end{align}
In the first line on the right, LG stands for
either symmetric log-Gaussian (sLG), used for fermionic operators, or centered log-Gaussian (cLG), used for bosonic operators (following Ref.~\cite{2021LeePRX}).
The log-Gaussian broadening smooths the logarithmically-spaced discrete data resulting from discretization.
The subsequent Fermi broadening removes artifacts at low frequency $\veps \sim T$.
The broadening width is proportional to the discrete frequency in the former and fixed in the latter.
Correspondingly, the broadening parameter $\sigma_{\rm LG}$ is dimensionless, while $\gamma_{\rm F}$ has the dimension of energy.

In Refs.~\cite{2021KuglerPRX,2021LeePRX}, a Lorentzian kernel
\begin{align} \label{eq:brodening_lorentzian}
    B_{\rm L}(\veps, \ul{\veps}) &= \frac{1}{\pi} \frac{(\sigma_{\rm L} \abs{\ul{\veps}} + \gamma_{\rm L})}{(\veps - \ul{\veps})^2 + (\sigma_{\rm L} \abs{\ul{\veps}} + \gamma_{\rm L})^2}
\end{align}
was used for 3p and 4p calculations because, for the aIE used there, log-Gaussian broadening turned out to yield seemingly under-broadened results.

\begin{figure}[tbp]
    \centering
    \includegraphics[width=1.0\columnwidth]{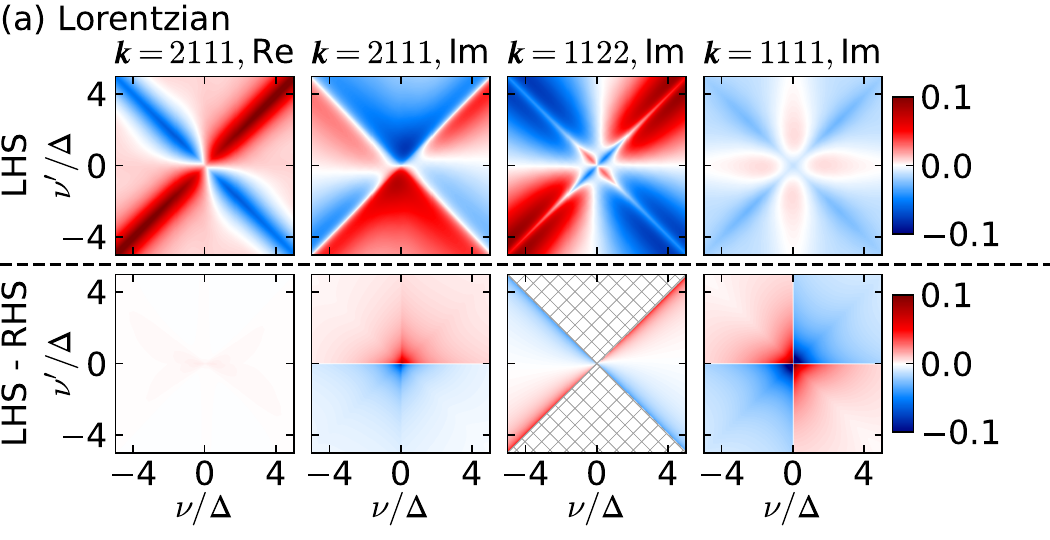}
    \includegraphics[width=1.0\columnwidth]{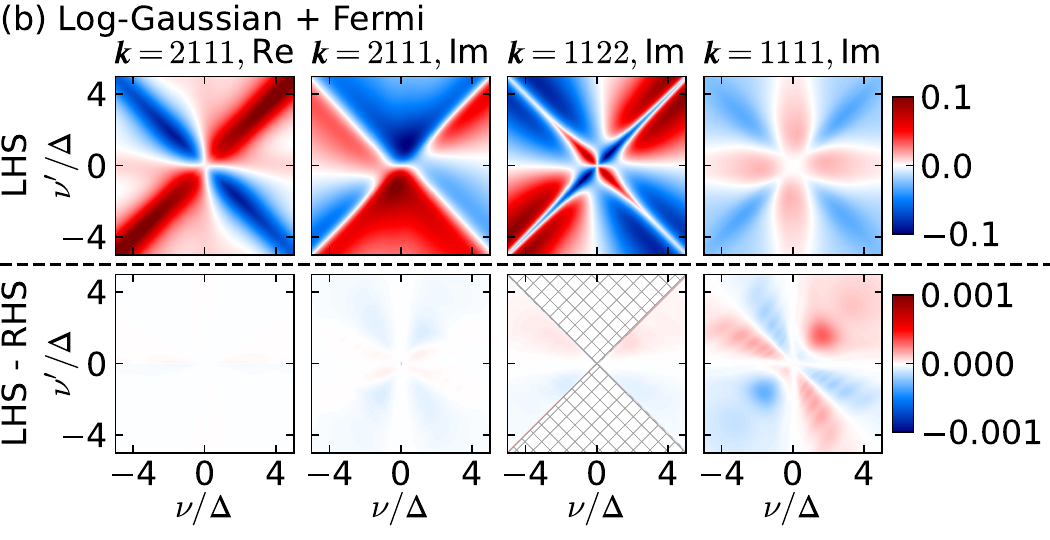}
    \caption{%
    FDRs of the KF 4p vertex, analogous to \Fig{fig:strong_FDR} but at weak interaction (parameters: same as for \Fig{fig:weak_vertex_overview}), obtained with sIEs and
    (a) a Lorentzian broadening \eqref{eq:brodening_lorentzian} with $\sigma_{\rm L} = 0.3$ and $\gamma_{\rm L} = 0.5T$, and (b) log-Gaussian and Fermi broadening \eqref{eq:brodening_log_gaussian} with $\sigma_{\rm LG} = 0.3$ and $\gamma_{\rm F} = 0.5T$.
    Note that the color scale for the panels showing FDR violation (lower row, LHS$-$RHS) is two orders smaller in (b) than in (a).
    }
    \label{fig:weak_FDT}
\end{figure}

Here, we find that, using the sIEs, log-Gaussian broadening does \textit{not} yield under-broadening.
On the contrary, we find that, as in the 2p case, log-Gaussian broadening is preferable to Lorentzian broadening for multipoint correlators because the Lorentzian kernel strongly over-broadens low-frequency features.
For example, \Fig{fig:weak_FDT} shows that the vertex obtained with Lorentzian broadening [\Fig{fig:weak_FDT}(a)] violates the FDT much more strongly than those obtained with log-Gaussian and Fermi broadening [\Fig{fig:weak_FDT}(b)].
Thus, we use the log-Gaussian and Fermi broadening \eqref{eq:brodening_log_gaussian} for all calculations.

In this work, the broadening parameters in \Eqs{eq:brodening_log_gaussian} were chosen as $\sigma_{\rm LG} \!=\! 0.3$ and $\gamma_{\rm F} \!=\! 0.5T$ throughout; these were the smallest values that remove the wiggles signalling under-broadening.
For the Lorentian broadening results, we used $\sigma_{\rm L} \!=\! 0.6$ and $\gamma_{\rm L} \!=\! 3T$ in Figs.~\ref{fig:weak_vertex_overview}(c) and \ref{fig:strong_vertex_overview}(c), which are the parameters used in Refs.~\cite{2021KuglerPRX,2021LeePRX}.

% =============================================
\subsection{Diagonalization of the kept density matrix}

As discussed in App.~B of Ref.~\cite{2021LeePRX}, the two sources of error in subtracting disconnected parts from the PSFs are
(i) finite off-diagonal elements of the density matrices in the kept sectors, and (ii) binning.
Effect of the latter is minimized by using the barycentric binning explained in App.~\ref{subsec:binning}.
To eliminate the former source of error, we re-diagonalize the kept-sector density matrices that are constructed by the standard full-density-matrix NRG.
The re-diagonalizing basis is taken as the kept energy eigenstates (overriding those obtained during the iterative diagonalization), and we identify the corresponding energy eigenvalues as the diagonal elements of the kept-sector Hamiltonian in the re-diagonalizing basis.

% =============================================
\subsection{Computational parameters} \label{sec:app_comp_params}
In NRG, the bath is discretized on a logarithmic grid with grid points $\pm D \Lambda^{-k-z}$, where $k \geq 0$ is an integer and $z \in (0, 1]$ is a shift parameter~\cite{Zitko2009NRG}.
PSFs computed with $z = 1/n_z, 2/n_z, \ccdots, 1$ are averaged to reduce discretization artifact.
In this work, we used $\Lambda=4$ and $n_z=4$, and kept $N_{\rm keep} = 200$ multiplets respecting U(1) charge and SU(2) spin symmetries in the iterative diagonalization of the Wilson chain, which amounts to keeping 440--479 states per iteration.

\begin{figure}[!htbp]
    \centering
    \includegraphics[width=1.0\columnwidth]{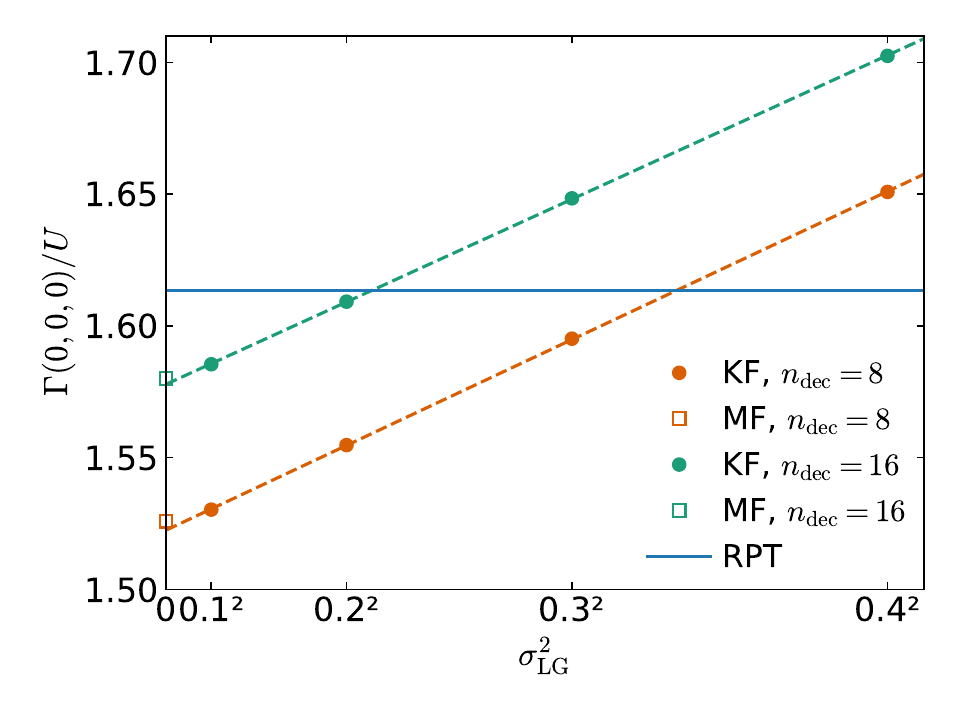}
    \caption{
    Dependence of the zero-energy KF vertex $\Gamma^{\rm causal}_{\uparrow\downarrow}(0,0,0)$ and the average of the MF vertices $\Gamma_{{\rm M,} \uparrow\downarrow}(\pm i\pi T,\pm i\pi T,0)$ in the AIM at strong interaction
    (parameters: same as for Fig.~\ref{fig:strong_vertex_overview}) on $\sigma_{\rm LG}^2$ and $n_{\rm dec}$.
    Dashed lines are linear fits of the KF results, shown in circles.
    Larger $n_{\rm dec}$ gives MF vertices in better agreement with RPT, while the KF vertices are further affected by broadening.
    }
    \label{fig:strong_FDT_compare_Hwidth}
\end{figure}

The transition energies are collected in logarithmically spaced bins~\cite{2021LeePRX}.
The bins are located at $\uline{\veps}[\pm m] = \pm 10^{(\abs{m} - 1) / n_{\rm dec}} \veps_{\rm min}$ for nonzero integer $m$ and $\uline{\veps}[0] = 0$.
We used $n_{\rm dec} = 8$ ($n_{\rm dec} = 16$) bins per decade for the AIM with weak interaction for \Figs{fig:weak_direct_vs_asym_vs_sym}, \ref{fig:weak_vertex_overview}--\ref{fig:weak_core_asym_vs_sym}, \ref{fig:weak_FDT} (strong interaction for \Figs{fig:strong_vertex_overview}--\ref{fig:strong_FDR}), and set $\veps_{\rm min} = T/20$.
Using $n_{\rm dec} = 16$ was needed to get the low-energy vertices for the AIM at strong interaction in better agreement with RPT as shown in Fig.~\ref{fig:strong_RPT}.
Figure~\ref{fig:strong_FDT_compare_Hwidth} shows that the agreement of the MF vertex at the lowest fermionic frequencies with the RPT vertex improves for the larger $n_{\rm dec} = 16$.

To evaluate an \lp\ sIE, one needs to compute $2^\ell$ \lp\ auxiliary correlators [\Eq{eq:4p_U4}], where each correlator involves $\ell !$ PSFs. For the 4p case, this amounts to $2^4 \times 4! = 384$ PSFs in total.
In practice, one can utilize symmetries to reduce the number of PSFs evaluated. Here, we use the SU(2) symmetry of the Hamiltonian and the permutation symmetry of the PSFs to reduce the number of independent 4p PSFs to 120.
These PSFs can all be computed in parallel.
Although we consider the single-orbital AIM only at half-filling, we did not exploit particle-hole symmetry.

\FloatBarrier % Forces bibliography to appear after all the figures.

\bibliography{main}

\end{document}